\documentclass{aa}  
\usepackage{amssymb}
\usepackage{mathtools}
\usepackage{cancel}
\usepackage{rotate}
\usepackage{lipsum}
\usepackage{graphicx}
\usepackage{txfonts}
\begin{document} 
   \title{The gas, metal and dust evolution in low-metallicity local and high-redshift galaxies}
\author{A. Nanni
          \inst{1}
          \and
          D. Burgarella\inst{1} \and P. Theulé\inst{1} \and B. Côté\inst{2} \and H. Hirashita\inst{3}
          }

   \institute{Aix Marseille Univ, CNRS, CNES, LAM, Marseille, France\\
              \email{ambra.nanni@lam.fr}
         \and
    Konkoly Observatory, Research Centre for Astronomy and Earth Sciences, MTA Centre for Excellence, Konkoly Thege Miklos 15-17, H-1121 Budapest, Hungary
\and
Institute of Astronomy and Astrophysics, Academia Sinica, Astronomy-Mathematics Building, AS/NTU, No. 1, Sec. 4,
Roosevelt Road, Taipei 10617, Taiwan
             \\
             }

   \date{Received...; accepted...}

% \abstract{}{}{}{}{} 
% 5 {} token are mandatory
 
  \abstract
  % context heading (optional)
  % {} leave it empty if necessary  
   {The chemical enrichment in the interstellar medium (ISM) of galaxies is regulated by several physical processes: star birth and death, grain formation and destruction and galactic inflows and outflows. Understanding such processes and their relative importance is essential in order to follow galaxy evolution and the chemical enrichment through the cosmic epochs, and to interpret the current and future observations.
Despite the importance of such topics, the contribution of different stellar sources to the chemical enrichment of galaxies, e.g. massive stars exploding as Type II supernovae (SNe) and low-mass stars, as well as the mechanisms driving the evolution of dust grains, e.g. grain growth in the ISM and destruction by SN shocks, remain controversial both on the observational and on the theoretical viewpoints.}
  % aims heading (mandatory)
   {In this work, we revise the current description of metal and dust evolution in the ISM of local low-metallicity dwarf galaxies and we develop a new description of Lyman Break Galaxies (LBGs) that are considered to be their high-redshift counterparts in terms of star formation, stellar mass and metallicity. Our goal is to reproduce the observed properties of such galaxies, in particular i) the peak in the mass of dust over the mass of stars (sMdust) observed within few hundreds Myrs; ii) the decrease of sMdust at later time.}
  % methods heading (mandatory)
   {The spectral energy distribution (SED) of dwarf galaxies and LBGs is fitted with the ``Code Investigating GALaxies Emission'' (CIGALE), through which the mass of stars, mass of dust and star formation rate are estimated. For some of the dwarf galaxies considered, the metal and gas content are available from the literature. We compute different prescriptions for metal and dust evolution in these systems (e.g. different initial mass functions for stars, dust condensation fractions, SN destruction, dust accretion in the ISM, inflow and outflow efficiency), and we fit the properties of the observed galaxies through the predictions of the models.}
  % results heading (mandatory)
   {Only some combinations of models are able to reproduce the observed trend and to simultaneously fit the observed properties of the galaxies considered.
   In particular, we show that i) a top-heavy initial mass function that favours the formation of massive stars and a dust condensation fraction for Type II SNe around 50\% or more help to reproduce the peak of sMdust observed after $\approx$100 Myrs since the beginning of the baryon cycle for both dwarf galaxies and LBGs; ii) galactic outflows play a crucial role in reproducing the observed decline in sMdust with age, and they are more efficient than grain destruction from Type II SNe both in local galaxies and at high-redshift; iii) a star formation efficiency (mass of gas converted into stars) of few per cent is required to explain the observed metallicity of local dwarf galaxies; iv) dust growth in the ISM is not necessary in order to reproduce the values of sMdust derived for the galaxies under study and, if present, the effect of this process would be erased by galactic outflows.}
  % conclusions heading (optional), leave it empty if necessary 
   {}


   \maketitle
%
%-------------------------------------------------------------------

\section{Introduction}
Understanding the metals and dust cycle in the interstellar medium (ISM) of galaxies is essential to study their properties and evolution, and to interpret different observations of local and very distant galaxies.
Indeed, dust shapes the spectra of galaxies by absorbing the stellar radiation in the ultraviolet (UV) and visible wavelengths and re-emitting it in the infrared (IR) bands, while the metals in the gas phase are responsible for various emission and absorption lines.
Different works have been devoted to the study of the dust content in galaxies in the Local Universe \citep{Lisenfeld98, Dwek98, Hirashita99, daCunha10, Zhukovska14, Schneider16, RR15, Popping17, Gioannini17, Devis17, McKinnon18, Ginolfi18, Devis19} as well as in a variety of high-redshift galaxies \citep{Graziani19, Burgarella20}, including quasars at $z>6$ \citep{Valiante14, Calura14, Mancini15, Aoyama17, Wang17} and sub-millimiter galaxies \citep{Rowlands14, Cai20}.

Dust is condensed in Type II supernovae (SNe II) remnants and in the dense outflows of low-mass stars at the end of their evolution, during the thermally pulsing asymptotic giant branch (TP-AGB) phase. 
Massive stars evolve on short time-scales ($< 30$ Myrs), while AGB stars need a longer time in order to start to enrich the ISM. Furthermore, massive and low-mass stars produce different types of elements and dust. 
Massive stars are factories of oxygen together with silicate and  dust \citep{Limongi18, Marassi19}, while low-mass stars can release large amount of carbon already at very low metallicities \citep{Nanni13, Nanni14}.
SN Ia, originating from exploding white dwarfs in binary systems, enrich the ISM predominantly with iron on a timescale comparable with the evolution of low mass stars that, contrary to massive stars, end their evolution as white dwarfs.
The role of SNe as dust producers is still debated. On the one hand, it has been argued that shock waves propagating after their explosion heat up the gas and can severely reduce the amount of dust condensed \citep{Nozawa11, Bocchio16, Bianchi07}. On the other hand, for SNe II, \citet{Gall14} have found evidence of the presence of 0.1--0.5~M$_\odot$ of large grains in SN2010jl, able to survive to the passage of the reverse shock.
In addition, recent observations from \citet{Matsuura19} of SN1987 have shown that dust might re-condense in the cooling gas that experienced the passage of the forward shock or survive to the passage of the reverse shock. In the former case, this might indicate that dust can be efficiently reformed after being destroyed by shocks. The presence of dust has been also suggested for Type Ia SNe \citep{Gomez12}.

Around TP-AGB stars, dust can be condensed in considerable amount \citep{Ventura12, Nanni13, Nanni14}, but since they evolve on longer time-scales their contribution may be irrelevant at early epochs.

In different investigations, dust growth in the ISM of galaxies is needed for explaining the large mass of dust observed that may be difficult to reproduce from stellar sources only \citep{Asano13, Zhukovska14, Michalowski15, Mancini15, Lesniewska19}. However, dust accretion in the ISM may be difficult to explain from the micro physical point of view \citep{Ferrara16}. Icy mantles can contribute in depleting elements from the gas phase, but they are expected to evaporate in presence of a strong radiation field \citep{Ferrara16}. Furthermore, the probability that two silicon-bearing molecules can encounter in the grain icy mantle is very low \citep{Ceccarelli18}. Some laboratory experiments showed that dust could accrete in low-temperature environments by addition of molecules on the surface of dust grains \citep{Rouille15}. This process might be prevented in space by the formation of icy mantles on the grain surface \citep{Ceccarelli18}, and grains may grow when the icy mantle evaporates \citep{Rouille20}. Theoretical studies and experiments also suggest charged grains and molecules could favour the process of grain growth \citep{Hollenstein00, Bleecker06}. In the astrophysical context, such a possibility has been investigated in the conditions typical of the Milky Way \citep{Zhukovska16} and further studies are needed. Grain growth also appears to be more efficient in galaxies which large stellar masses and related and to their environment (e.g. more efficient grain growth fro larger gas densities) \citep{Mancini15, Schneider16, Popping17, Graziani19}, or enhanced if the ISM is turbulent \citep{Mattsson19,Mattsson20}.
Because of these controversial results, it is important to critically examine dust growth in comparison with observational data.

From the Atacama Large Millimiter Array (ALMA) detection, in \citet{Burgarella20} we have proposed a template for the dust emission of Lyman Break Galaxies (LBGs) in the end of reionization that might be the first dust grains in the Universe. 
Along with rest-frame UV and optical data, this template is used to fit and model the multi-wavelength emission of these objects. 
From the derived physical parameters, we build a chemical evolutionary model that explains the (sub-)mm detection (or non-detection) 
of high-redshift LBGs in the end of reionization without any dust growth in the ISM. This model successfully explains different diagnostic diagrams (e.g. the evolution of the dust content in these galaxies).
The non-detection is explained by dust destruction from shocks produced by SNe plus removal in the circum- and intergalactic media by outflows.

In this paper, we interpret the observations of low-metallicity dwarf galaxies \citep{RR15} in the Local Universe from the Dwarf Galaxy Survey \citep[DGS][]{Madden13}, together with the LBGs which share similar properties in terms of star formation activity, metallicity and stellar mass. The spectral energy distribution (SED) of these galaxies is fitted as discussed in \citep{Burgarella20}. In particular, we aim to simultaneously reproduce the available constraints for such galaxies: total dust mass, star formation rate (SFR), metallicity, dust-to-gas ratio, gas fraction, age and, for few local galaxies, circumgalactic dust. By doing this, we identify the critical processes that drive the evolution of baryons in these galaxies: star formation efficiency, metal and dust enrichment from stellar sources (SNe II, TP-AGB stars and SNe Ia), dust destruction from SNe, galactic outflows, and the role of dust accretion in the ISM.

The paper is organised as follows. In Section \ref{Sample} we describe the two samples of galaxies. In Section~\ref{Sec:model} we explain the prescriptions adopted for following the gas and dust evolution in the ISM of galaxies. In Section~\ref{sec:method} we present the method for fitting the SED of galaxies and for comparing the model predictions with the properties derived. In Section~\ref{Sec:results} the results are provided, while Section~\ref{Sec:discussion} is devoted to discuss the results of this work in comparison with the ones in the literature. Our conclusions are provided in Section~\ref{Sec:conclusions}.

\section{Selected sample}\label{Sample}
In this work, we analyse two data samples of DGS galaxies and LBGs for which the SED fitting has been performed in \citet{Burgarella20} by means of the code \textsc{cigale} \citep{Boquien19}. 
In \textsc{cigale} the synthetic spectra are calculated from the energy balance between the photons absorbed and re-emitted by dust. Emission and absorption lines of gas are also included. The physical properties of the galaxies are estimated through the analysis of the likelihood distribution.
The DGS galaxies form a low-metallicity ($12 + \log(O/H) < 8.5$) sub-sample of the galaxies presented in \citet{RR15} for which the photometric data from the infrared to the sub-mm range are available. The SED is then built by adding the photometry at shorter wavelengths (UV, B and R) taken from the NASA Extragalactic Database (NED). We take care of using the most recent data corresponding to the integrated photometry of the galaxy. Employing data at short wavelengths allows us to better constrain the stellar mass and the SFR of the selected galaxies. LBGs were selected in the rest-frame UV and observed by ALMA. \citet{Burgarella20} used the detections to build an infrared SED template that is later used to re-fit all the objects individually. The SED fitting has been performed employing a Chabrier initial mass function (IMF) in \textsc{cigale} with stellar mass in the range $0.1 \leq M_{*} \leq 100$ M$_\odot$ \citep{BC03}. A top-heavy IMF is not available in the current version of the \textsc{cigale} and will be a future improvement in the code. 
For the galaxies selected, the code \textsc{cigale} provides a good fit between the synthetic and observed photometry (reduced $\chi^2\leq 5$) for 31 out of 48 DGS galaxies and for 18 LBGs which represents the two samples studied here.

%--------------------------------------------------------------------
\section{Model for chemical evolution of galaxies}\label{Sec:model}
In order to model the chemical evolution of the ISM of galaxies, we need to understand how the gas, metals and dust evolve with time, given the IMF of the stars, the star formation history (SFH) of the galaxy, and the theoretical yields from the stars.
\subsection{Basic assumptions and parameters}
The stars in the galaxies considered are assumed to form with a certain SFR.
Consistently with the SED fitting performed in \textsc{cigale}, the functional form of the SFR used through this work is the ``delayed'' one:
\begin{equation}\label{Eq:SFR}
    SFR = C_{\rm SFR} \frac{t}{\tau^2}e^{-t/\tau},
\end{equation}
where the SFR is given in $M_\odot yr^{-1}$, $\tau$ is a characteristic time-scale and $C_{\rm SFR}$ a constant defined in such a way that the final mass of stars is normalised to one solar mass:
\begin{equation}\label{star_norm}
C_{\rm SFR}=\frac{1 M_\odot}{\int_{0}^{t_{\rm end}} SFR(t^\prime) dt^\prime - M_{\rm ev}},
\end{equation}
where $M_{\rm ev}$ is the total mass released into the ISM from the evolving stars and $t_{\rm end}=13$ Gyrs. 

In this work we tested two different types of IMF for the stars: the Chabrier IMF \citep{Chabrier03} and a ``top-heavy'' IMF defined as:
\begin{equation}\label{eq:top-heavy_IMF}
    IMF= C_{\rm IMF} m^{-\alpha},
\end{equation}
with $\alpha=1, 1.35, 1.5$. The constant $C_{\rm IMF}$ is such that the integral of the IMF is normalised to one solar mass:
\begin{equation}
C_{\rm SFR}=\frac{1 M_\odot}{\int_{M_{\rm min}}^{M_{\rm max}} IMF(m^\prime) dm^\prime},
\end{equation}
where $M_{\rm min}$ and $M_{\rm max}$ are the minimum and maximum masses within which the theoretical metal yields are computed. Different metal yields available in the literature (characterised by diverse values of $M_{\rm min}$ and $M_{\rm max}$) are  tested in this work (see Section \ref{Sec:test_yields}).
A top-heavy IMF implies a higher number of SNe II: the lower is the value of $\alpha$ the larger is the number of SNe II.

\subsection{Gas and metals}
Given the gas and metal yields for stars, the IMF, the SFR and outflows, we are able to describe the evolution of the mass of gas and chemical element, $i$, in the ISM.
Under the aforementioned assumptions, the evolution of the mass of gas and metals is:

\begin{equation}\label{Mgas}
\begin{multlined}
    \frac{dM_{\rm g}}{dt}=\int_{M_{\rm L}}^{M_{\rm U}} IMF\times SFR (t-t^\prime) f_{\rm g}[M_*(t^\prime), Z(t-t^\prime)] dM - \\ -SFR(t)- ML\times SFR(t)+ ML\frac{I}{O}\times SFR(t),
\end{multlined}
\end{equation}

\begin{equation}\label{MZ}
\begin{multlined}
    \frac{dM_{\rm i}}{dt}=\int_{M_{\rm L}}^{M_{\rm U}} IMF\times SFR (t-t^\prime) f_{\rm i,g}[M_* , Z(t-t^\prime)] dM- \\-SFR(t) f_{\rm i, g}(t)- ML\times SFR(t) f_{\rm i, g}(t),
    \end{multlined}
\end{equation}

where the first term is the integral between the minimum ($M_{\rm L}$) and the maximum ($M_{\rm U}$) mass of stars having life-time $t^\prime$ that are enriching the ISM at time $t$ with a mass gas fraction or metal fraction $f_{\rm g}$ and $f_{\rm i, g}$, respectively. The stars enriching the ISM at a certain time were formed at $t-t^\prime$ and have metallicity $Z(t-t^\prime)$. The fractions of gas and metal ejected into the ISM depend on the metallicity and on the stellar mass $M_*$. The contribution to the ISM enrichment is weighed for the IMF. The second term of the equations is the
astration of gas and metals due to formation of stars. 
The chemical enrichment of the ISM and astraction due to star formation (given by the first and second terms of Eqs. \ref{Mgas}, \ref{MZ}) are computed by means of the \textsc{omega} code \citep[][One-zone Model for the Evolution of GAlaxies]{Cote17, Ritter18}.
Starting from these calculations, the overall evolution of gas and metals is followed by an external routine. At each time-step the galactic outflow is assumed to be proportional to the SFR through the ``mass-loading factor'' (ML), which assumes the outflow to be due to the stellar feedback in the ISM \citep{Murray05}. A corresponding mass of gas and metals is removed from the ISM. The code allows the possibility of introducing a galactic inflow of pristine gas provided by the last term of Eq. \ref{Mgas}. Such a term is proportional to ML and regulated by the ratio between the mass of inflowing and outflowing gas (I/O). The total initial mass of baryons in the galaxy, M$_{\rm bar}$, is another input quantity of our model. This quantity is entirely composed by gas at the beginning of the simulation, and it is partially converted into stars according to Eq. \ref{Eq:SFR}. We set $M_{\rm bar}$ to a multiple of the stellar mass (that we call $M_{\rm gas}$) which is normalised to 1 M$_\odot$ at 13 Gyrs, and we add to this quantity the mass of gas ejected by all the stars during their evolution, $M_{\rm ev}$ in Eq. \ref{star_norm}. The value of $M_{\rm ev}$ is evaluated with the \textsc{omega} code.
\subsection{Dust}
We assume that the dust formed in supernovae remnants, around evolved low-mass stars and in the ISM is composed predominantly by silicates (olivine and pyroxene), amorphous carbon dust (characterised by different fractions of sp$^2$ and sp$^3$ bonds) and metallic iron.
Carbon is also partially locked in hydrocarbons, such as in polycyclic aromatic hydrocarbons.
Iron dust can be partially included in silicates or being in different chemical forms (e.g. iron oxides, iron carbides, iron sulphides, pyrrhotite). Throughout this work we consider iron-free silicates (olivine and pyroxene), amorphous carbon dust and metallic iron. Since we are interested in estimating the total mass of dust rather than its specific chemical composition, the presence of dust species of chemical composition different from the one here considered (such as iron-rich silicates, or other dust species that includes carbon or iron atoms) are not expected to significantly modify the predicted mass of dust with respect to our estimates.

Starting from the calculations from \textsc{omega}, we then follow the evolution of the dust content in the ISM, by assuming at each time step that a fraction of the available metals is condensed into dust grains. In addition, we take into account different physical processes that change the mass of dust as a function of time: outflows, dust destruction from SN shock waves that propagate in the ISM and dust growth in the ISM. Consistently with the calculations of gas and metal evolution, we assume that the inflow material is composed by pristine gas that does not change the total content of dust.
For each dust species, the variation of the dust content in the galaxy will be given by:
\begin{equation}\label{eq:dust}
    \frac{dM_{\rm d, i}}{dt}=f_{\rm i, d}\frac{dM_{\rm i,d}}{dt} - ML\times SFR(t) \delta_{\rm i, d}(t) - \frac{dM_{\rm d, i, destr}}{dt}+\frac{dM_{\rm d, i, gr}}{dt},
\end{equation}

where the first term of the equation takes into account the metal enrichment from the composite stellar population and the astraction computed with \textsc{omega} and $f_{\rm i, d}$ is the condensation fraction of dust\footnote{We define the condensation fraction of a dust species as the ratio between the number of atoms condensed into the dust grain over the total that can condense for the least abundant among the atomic species that forms a certain dust species.}. Different condensation fractions for SNe and TP-AGB stars are assumed. The quantity $dM_{\rm i,d}/dt$ in the first term is the maximum mass of each dust species that can be formed and it is computed by dividing $dM_{\rm i}/dt$ in Eq. \ref{MZ} by the mass of each element and by multiplying for the mass of the nominal monomer of dust given its stoichiometric formula. A fraction of dust is lost due to outflows and it is provided by the second term in the right hand of Eq. \ref{eq:dust}, where $\delta_{\rm i, d}(t)$ is the dust-to-gas ratio at time ``t''.
The terms $dM_{\rm d, i, destr}/dt$ is the variation of the mass of dust due to the dust destruction from SN shocks and $dM_{\rm d, i, gr}/dt$ is dust growth in the ISM. We discuss the different mechanisms separately in the next sections. 
\subsubsection{Dust condensation fractions for TP-AGB stars and SNe}
For TP-AGB stars, the chemistry of dust is mainly determined by the number of carbon atoms in the stellar atmosphere over the number of oxygen ones (C/O ratio).
For $C/O<1$ all the carbon atoms are locked in the very stable carbon monoxide molecules (CO) and the remaining oxygen is present in the atmosphere to form molecules and dust. The star is classified as oxygen-rich. For $C/O>1$ an excess of carbon is present in the atmosphere from which molecules and dust are formed. The star is classified as carbon-rich. Normally, the value of the $C/O$ is less than one, but during the TP-AGB evolution values of $C/O>1$ are attained as a consequence of third dredge-up events that follow the thermal pulse \citep{Herwig05}.
Around oxygen-rich TP-AGB stars, a certain fraction of silicon, oxygen and magnesium are used to build silicate dust (mainly olivine and pyroxene) and, in smaller amounts, metal oxides, while during their carbon-rich phase, the main dust species produced is carbon dust. Other species such as silicon carbide and metallic iron are also produced but in smaller amounts \citep{Hofner18}.
The chemistry of the dust formed around SNe II is instead more uncertain, since it depends on the mixing between the C and O layers which determines the amount of such species locked into CO molecules \citep{Nozawa03}. 

We here consider two different sets of condensation fractions for the different species formed around TP-AGB stars or in SN remnants. Around low-mass stars, the condensation fraction of silicate and carbon dust can reach up to 50-
60\% (or slightly more) during the superwind phase when the star loses most of its mass \citep{Ventura12, Nanni13, Nanni14}, while for SN remnants the dust-to-metal mass fraction is estimated to be between 30 and 60\% without considering the effect of dust destruction from the reverse shock and/or possible dust reformation \citep{Marassi19}. 

Following \citet{Nanni13}, for each time-step we assume the following condensation fractions for the different species for TP-AGB stars: $f_{\rm py}= 0.3$ for pyroxene, $f_{\rm ol}= 0.3$ for olivine, $f_{\rm car} = 0.5$ for carbon dust and $f_{\rm ir} = 0.01$ for metallic iron. 
Different condensation fractions for the dust species formed in SNe II are considered. Given the uncertainties in the dust condensation fraction for SNe II, we tested a range of values between $0.1$ and close to $1$ for all the dust species. In case the condensation fraction is close to the maximum, we set the condensation fraction of carbon equal to $0.5$, since some should be available in the atmosphere to form CO molecules \citep{Spyromilio01}.

Despite being a simplified approach, since dust yields are not consistently computed in the circumstellar envelopes of TP-AGB stars and in SN remnants for any of the metal yields considered in this work, such a choice allows us to test the results for different choices of the metal yields available in the literature.
\subsubsection{Dust destruction from SN shock waves}
We adopt a macroscopic description for dust destruction due to the passage of SN forward shocks in the ISM. We assume that all the dust species are destroyed with the same efficiency by such a process.
The time-scale which regulates the destruction from SN shock waves is given by:
\begin{equation}\label{eq:tdestr}
    \tau_{\rm d}= \frac{M_{\rm g}(t)}{\epsilon R_{\rm SN}(t)M_{\rm swept}},
\end{equation}
where $M_{\rm g}$ is the gas mass in the ISM which evolves according to equation \ref{Mgas}. The quantity $M_{\rm swept}$, given in M$_\odot$, is a model parameter that provides how much of the ISM mass is swept by each SN event, and $\epsilon$ is the destruction efficiency. The quantity $R_{\rm SN}$ is instead the SN rate that depends on the SFH and the IMF. 
For each dust species $i$ the destruction term is therefore given by:
\begin{equation}\label{eq:rdestr}
\frac{dM_{\rm d, i, destr}}{dt}=\frac{M_{\rm d, i}}{\tau_{\rm d}}.
\end{equation}
In this work we consider two  values of $M_{\rm swept}=1000, 6800$ M$_\odot$, while $\epsilon=0.1$ in all the cases \citep{McKee89}. $M_{\rm swept}=6800$ M$_\odot$ represents the standard case in the literature \citep[e.g.][]{Dwek07}, while the assumption $M_{\rm swept}=1000$ M$_\odot$ has been also considered on the basis of the calculations performed by \citet{Nozawa06}.  
\subsubsection{Dust growth in the ISM}\label{Section:dust_growth}
\begin{table*}
\centering
\caption{Overall reactions adopted for modelling dust accretion in the ISM. The molecules and atoms in the gas phase
from which each dust species is formed are also listed.}
\begin{tabular}[width=0.95\textwidth]{l c c}
\hline
Species & Reactions & Molecules \\
\hline
Olivine & $ 2 {\rm Mg} + {\rm SiO} + 3{\rm H_2O} \rightarrow {\rm Mg}_2 {\rm SiO}_4(s) + 3{\rm H}_2$ & Mg, SiO, H$_2$O \\
Pyroxene & $ {\rm Mg} + {\rm SiO} + 2{\rm H}_2{\rm O}\rightarrow {\rm Mg} {\rm SiO_3(s)} + 2{\rm H}_2$  & Mg, SiO, H$_2$O   \\
Fe(s) & $\rm {Fe\rightarrow Fe(s)}$ & Fe \\
C(s) & $\rm{C_2H_2 \rightarrow 2C(s) + H_2}$ &  C$_2$H$_2$ \\
\hline
\end{tabular}
\label{Table:reactions}
\end{table*}
We here analyse the efficiency of the dust growth process in the standard framework where molecules are added from the gas phase to the grain, forming the bulk of silicates (olivine or pyroxene), amorphous carbon or metallic iron. No icy mantle formation is considered in this work. Different parameters are adopted in order to model such a process. Due to the lack of detailed laboratory measurements of grain growth at low temperature, many of these parameters are uncertain.
We treat such a process following grain growth and the variation of the size of each dust species as a function of time, as calculated in \citet{Nanni13, Nanni14}. We assume accretion to start on pre-existing grains (seed nuclei) on which other molecules from the gas phase are added to form the bulk of the grain. Such grains can be reprocessed in the ISM by SN shocks or aggregate in dense clouds modifying their size distribution. 
For each of the dust species the overall reaction for the formation of the grain needs to be chosen a priori. The reactions assumed in this work are provided in Table \ref{Table:reactions}.
Carbon dust is formed from acetylenic radicals (C$_2$H$_{(0<=y<=4)}$) on the grain surface. We here only consider the formation of carbon through the addition of C$_2$H$_2$ since we are interested in estimating the total amount of dust formed in the ISM. The exact pathway of grain formation is uncertain and such an investigation is beyond the scope of this paper.
The possible destruction of dust due to the sputtering of the grains by $H_2$ molecules is neglected in this work, since this process is efficient above $\approx 1000$~K \citep{Gail99, Nanni13}. 
The variation of the grain size is regulated by the equation:
\begin{equation}
    \frac{da_{\rm i}}{dt}=V_{\rm 0,i} J^{\rm gr}_{\rm i}, 
\end{equation}
where we assume the initial size of dust grains to be 0.01 $\mu$m \citep{Hirashita11} which is derived from the size distribution of grains in our Galaxy \citep{Mathis77}. If smaller grains are initially available in the ISM dust growth in the ISM may occur more rapidly \citep{Hirashita12}. The quantity $V_{\rm 0,i}$ is the volume of the nominal monomer of each dust species and the growth rate, $J^{\rm gr}_{\rm i}$, is instead evaluated as:
\begin{equation}
    J^{\rm gr}_{\rm i}=\min\Big(\frac{s_i}{s_j}\alpha_{\rm i} n_{\rm j} v_{\rm th, j}\Big),
\end{equation}
where $s_{\rm j}$ and $s_{\rm i}$ are the stoichiometric coefficient in the reaction for the species in the gas phase and for the monomer of dust, respectively. The quantity $\alpha_{\rm i}$ is a unitless parameter known as ``sticking coefficient'' and represents the probability that a species in the gas phase sticks on the grain surface. The value of the sticking coefficient varies from 0 to 1. On the basis of recent experiments at low temperature, the reaction between SiO molecules and the grain surface is expected to have zero energy barrier that corresponds to a sticking coefficient equal to 1 \citep{Rouille15}. We here assume $\alpha_{\rm i}=1$ for all the dust species. Such an assumption, however, represents an upper limit of the efficiency of grain accretion.
The quantity $n_{\rm j}$ is the number density of the species $j$ in the gas phase involved in the formation of the dust grain, computed as:
\begin{equation}
n_{\rm j}=\epsilon_{\rm j}n_{\rm H},
\end{equation}
where $\epsilon_{\rm j}$ is the abundance of the element $j$ and we adopt $n_{\rm H}=10^3$~cm$^{-3}$ \citep{Hirashita00, Hirashita12}. 
For larger values of the gas density grain growth would proceed more rapidly.
At each time-step the quantity $\epsilon_{\rm j}$ in the gas phase is computed by subtracting the fraction of each element condensed into dust grains from the total available. The incorporation of CO molecules in grains is not considered. Carbon monoxide is a very stable molecule that does not contribute to the accretion of dust grains. The fraction of carbon monoxide in the gas phase is estimated to be around 40\% of the total amount of carbon \citep{Lodders10,Agundez13}. The abundance of SiO, Mg, H$_2$O, C$_2$H$_2$ and Fe in the gas phase is therefore estimated respectively as the abundance of Si, Mg, O, C and Fe neither locked in dust grains nor in CO molecules.
The quantity $v_{\rm th, j}$, is its thermal velocity, given by:
\begin{equation}
    v_{\rm th, j}= \sqrt{\frac{k_{\rm B} T_{\rm gas}}{2\pi m_{\rm j}}},
\end{equation}
where $k_{\rm B}$ is the Boltzmann constant, $T_{\rm gas}=25$~K the gas temperature and $m_{\rm j}$ the mass of the gas species.
From the grain size it is possible to directly compute the total amount of dust produced in the dense phase of the ISM.
\begin{equation}
    \frac{dM_{\rm growth, i}}{dt}=\frac{4\pi}{3}\frac{da_{\rm i}}{dt}a_{\rm i}^2 \rho_{\rm i} n_{\rm s, i},
\end{equation}
where $\rho_{\rm i}$ is the mass density of the dust species $i$, $n_{\rm s, i}$ is the number of seed nuclei. We estimate the total number of seed particles $n_{\rm s, i}$ by dividing the mass of each dust species by the mass of a dust grain which depends on its size, as done in \citet{Asano13}. With this choice we implicitly assume that all the dust grains already present in galaxies can act as seed particle on which other molecules can accrete. The number of seed nuclei is expected to change as the ISM is progressively enriched by newly formed dust grains of a certain typical size and will depend on the different physical processes that modify the number of grains (e.g. shattering, coagulation). The larger is the number of seed nuclei the faster dust growth in the ISM is. If star formation mainly occurs at the beginning of the cycle (which is the case for the galaxies considered here), a further increase of the mass of dust in the ISM is mainly due to grain growth in the ISM.
In this regime, the number of seed nuclei can be assumed to be roughly constant (if new seeds are not formed in the ISM) and equal to the total mass of a certain dust species divided by the current mass of the grain for each dust species, $i$ as computed in \citet{Asano13}.  

\section{Method}\label{sec:method}
The observed photometry for galaxies is fitted with the code \textsc{cigale}. The features of the code are explained in \citet{Boquien19} while the assumptions adopted for computing the synthetic spectra are discussed in \citet{Burgarella20}.
The SED fitting procedure of the sources described in Section~\ref{Sample} allows us to derive the best value for the parameter $\tau$ in Eq.~\ref{Eq:SFR} for each of the galaxies, as well as the total mass of stars and dust, and the SFR. 
Furthermore, throughout this paper we use the specific mass of dust (sMdust), i.e. the mass of dust divided by the mass of stars, specific SFR (sSFR), i.e. the SFR divided by the mass of stars, age. For DGS galaxies other properties are available from the literature: metallicity, gas fraction, dust-to-gas ratio, circumgalactic dust fraction. 

The properties derived are compared with the output of a set of chemical evolutionary tracks obtained by varying the input parameters described in Section~\ref{Sec:model}.
The different combinations of the parameters adopted in Section~\ref{Sec:model} (IMF, condensation fraction, stellar mass range of the SN progenitors, efficiency of the galactic outflow, SN destruction, and initial gas content in the ISM) are listed in Table~\ref{Table:models}.
For all the models (one at each time-step) inside the different chemical evolution tracks, we find the residual between the properties of the galaxies predicted by the chemical evolution calculation and the ones derived from the SED fitting of \textsc{cigale}:
\begin{equation}\label{chi2}
    R^2_{\rm gal}= \sum_{\rm k}^{N_{\rm k}} \frac{(f_{\rm obs, k}-f_{\rm th, k})^2}{\sigma_{\rm k}^2},
\end{equation}
where $k$ represents the property of the galaxy. 
 For the error on the age we adopted the maximum value between the uncertainty given by \textsc{cigale} and 10\% of the age. The metallicity and the gas mass are taken from \citet{RR13, RR14} for DGS galaxies. We compute the total mass of gas by selecting from \citet{RR14} the atomic and molecular hydrogen mass. The total hydrogen gas is then multiplied by the mean atomic weight of the galaxy ($\mu_{\rm gal}$). The molecular hydrogen mass is derived from a metallicity-dependent conversion factor between the CO and H$_2$ masses (M$_{H2, z}$). All the data adopted are provided in Table \ref{Table:gal_prop}. In case only upper limits are available, we adopt the maximum possible value. The error associated to the gas fraction and to the dust-to-gas is assumed to be 100\% given the uncertainties affecting the conversion between CO and H$_2$ abundances.
  For LBGs, we do not have direct measurements of the metallicity and of the gas content, and therefore we do not use such quantities for fitting the galaxies. 
 The estimates of the circumgalactic dust fraction are taken from \citet{McCormick18} for three of the DGS galaxies in the sample that are He2-10, NGC 1569, NGC 5253.

From the residual $R_{\rm gal}$ we calculate the probability density distribution (in analogy with the $\chi^2$ distribution), $p=\exp{(-R^2_{\rm gal}/2)}$. Each of the properties of the galaxies is then estimated through the likelihood analysis by computing its average value and standard deviation by using as weights the probability density distribution among all the calculated models. The predicted values in different diagnostic diagrams (e.g. sMdust vs sSFR) are compared with the ones derived from the SED fitting with \textsc{cigale} or taken from the literature. This procedure allows to constrain the properties of individual galaxies.
\begin{table*}
\begin{center}
\caption{List of parameters adopted in the simulations of metal and dust evolution described in Section~\ref{Sec:model}. First tests are run in order to select the reference parameters adopted to run systematic calculations. The stellar mass produced after 13 Gyrs is always normalised to 1~M$_\odot$. Different theoretical metal yields are tested in Section~\ref{Sec:test_yields}.}
\label{Table:models}
\begin{tabular}{l l l}
\hline
First tests &  &\\
\hline
 $\tau$ [Myrs] & 83, 300 & \\
 M$_{\rm stars}$ [M$_\odot$] & 1 & \\
 M$_{\rm gas}$ [M$_\odot$] & (2,10,20,100) $\times M_{\rm star}$ &\\
 M$_{\rm bar}$ [M$_\odot$] & $M_{\rm gas} + M_{\rm ev}$ &\\
 M$_{\rm swept}$  [M$_\odot$] & 1000, 6800 & \\
ML & (0, 0.5, 0.6, 0.65, 0.8)$\times M_{\rm gas}$ & \\
I/O & 0, 0.2, 0.5 & \\
 IMF    & Chabrier & \\
  & $\propto$ M$^{-\alpha}$, $\alpha=-1,-1.35,-1.5$ & \\
 SN condensation fraction &   $f_{\rm py}=1$, $f_{\rm ol}=0$, $f_{\rm ir}=1$, $f_{\rm car}=0.5$ & \\
  & $f_{\rm py}=0.5$, $f_{\rm ol}=0$, $f_{\rm ir}=0.5$, $f_{\rm car}=0.5$ & \\
  & $f_{\rm py}=0.25$, $f_{\rm ol}=0$, $f_{\rm ir}=0.25$, $f_{\rm car}=0.25$ & \\
   & $f_{\rm py}=0.10$, $f_{\rm ol}=0$, $f_{\rm ir}=0.10$, $f_{\rm car}=0.10$ & \\
% & $f_{\rm py}=0.5$, $f_{\rm ol}=0$, $f_{\rm ir}=0.5$, $f_{\rm car}=0.5$ & \\
 TP-AGB condensation fraction & $f_{\rm py}=0.3$, $f_{\rm ol}=0.3$,  $f_{\rm ir}=0.01$, $f_{\rm car}=0.5$ & \\
Dust growth in the ISM & YES, NO & \\
\hline
Theoretical metal yields &  & \\
%\hline
Stellar source & Data set and denomination  & mass range in M$_\odot$\\
\hline
Type II SNe & \citet{Kobayashi06} - K06 & [13-40] \\
 & \citet{Nomoto13} - N13 & [13-40] \\
&  \citet{Limongi18} - LC18 &  [13-120] \\
&   \citet{Ritter18} - R18 &   [12-25] \\
TP-AGB & \citet{Cristallo15} - C15 &  [1-7] \\
 &  \citet{Karakas10} - K10 &  [1-6] \\
 &  \citet{Ritter18} - R18 &  [1-7] \\
Pop III stars & \citet{Heger10} & [10-100]  \\
  &  \citet{Nomoto13} & [13-300]  \\
Type Ia SN & \citet{Iwamoto99} &  - \\
\hline
Systematic calculations & &\\
\hline
 $\tau$ [Myrs] & 83, 300 & \\
 M$_{\rm stars}$ [M$_\odot$] & 1 & \\
 M$_{\rm gas}$ [M$_\odot$] & [10,100] $\times M_{\rm star}$, spacing 10 & \\
 M$_{\rm bar}$ [M$_\odot$] & $M_{\rm gas} + M_{\rm ev}$ & \\
 M$_{\rm swept}$  [M$_\odot$] & 1000, 6800 & \\
 ML & [0, 0.95]$\times$M$_{\rm gas}$, spacing 0.05 & \\
 I/O & 0 & \\
 IMF & $\propto$ M$^{-\alpha}$, $\alpha=-1,-1.35,-1.5$ & \\
 SN condensation fraction & $f_{\rm py}=0.25$, $f_{\rm ol}=0$, $f_{\rm ir}=0.25$, $f_{\rm car}=0.25$ & \\
  & $f_{\rm py}=0.5$, $f_{\rm ol}=0$, $f_{\rm ir}=0.5$, $f_{\rm car}=0.5$ & \\
 & $f_{\rm py}=1$, $f_{\rm ol}=0$, $f_{\rm ir}=1$, $f_{\rm car}=0.5$ & \\
 AGB condensation fraction & $f_{\rm py}=0.3$, $f_{\rm ol}=0.3$,  $f_{\rm ir}=0.01$, $f_{\rm car}=0.5$ & \\
Dust growth in the ISM & NO & \\
\hline
\end{tabular}
\end{center}
\end{table*}
\section{Results}\label{Sec:results}
\subsection{Properties of galaxies derived from the SED fitting}\label{Sec:prop}

\begin{figure}
\centering
\includegraphics[scale=0.5]{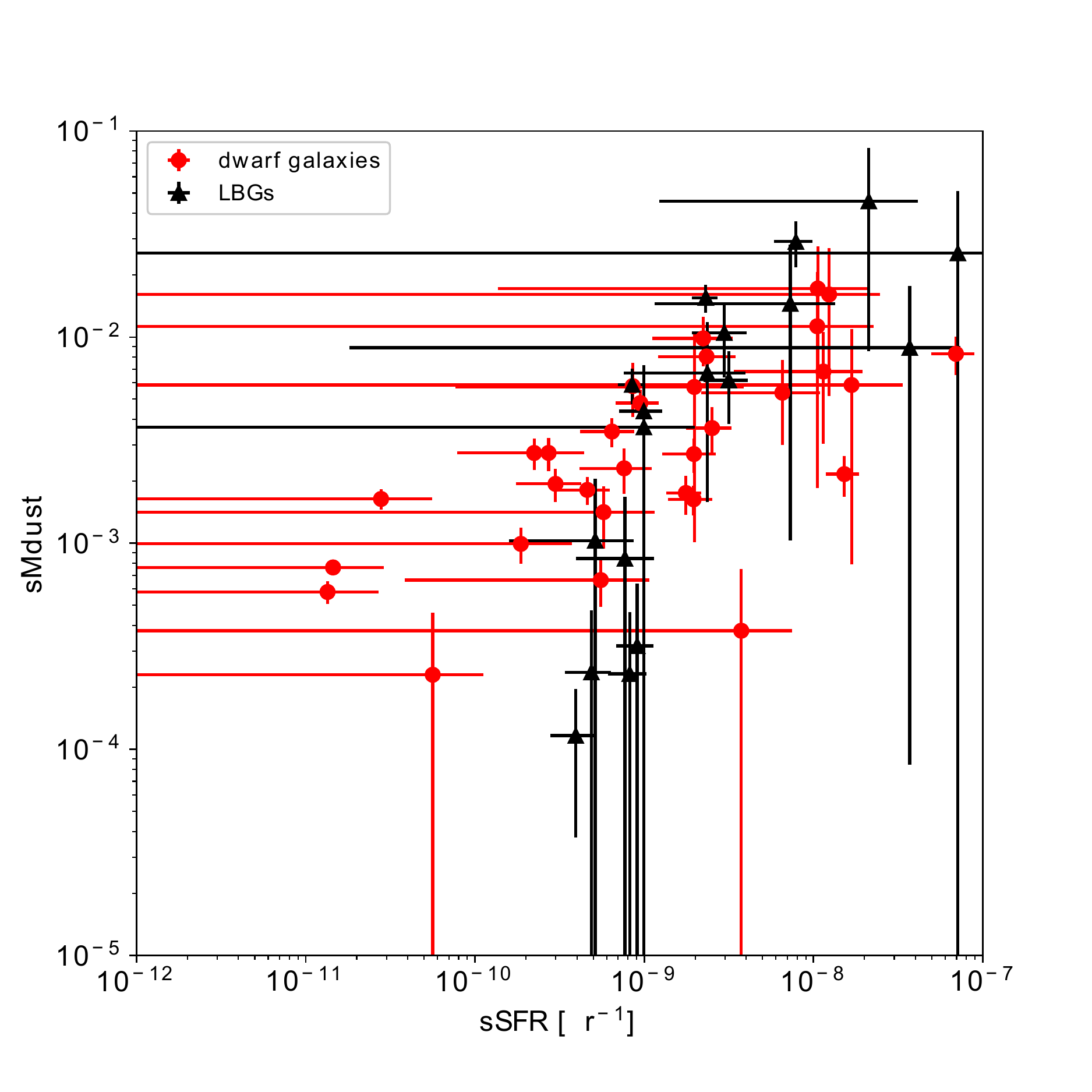}
\caption{sMdust against sSFR derived from the best fit performed with the code \textsc{cigale} for DGS galaxies (red dots) and LBGs (black triangles). See also \citet{Burgarella20} for all the details.}
\label{Fig:obs_DGS_LBGs}
   \end{figure}

\begin{figure}
\centering
\includegraphics[scale=0.35]{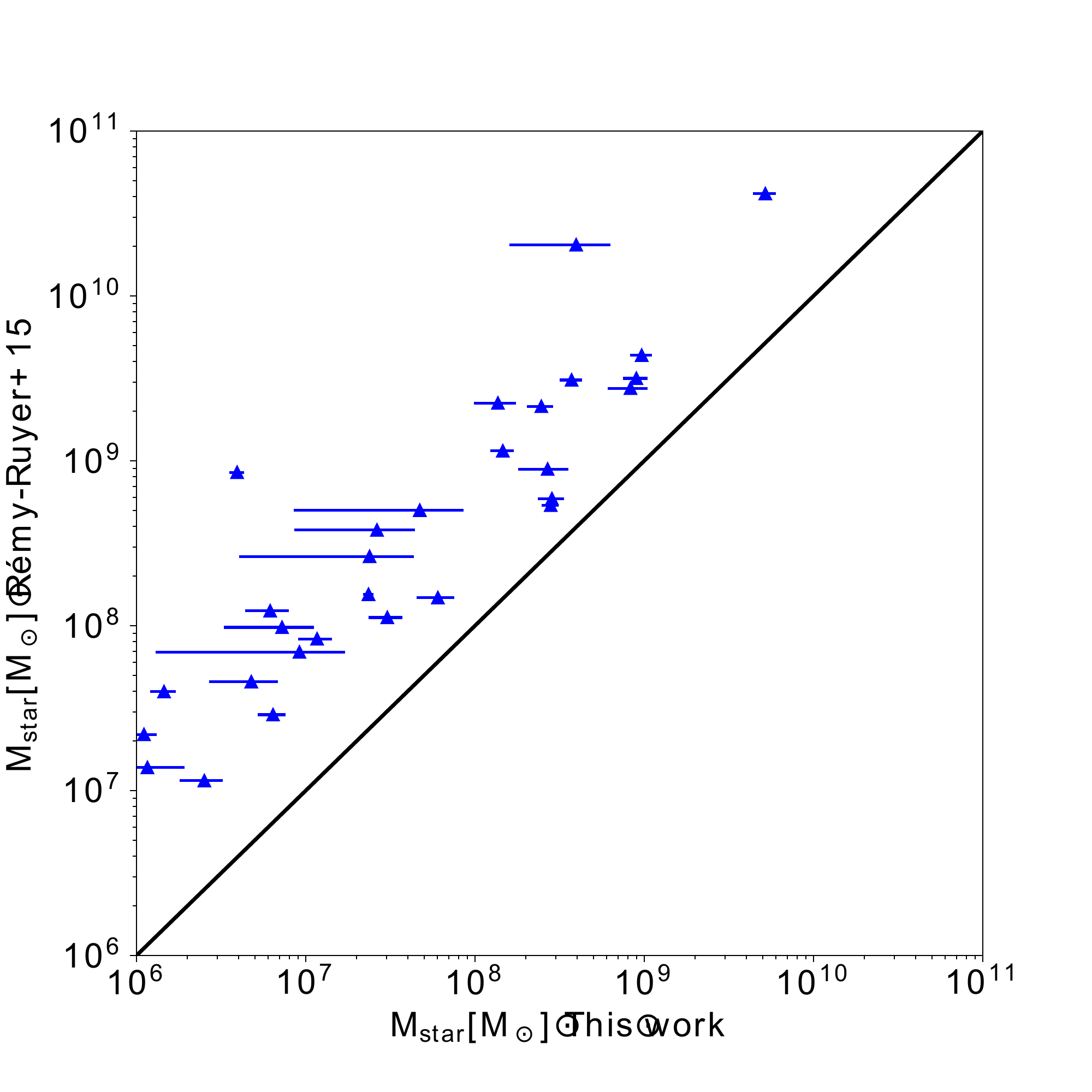}
\includegraphics[scale=0.35]{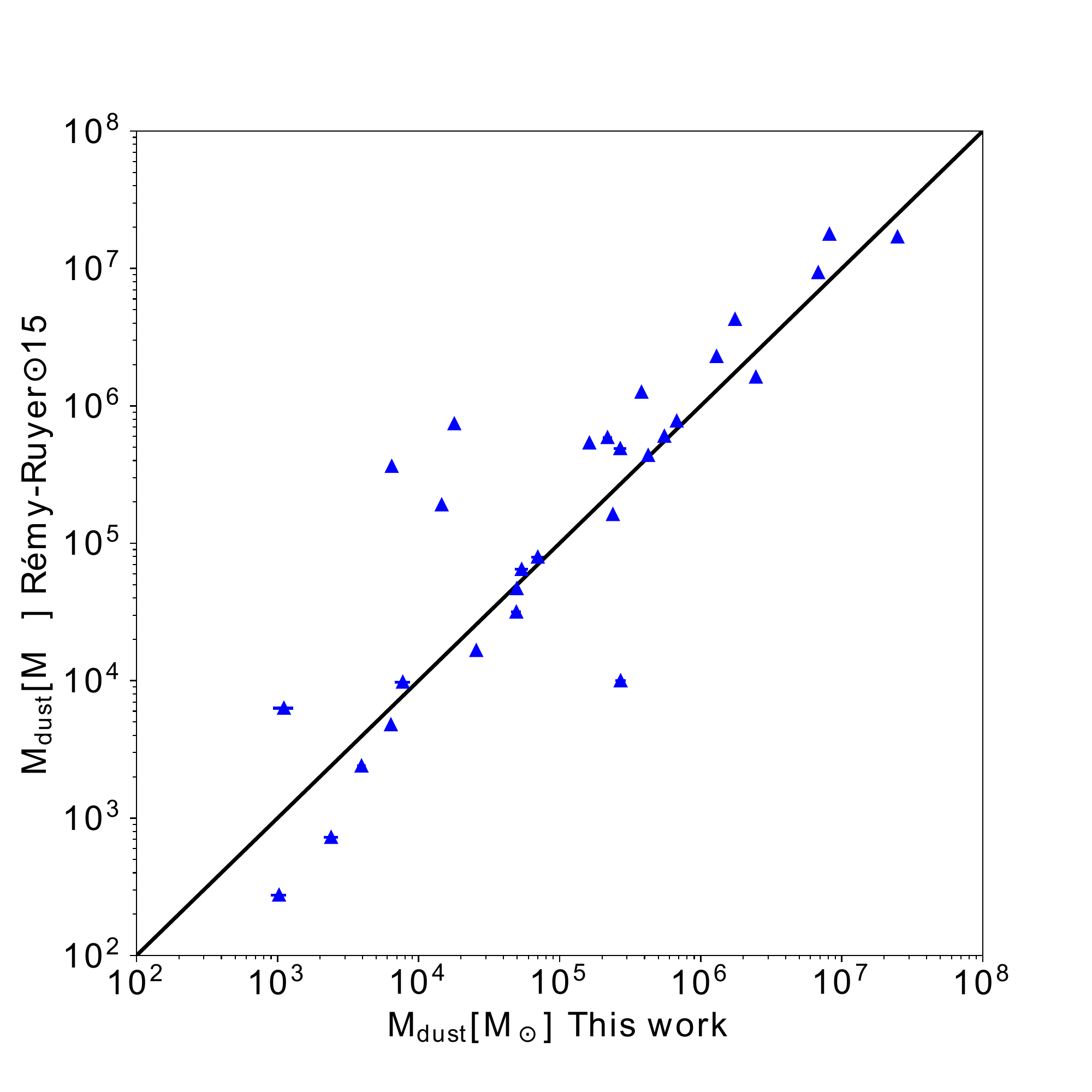}
\includegraphics[scale=0.35]{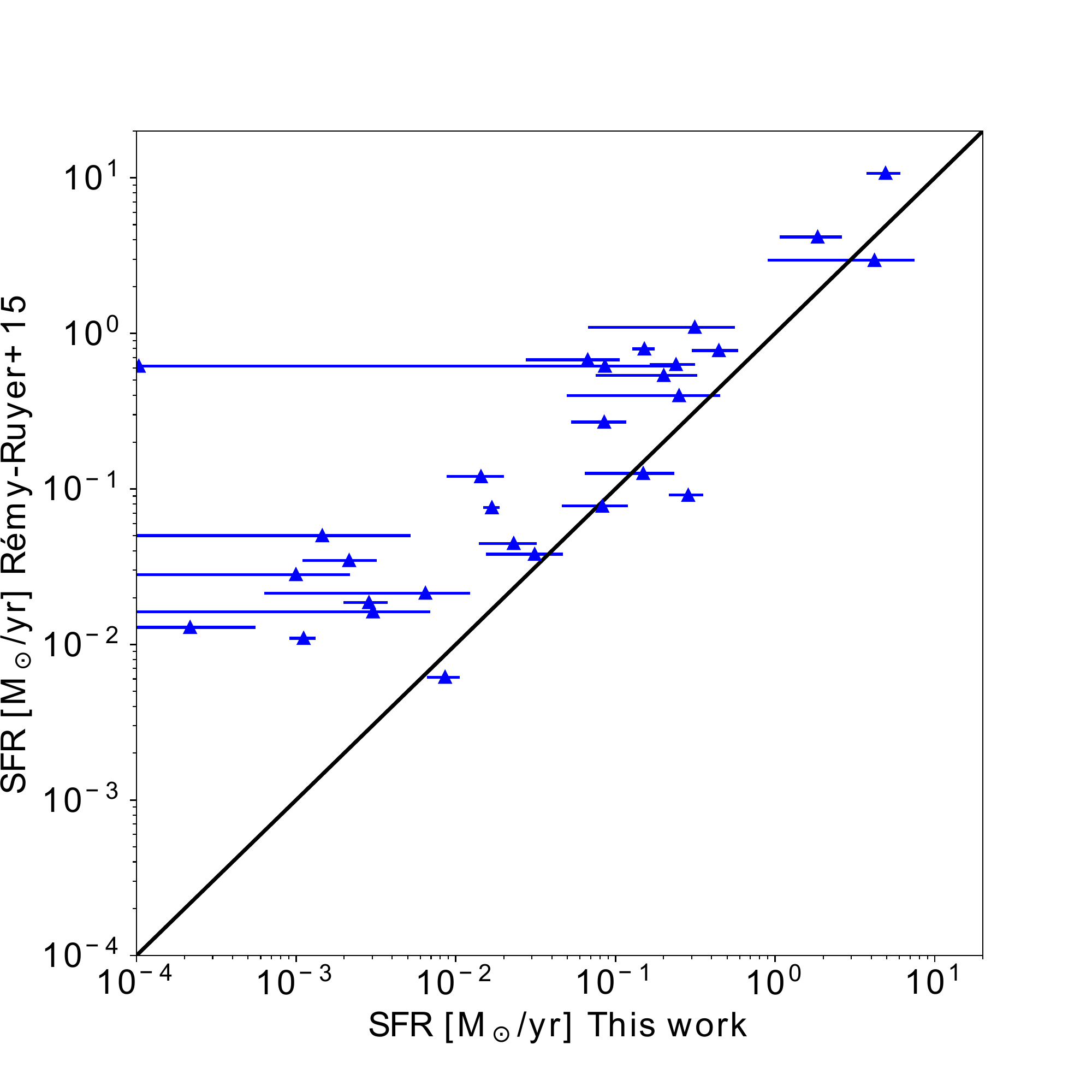}
\caption{Upper panel: stellar mass derived \citet{RR15} against the one derived in \citet{Burgarella20}. Middle and lower panels: same as in the upper panel but for the mass of dust and SFR, respectively.}
\label{cfr}
   \end{figure}
   
   \begin{figure}
\centering
\includegraphics[scale=0.5]{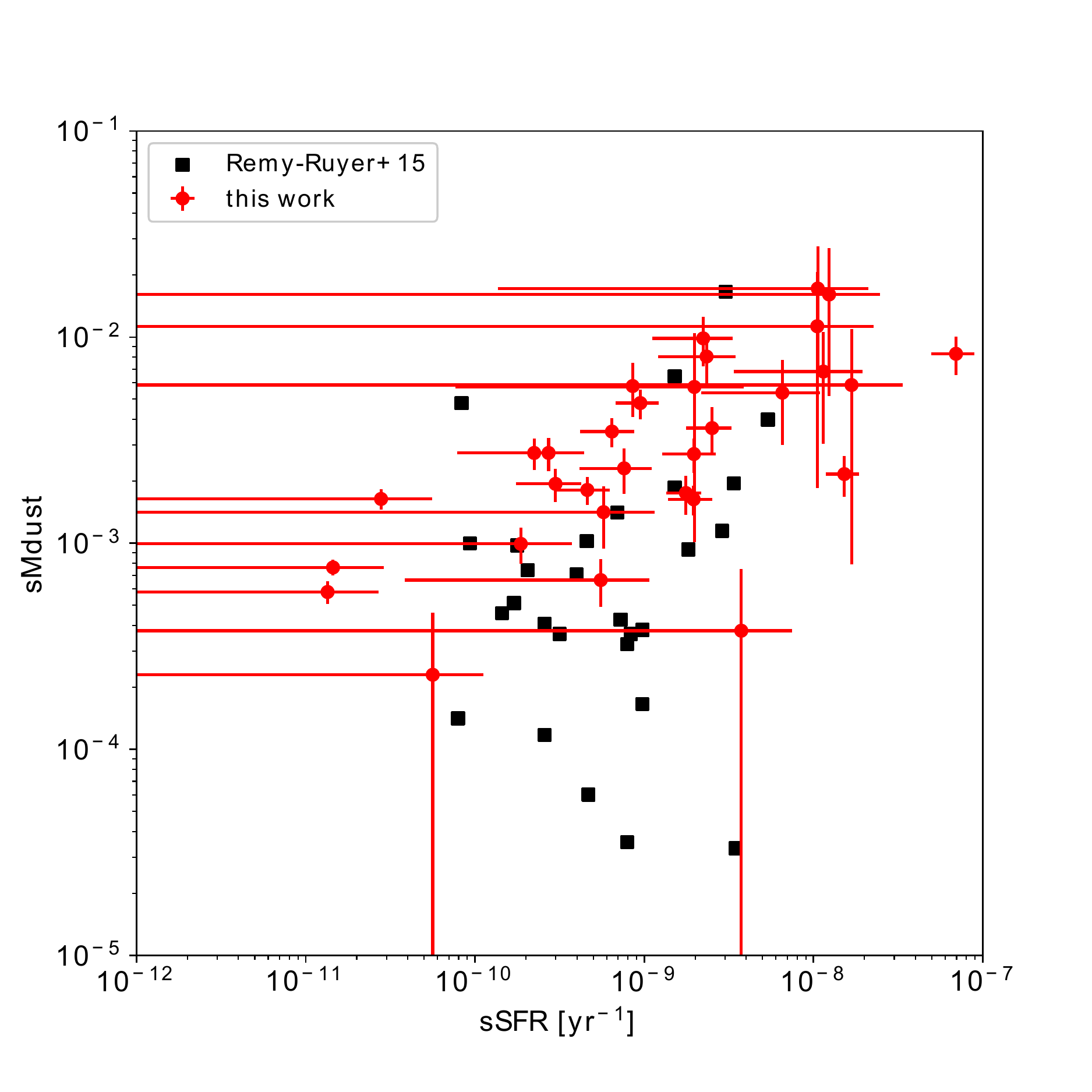}
\caption{sMdust versus the sSFR for the DGS galaxies selected and analysed in this work (red circles) and in \citet{RR15} (black squares). }
         \label{cfr:sMd_ssfr}
   \end{figure}
We here discuss the main properties derived for the galaxies considered from the SED fitting procedure.  
The SED fitting performed for the DGS galaxies in this work in which the infrared data from \citet{RR15} are combined with UV and optical photometry (see Section~\ref{Sample}) allow us to better constrain the stellar mass and SFR of these galaxies with respect to the ones in \citet{RR15} where only infrared data have been employed.

The average value of $\tau$ is estimated to be of $83$~Myrs for LBGs, while for DGS galaxies, we selected two representative values to run the chemical evolutionary models which are of $\tau=83$~Myrs for those objects with $\tau<100$~Myrs (5 galaxies), and the average value of $\tau=300$~Myrs for the others 26 galaxies. 
LBGs are characterised by stellar population no older than $\approx 700$~Myrs, while stellar populations up to $\approx 800$~Myrs are derived for DGS.
As shown in Fig. \ref{Fig:obs_DGS_LBGs} the sSFR of DGS galaxies is between $\approx10^{-11}$ and $\approx10^{-7}$~yrs$^{-1}$, while for LBGs the range is narrower, between $\approx10^{-9}$ and $\approx10^{-7}$~yrs$^{-1}$. The sMdust is between $\approx10^{-4}$ and $\approx 10^{-2}$ for DGS galaxies, while LBGs attain larger values up to $\approx 6\times10^{-2}$.
In the plot shown in Fig. \ref{Fig:obs_DGS_LBGs} the evolutionary time increases from right to left. This represents a relevant diagnostic diagram largely employed throughout this work and in the literature.
The main features for both the DGS galaxies and LBGs in the sMdust against sSFR (and age) shown in Fig. \ref{Fig:obs_DGS_LBGs}, is i) a peak at the beginning of the cycle; ii) a later decrease. 
LBGs attain a value of sMdust larger than DGS galaxies, while the decline of sMdust is faster. The large error bars in of sMdust obtained for some of the galaxies are due to the non detection for these objects of the dust continuum emission for which only upper limits are available.

For DGS galaxies we have additional information on the metallicity and on the gas content \citep[See][and references therein]{RR13, RR14}. The metallicity values are $7.1 \lessapprox 12+\log(O/H)\lessapprox 8.4$, the gas fraction $0.4\lessapprox f_{\rm gas}\lessapprox 1$ and the dust-to-gas ratio $10^{-5}\lessapprox D/G \lessapprox 10^{-3}$, where this latter quantity has been estimated by considering the dust mass derived in the analysis here presented.
All the properties of DGS galaxies derived from the SED fitting performed with \textsc{cigale} are provided in Table~\ref{Table:gal_prop} in the Appendix together with all plots showing the best fitting spectra. For LBGs properties and SED fitting we refer to the work of \citet{Burgarella20}.

In Fig.~\ref{cfr} we show the comparison between the mass of dust, stellar content and SFR derived in \citet{Burgarella20} and the ones in \citet{RR15}. We find that the stellar mass and SFR derived by \citet{RR15} are systematically larger than the one derived in this work, while the mass of dust, constrained by the infrared emission is comparable.
In Fig.~\ref{cfr:sMd_ssfr} we compare the sMdust  against the sSFR of the work by \citet{RR15} with the ones derived from the SED fitting with \textsc{cigale} for the same galaxies. Our data points appear to be shifted at higher sMdust and a few of them present high sSFR. We will discuss in the following sections the consequences for our analysis. 
The trend shown has been derived through the SED fitting performed with a Chabrier IMF since a top-heavy one is not yet available in the SED fitting code \textsc{cigale}. By employing a top-heavy IMF in the SED fitting we do not expect significant variations in the stellar and dust masses derived, while the predicted age of the stellar population will tend to be older with lower values of sSFR \citep{Pforr12}. Since the values of sMdust would remain approximately the same, we expect that the trend between sMdust and sSFR will still be present, but shifted to lower sSFR.
 \subsection{Reproducing the gas, metal and dust content of galaxies}\label{Section:Z_D}
We here present the results obtained from several tests from which we derive the suitable parameters to perform the systematic calculations and to build a grid of models through which we fit the properties of galaxies derived from \textsc{cigale}. The parameters adopted for these calculations and for computing the grid are listed in Table~\ref{Table:models}. We use different diagnostic diagrams such as sMdust, metallicity, gas fraction as a function of the sSFR (or age) to investigate the metal and dust enrichment in galaxies, as well as the gas and dust removal from their ISM.
\subsubsection{Initial mass function and dust condensation fraction for SNe II}\label{Sec:IMF}
\begin{figure}
\centering
\includegraphics[scale=0.5]{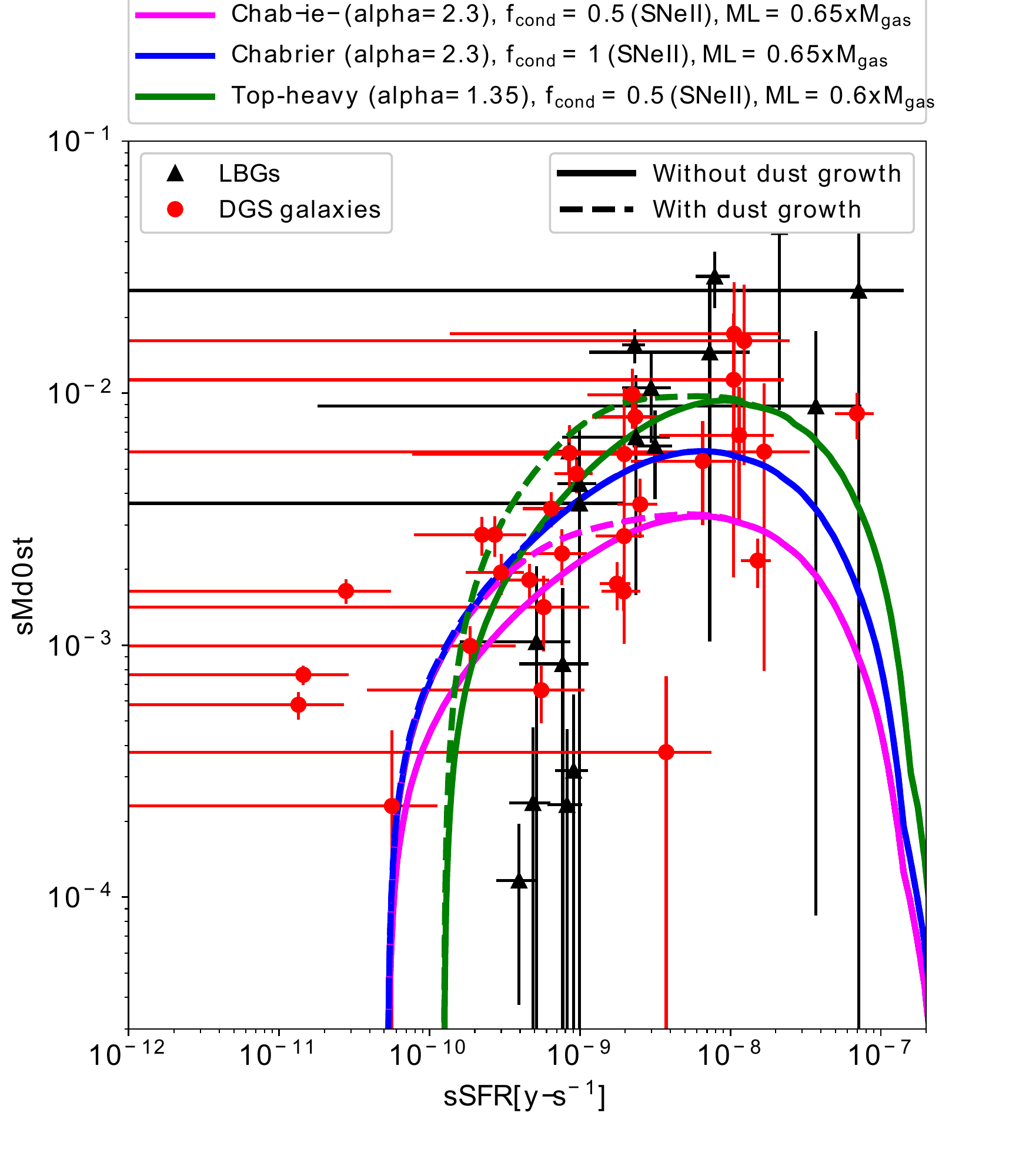}
\includegraphics[scale=0.5]{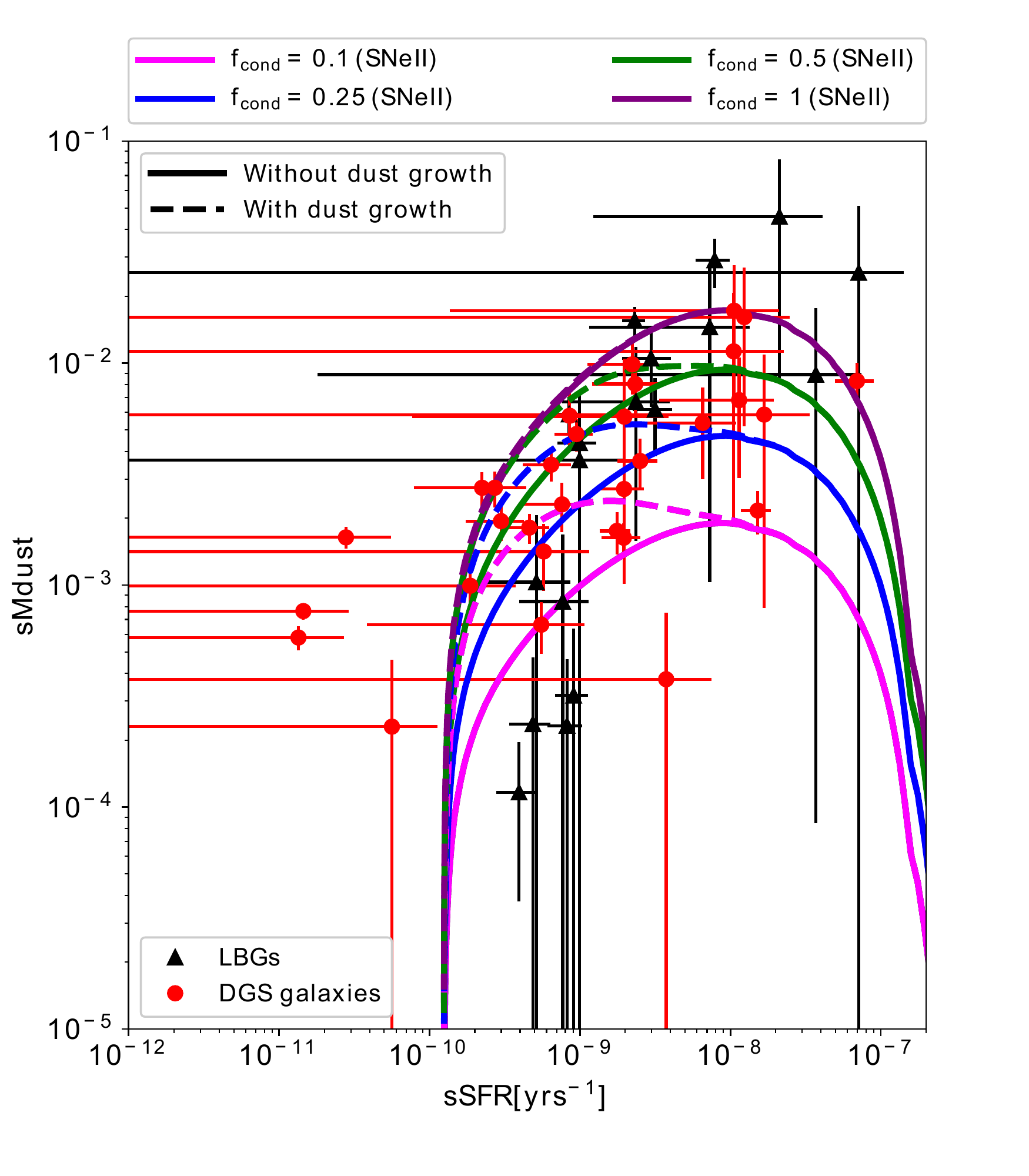}
\caption{Upper panel: sMdust against sSFR for DGS galaxies (red dots) and for LBGs (black triangles). Different chemical evolutionary models computed with a Chabrier IMF or with a top-heavy IMF ($\alpha=1.35$) and different condensation fractions as indicated in the legend. For all the models shown we select a value of $\tau=300$~Myrs in Eq.~\ref{Eq:SFR} and $M_{\rm swept}=1000$~M$_\odot$ in Eq.~\ref{eq:tdestr} and $M_{\rm gas}=100\times M_{\rm stars}$. Lower panel: the same as in the lower panel but overplotting models with a top-heavy IMF and different condensation fractions indicated in the legend.}
         \label{fig:IMF_test}
   \end{figure}
The first constrain that we are able to provide to our models is on the IMF, since not all the choices are able to reproduce the largest values of sMdust at large sSFR (corresponding to the beginning of the cycle).
In the upper panel of Fig.~\ref{fig:IMF_test} we show an example of calculation adopting the Chabrier IMF and the top-heavy IMF with $\alpha=1.35$. The cases with $\alpha=1, 1.5$ yield results close to this latter case.
We focus our investigation to models with $\tau=300$~Myrs, $M_{\rm gas}=100\times M_{\rm stars}$ and $M_{\rm swept}=1000$~M$_\odot$. Such a choice of parameters allows to minimise the effect of dust destruction in the ISM (see Eq.~\ref{eq:tdestr}) that may prevent the models to attain the largest values of sMdust at the beginning of the baryon cycle. For each of the models shown the efficiency of the outflow reasonably reproduces the observations.
Similar trends are recovered by employing $\tau=83$~Myrs.
The observed values of sMdust at the beginning of the cycle are not well reproduced by a Chabrier IMF with 50\% condensation fraction for SNe. This holds both in case dust growth in the ISM is considered or neglected. The comparison improves by increasing the condensation fraction close to 100\% for SNe which represents an extreme case, and still does not reproduce the sMdust for most of the galaxies at the beginning of the cycle. The observations are better reproduced if a top-heavy IMF is assumed. For such choice of the IMF the ISM is rapidly enriched with metals and dust: a large value of sMdust ($\approx 10^{-2}$) is attained within 100 Myrs ($\approx sSFR\approx10^{-8}$ yr$^{-1}$) by assuming 50\% condensation fraction for SNe. In the lower panel of Fig. \ref{fig:IMF_test} we also show the effect of changing condensation fraction for SNe II. Even when dust growth in the ISM is included, a low condensation fraction of dust in SNe ($\approx10\%$) prevents the models from reproducing the largest values of sMdust attained around $sSFR\approx10^{-8}$ yr$^{-1}$. Therefore, a larger amount of dust needs to be condensed (25-50\%). In the same panel we also show the case with condensation fraction for SNe II close to the maximum. Also for this extreme value of the IMF and of the condensation fraction of dust the largest values of sMdust for some LBGs is not attained. This discrepancy between theoretical calculations and observations may depend on several factors, as for example, the chemical composition of dust assumed to derive the mass of dust from the far-infrared emission, which is affected by uncertainties of a factor of 10 \citep{Ysard19}.

In summary: a top-heavy IMF helps to reproduce the largest sMdust derived from observations. With such a choice a dust condensation fraction around 50\% is required. In this work, we consider condensation fractions between 25\% and $\approx$100\% for the systematic calculations.
\subsubsection{Gas reservoir and star formation efficiency}\label{Sec:metallicity}
\begin{figure}
\centering
\includegraphics[scale=0.5]{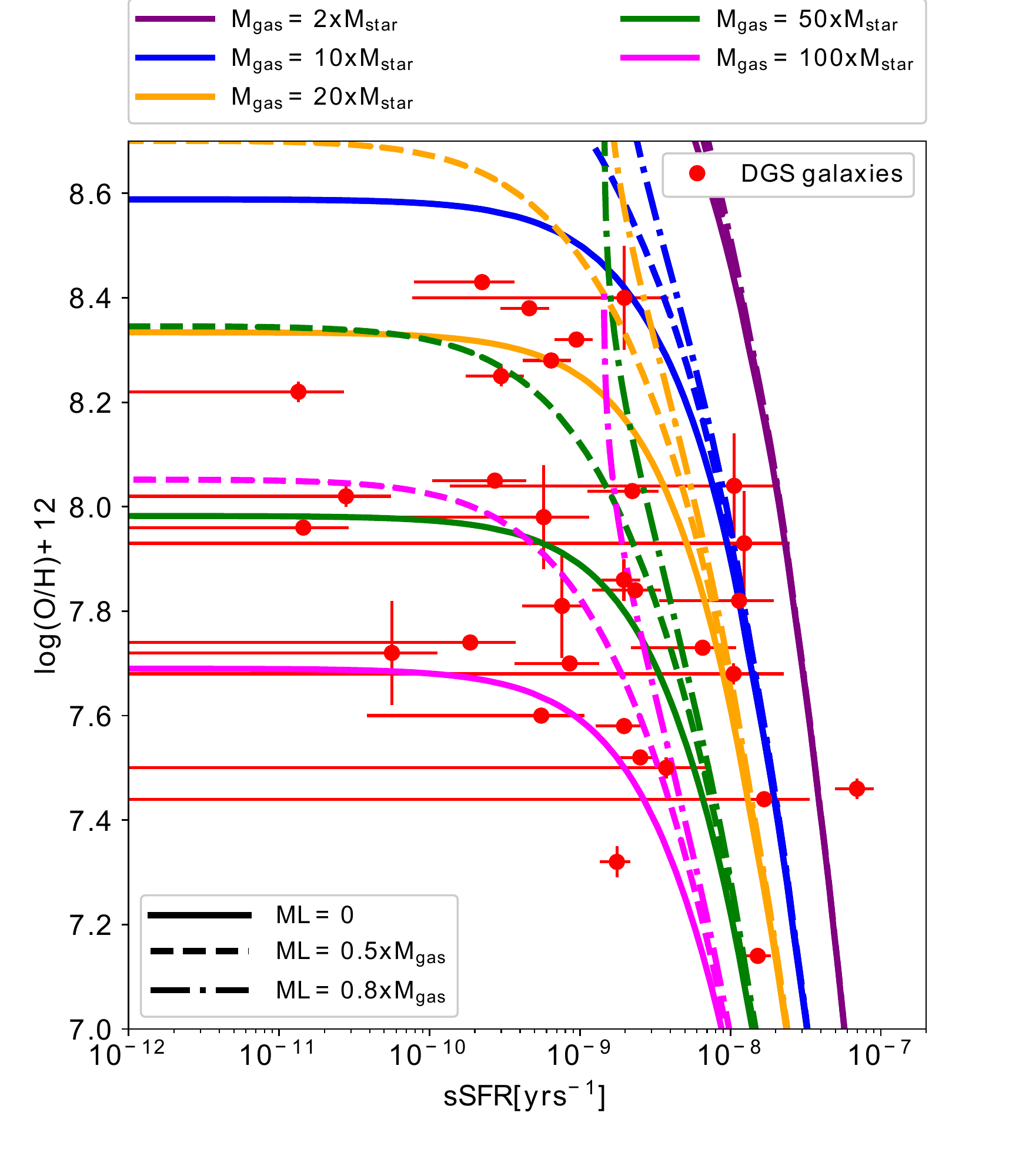}
\caption{Metallicity against sSFR for DGS galaxies (red dots). Different chemical evolutionary models characterised by diverse $M_{\rm gas}$ and different mass-loading factors as mentioned in the legend are shown. For all the models we select a value of $\tau=300$~Myrs in Eq.~\ref{Eq:SFR}, $\alpha=1.35$ for the top-heavy IMF in Eq.~\ref{eq:top-heavy_IMF}} and $M_{\rm swept}=1000$ M$_\odot$ in Eq.~\ref{eq:tdestr} and dust condensation fraction for SNe II of 50\%. No dust growth in the ISM is considered.
         \label{fig:O_gas_test}
   \end{figure}
For DGS galaxies it is possible to constrain the star formation efficiency (i.e. the mass of gas converted into stars) by studying the trend between their metallicity and the sSFR (or age). 
The comparison between model predictions and observations is shown in Fig.~\ref{fig:O_gas_test} where the metallicity ($\rm\log (O/H)+12$) is plotted as a function of the sSFR. As representative case, we study the behaviour of models with $\tau=300$ Myrs, $M_{\rm swept}=1000$ M$_\odot$, condensation fraction of dust for SNe II of 50\%, a top-heavy IMF with $\alpha=1.35$, different mass-loading factors and $M_{\rm gas}=2-100 \times M_{\rm star}$.
The initial mass of baryons is composed only by gas at the beginning that will then form stars according the star formation law define by Eq.~\ref{Eq:SFR}. The quantity ($\log (O/H)+12$) is the abundance of oxygen in the gas phase, which is obtained in the models by subtracting the oxygen condensed into dust grains (silicates) and in CO molecules from the total. From this plot we conclude that the initial mass of gas in the galaxy should be between 10 and 100 times the final stellar mass. For lower mass of gas, the ISM is enriched too quickly with respect to the observations.
In absence of outflow, the metallicity increases for a given choice of the mass of baryons, until a plateau is reached. If some galactic outflow is present, the final metallicity attained is larger and it rises more quickly. This is due to the fact that the outflow remove simultaneously both metal and hydrogen from the ISM, however, since the mass of hydrogen is larger than the mass of oxygen, hydrogen is removed in larger amounts, and the ratio decreases with time (see also Eq.~\ref{MZ}).
Different combinations of initial baryon mass and outflow are able to cover the observed range of metallicity.

In summary: an initial gas reservoir between $\approx$10 and $\approx$100 times larger than the final stellar mass is needed to reproduce the observed metallicity of DGS galaxies. Different choices of the mass-loading factor allow us to cover the observed range of metallicity values. This constraint is used to set the input parameters for the evolution of LBGs for which the measurements of the metallicity are not available.
\subsubsection{Dust destruction by SNe, star formation and removal from galactic outflows}
\begin{figure}
\centering
\includegraphics[scale=0.5]{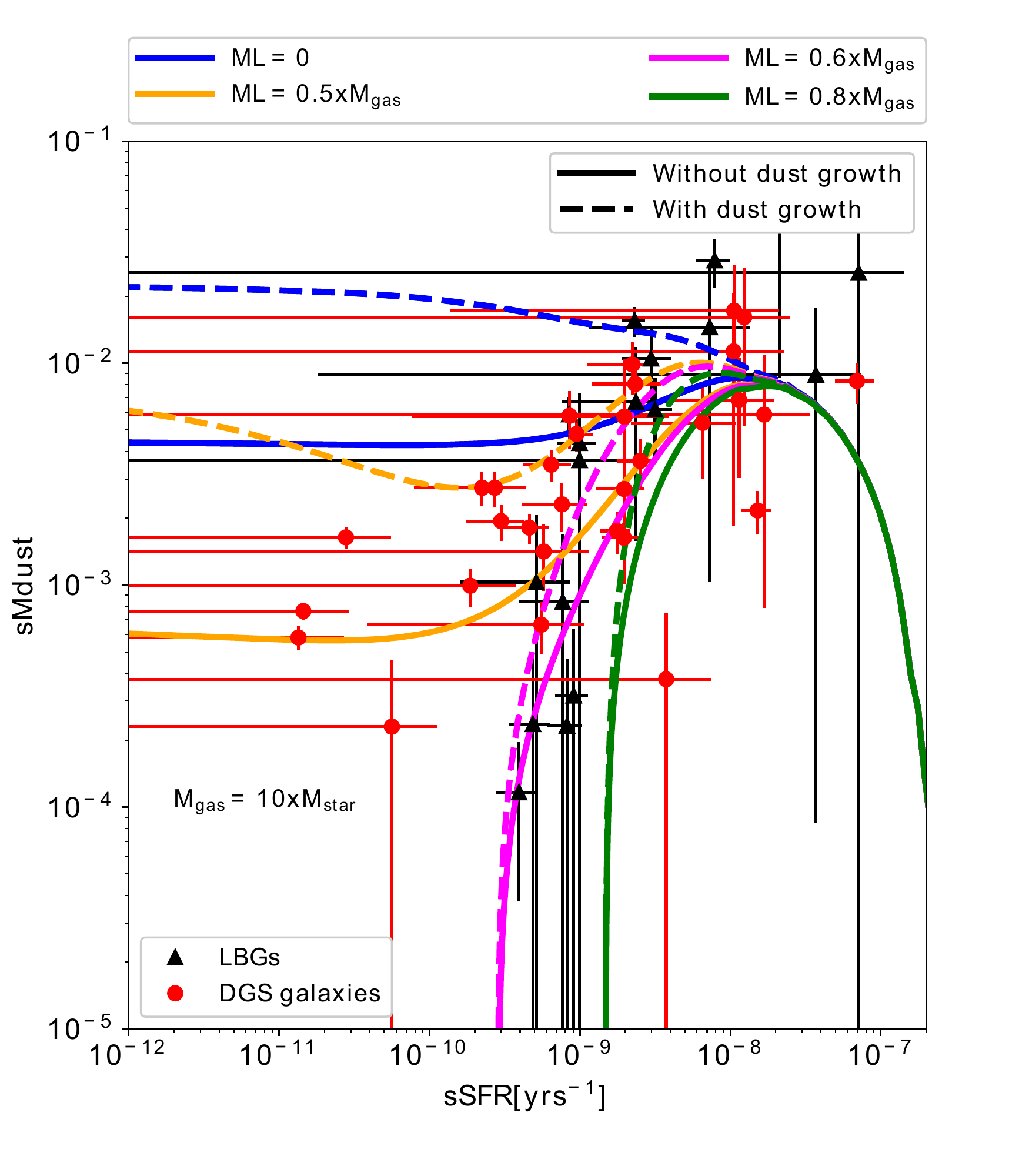}
\includegraphics[scale=0.5]{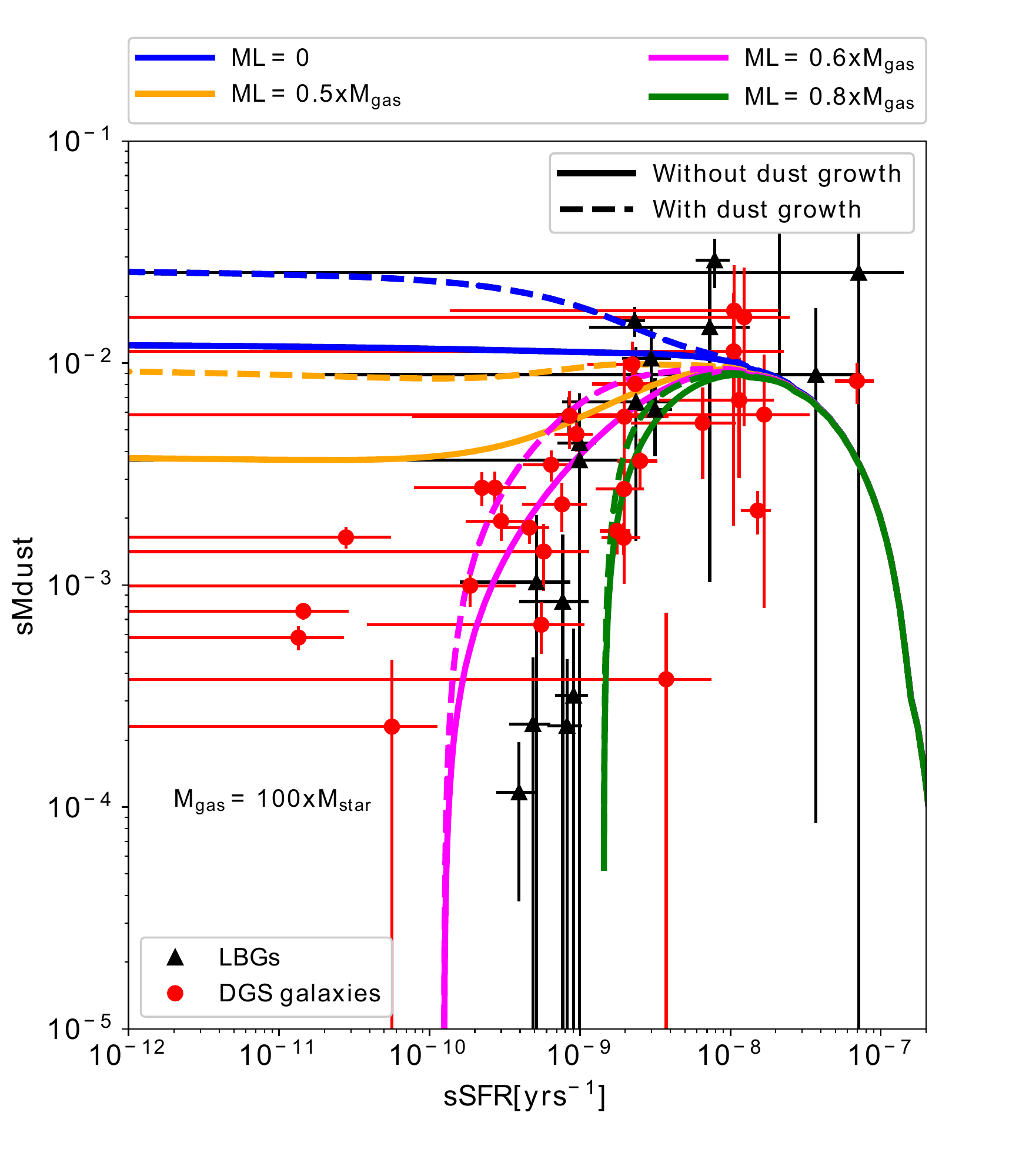}
\caption{sMdust against sSFR for DGS galaxies (red dots) and LBGs (black tringles) overplotted with different models with two extreme values of $M_{\rm gas}$: equal to $10\times M_{\rm stars}$ (upper panel) and to $100\times M_{\rm stars}$ (lower panel). Different combination for the outflow efficiency are selected as mentioned in the legend. Both the case with and without dust growth in the ISM are shown. For all the models we select a value of $\tau=300$~Myrs in Eq.~\ref{Eq:SFR}, $\alpha=1.35$ for the top-heavy IMF in Eq.~\ref{eq:top-heavy_IMF} and $M_{\rm swept}=6800$ M$_\odot$ in Eq.~\ref{eq:tdestr} and dust condensation fraction for SNe II of 50\%.}
         \label{fig:destr_test}
   \end{figure}
\begin{figure}
\centering
\includegraphics[scale=0.5]{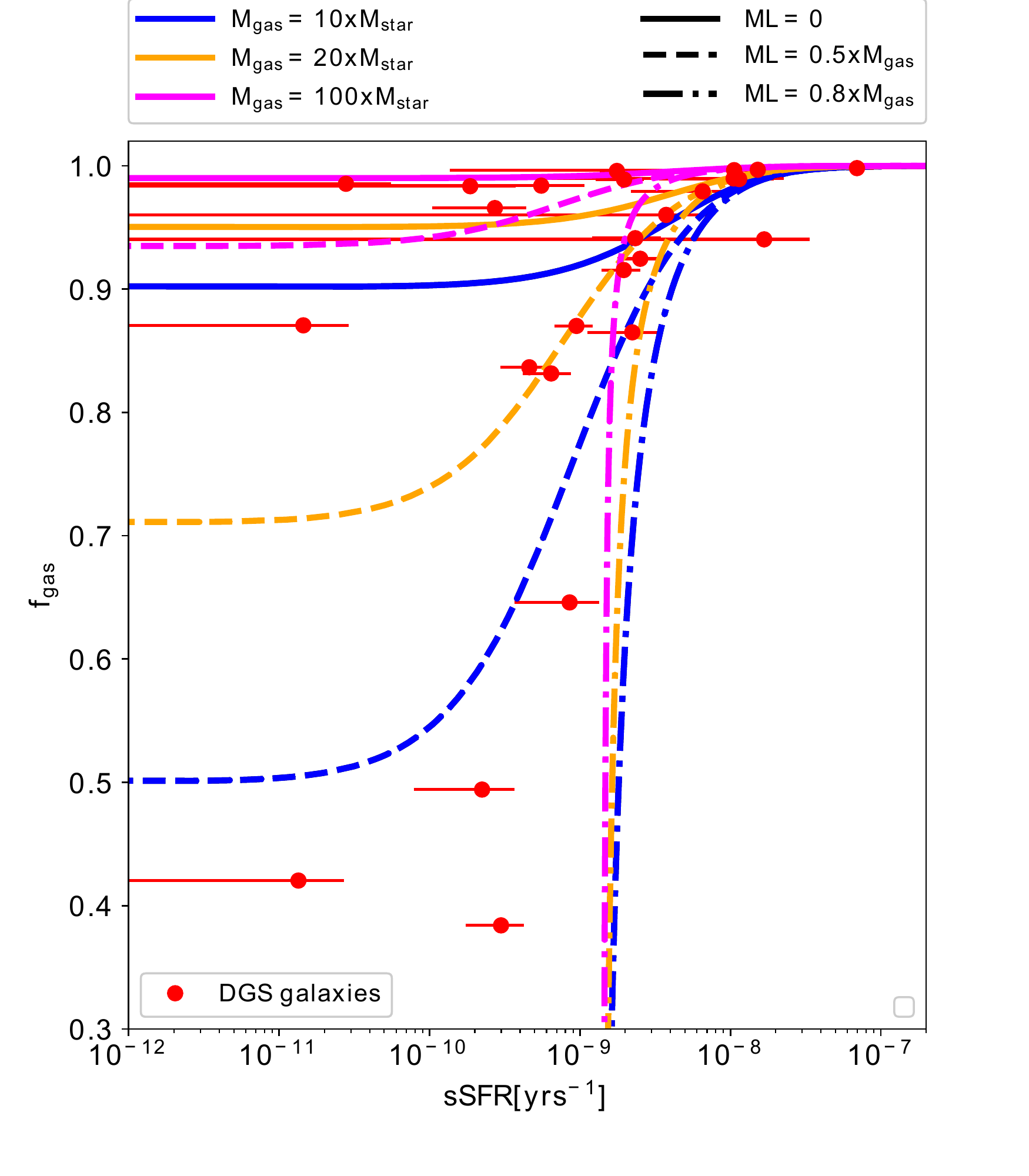}
\caption{Gas fraction against sSFR for DGS galaxies (red dots) overplotted with models with different choices of $M_{\rm gas}$ and mass-loading factors. For all the models we select a value of $\tau=300$~Myrs in Eq.~\ref{Eq:SFR}, $\alpha=1.35$ for the top-heavy IMF in Eq.~\ref{eq:top-heavy_IMF} and $M_{\rm swept}=6800$ M$_\odot$ in Eq.~\ref{eq:tdestr} and dust condensation fraction for SNe II of 50\%.}
         \label{fig:fgas_test}
   \end{figure}   
   
After a peak in sMdust, a decline is observed for decreasing sSFR or increasing age, as shown in Fig.~\ref{Fig:obs_DGS_LBGs}.
In general, this effect can be due to combination of (i) star formation which increases the total stellar mass, (ii) dust astraction by star formation, (iii) dust destruction from SN shocks, (iv) galactic outflows.
In the two panels of Fig.~\ref{fig:destr_test} we show the behaviour of different models computed with top-heavy IMF, with the maximum value of $M_{\rm swept}=6800$~M$_\odot$, with different choices of the outflow for two extreme values of $M_{\rm gas}=10\times M_{\rm stars}$ (upper panel) and $M_{\rm gas}=100\times M_{\rm stars}$ (lower panel) selected as described in Sections~\ref{Sec:metallicity} and \ref{Sec:IMF}. From Fig. \ref{fig:destr_test}, it is clear that if outflow is not included, only in the case with the lowest mass of gas and no grain growth in the ISM, the combination of dust destruction from SNe and astraction due to star formation partially reduces sMdust. In any case, all the models without outflow remain too flat for decreasing values of the sSFR. This trend indicates that dust astraction and destruction from SNe are not sufficient to decrease sMdust for decreasing sSFR, even for the maximum efficiency of dust destruction from SNe ($M_{\rm swept}=6800$~M$_\odot$).
Indeed, since for most of DGS galaxies a large amount gas is required in order to be able to reproduce their metallicity trend, dust destruction by SNe is not efficient by construction (see Eq. \ref{eq:rdestr}).
By employing different values of $ML$ combined with $M_{\rm gas}$ we are able to reproduce different sMdust in the galaxies.
Therefore, we conclude that galactic outflows are required in order to reproduce the observed trend between sMdust and the sSFR for $sSFR<10^{-8}$~yr$^{-1}$ for both DGS galaxies and LBGs.
We also notice that despite the fact that a combination with a mass of baryons equal to 100 the final mass of the stars without galactic outflow will be suitable to reproduce the observed metallicity of galaxies with $log(O/H)+12<7.8$ and $sSFR<10^{-9}$ yrs$^{-1}$ the corresponding sMdust which is less than few $10^{-3}$ at those sSFR is not reproduced. 

Galactic outflows are also needed in order to reproduce the observed gas fraction as a function of the sSFR for DGS galaxies. In Fig. \ref{fig:fgas_test} we show the observations for DGS galaxies overplotted with the evolutionary models characterised by different $M_{\rm gas}$ and outflow efficiencies. In the case without outflow, only astraction is at work in reducing the amount of available gas.
As expected, gas consumption by star formation only mildly affects the gas fraction in case $M_{\rm gas}=10\times M_{\rm stars}$, while the effect is negligible for larger gas content. Galactic outflows are therefore necessary to efficiently decrease gas fractions for decreasing sSFR.

In summary: galactic outflow is an essential feature to reproduce the decline in sMdust for $sSFR\gtrapprox10^{-8}$ yr$^{-1}$ observed for both DGS galaxies and LBGs, and for reproducing the trend between the observed gas fraction and the sSFR in DGS galaxies. We therefore considered different efficiency for this process in the systematic calculations.
\subsubsection{Efficiency of dust growth in the ISM}
We here discuss the efficiency of dust growth in the framework introduced in Section~\ref{Section:dust_growth} in order to assess the relevance of this process in the galaxies under study.
By looking at Figs.~\ref{fig:IMF_test} and \ref{fig:destr_test} it is possible to compare the efficiency of dust growth for different physical assumptions adopted in the simulations.
In all the models shown, the chemical evolutionary tracks computed by including dust growth overlap with the ones without dust growth between sSFR of $\approx10^{-8}$~$yr^{-1}$ and $10^{-7}$~$yr^{-1}$ which corresponds to the early dust enrichment from SNe II. 
Around sSFR of $\approx 10^{-8}$~$yr^{-1}$ the mass of dust tends to increase due to the effect of grain growth in the ISM. When the outflow is not included, the theoretical curves rise before reaching a plateau which correspond to the maximum possible condensation of dust from the available metals (Fig. \ref{fig:destr_test}). This trend is qualitatively similar to the ones obtained by \citet{Asano13}. 
In the models without outflow, the increase of the dust in the theoretical tracks due to grain growth does not reproduce the observed decrease in sMdust, as already noticed for the models without grain growth.
The models with and without dust growth become more similar for increasing efficiency of the outflow, which is an essential feature to reproduce the observed trends. In this case, the effect of grain growth in the ISM tends to be cancelled by dust removal.
Furthermore, the effect of dust growth combined with a more efficient outflow is degenerate with the models without grain growth and less outflow.

From Fig. \ref{fig:IMF_test} we also notice that for low dust condensation fractions of SNe II and/or an IMF different from the top-heavy one, the galaxies with the largest sMdust observed at the beginning of the cycle are not reproduced even if grain growth in the ISM is included in the calculations. This might indicate that such a process is not the main one responsible for dust production at the beginning of the baryon cycle, unless we assume that grain growth occurs at larger densities ($> 10^3$ cm$^{-3}$) or/and the dust grains initially present in the ISM are small ($<10^{-6}$ cm).

In summary: dust growth in the ISM is not dominant compared to outflow. Thus, it is not necessary to include it in order to reproduce the trend between sMdust and the sSFR for both DGS galaxies and LBGs. If present, the effect of such a process is not evident because combined with the one of galactic outflow.
\subsubsection{Galactic inflows}
\begin{figure}
\centering
\includegraphics[scale=0.5]{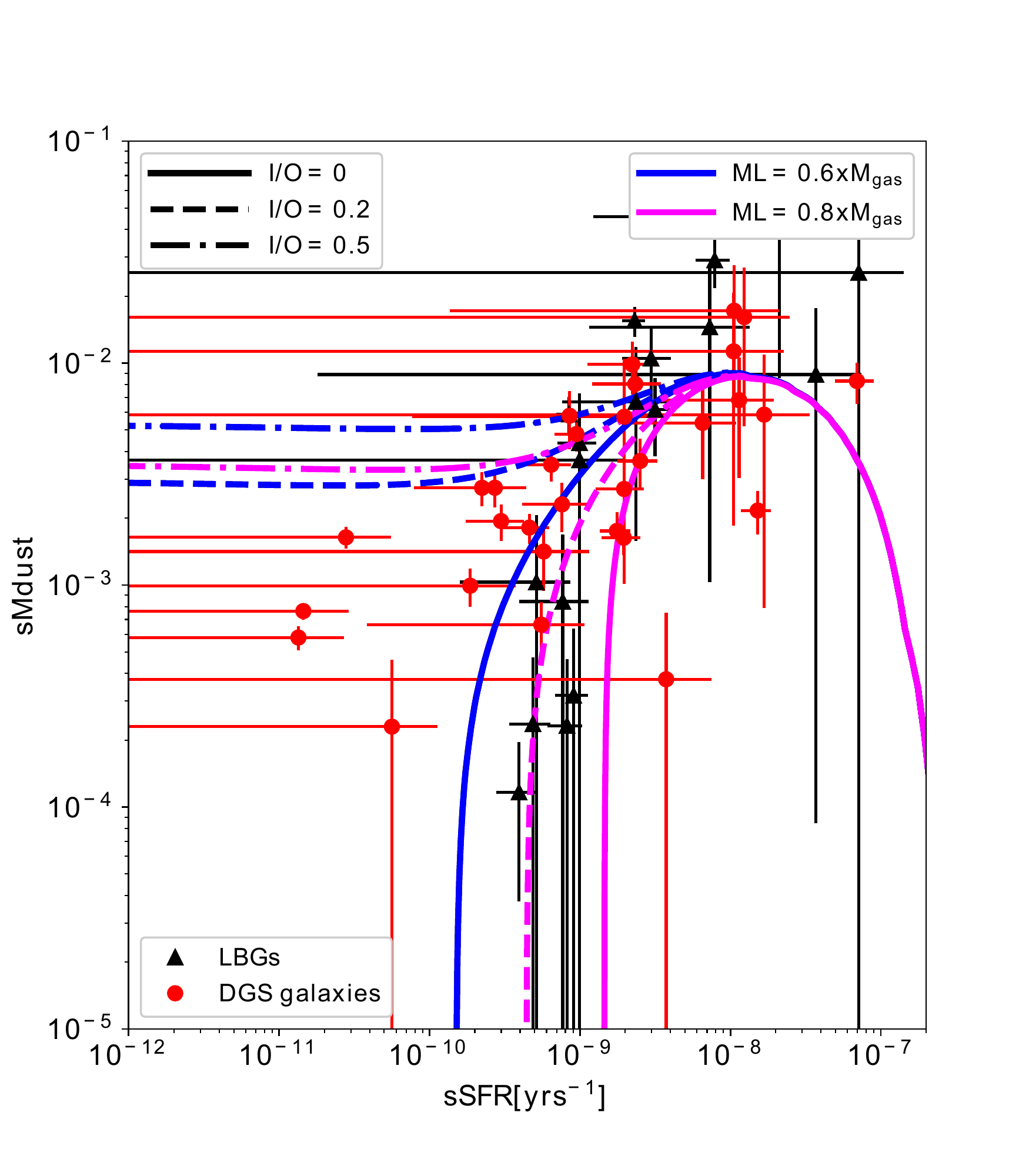}
\includegraphics[scale=0.5]{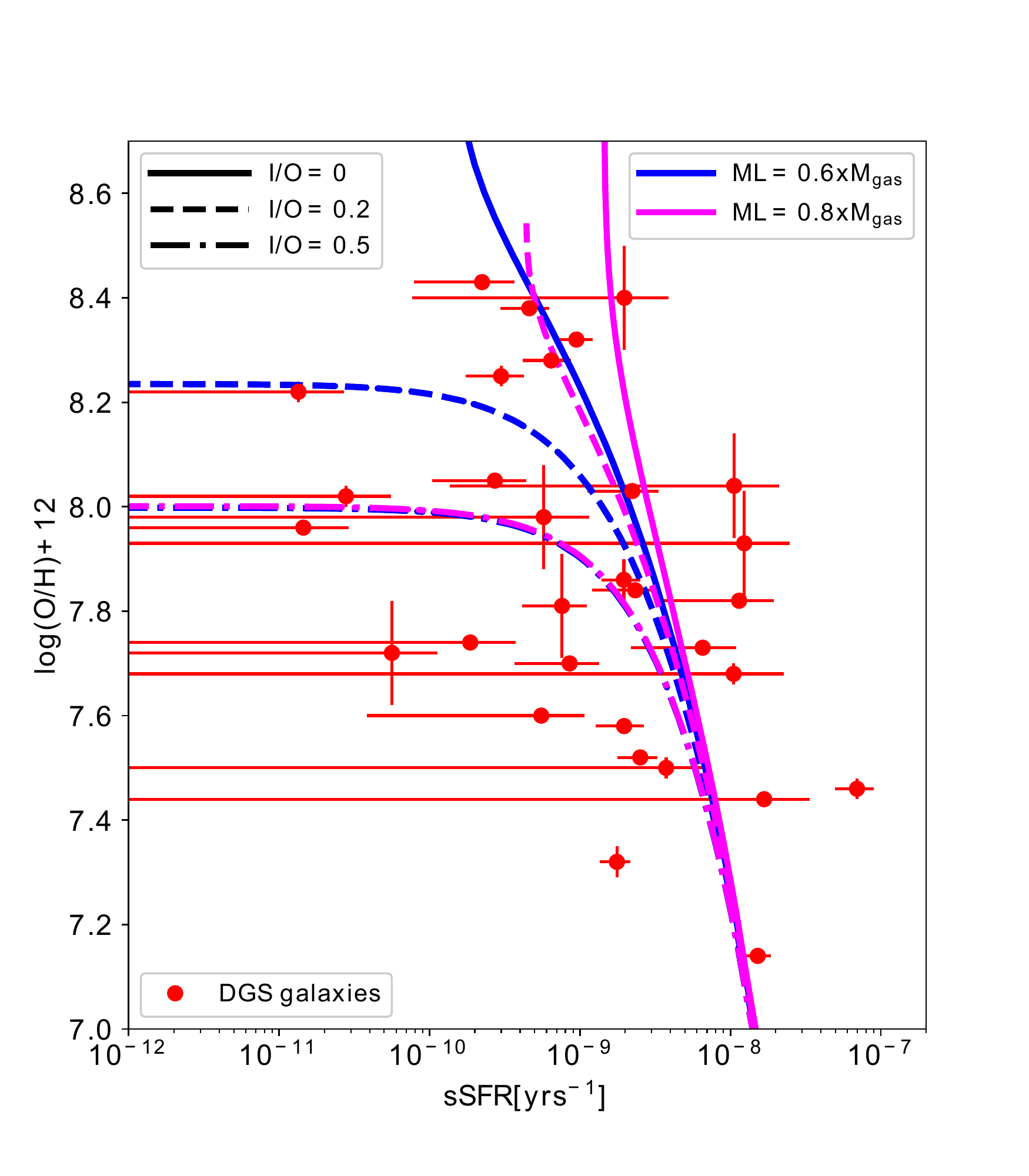}
\caption{Upper panel: sMdust vs sSFR for DGS galaxies (red dots) and LBGs (black tringles) overplotted with models characterised by different choices the mass-loading factor and inflows as indicated in the legend.  For all the models we select a value of $\tau=300$~Myrs in Eq.~\ref{Eq:SFR}, $\alpha=1.35$ for the top-heavy IMF in Eq.~\ref{eq:top-heavy_IMF} and $M_{\rm swept}=6800$ M$_\odot$ in Eq.~\ref{eq:tdestr},  $M_{\rm gas}=50\times M_{\rm stars}$ and dust condensation fraction for SNe II of 50\%. Lower panel: metallicity vs sSFR for DGS galaxies overplotted with the same models as in the upper panel. }
         \label{fig:inflow}
   \end{figure}
We here explore the possible effect of introducing galactic inflows in our models in addition to the outflow required to reproduce different observations.
In the two panels of Fig. \ref{fig:inflow} we show as example the effect introduced by including different amount of inflow besides galactic outflow in the sMdust against sSFR and in the metallicity against sSFR plots for a few selected models with $\tau=300$ Myrs, $M_{\rm gas}=50\times M_{\rm stars}$, $M_{\rm swept}=6800$ M$_\odot$ and different choices of the galactic inflows and outflows.
Since the dust and gas need to be removed from the galaxy I/O in Eq. \ref{Mgas} needs to be $<1$. We select as test cases $I/O=0.2, 0.5$ combined with efficient outflow ($ML=0.6, 0.8\times M_{\rm gas}$).
As shown in Fig. \ref{fig:inflow}, in those models in which  galactic inflows is include an efficient outflow is needed in order to remove the dust in the galaxies and to reproduce the observations. The difficulty in removing the dust in the ISM in presence of an inflow is caused by the fact that the dust is diluted in the ISM. This process decreases the dust fraction $\delta_{\rm i, d}$ in Eq. \ref{eq:dust} and dust is removed less efficiently from the galaxy. From the lower panel in the same figure, it is instead possible to appreciate the slower metal enrichment due to the inflow of pristine gas, and the corresponding decrease of the fraction of metals ejected in the outflow (due to the decrease of $f_{\rm i,g}$ in Eq. \ref{MZ}).
Furthermore, the various combinations of galactic inflow and outflow introduce some degeneracy in the models with the ones discussed in the previous sections. 
We chose not to introduce any further degeneracy in our systematic calculations and to assume that all the gas is already present in the galaxy before the beginning of star formation.
\subsubsection{Metals from different theoretical data sets and the contribution of different sources}\label{Sec:test_yields}
\begin{figure}
\centering
\includegraphics[scale=0.5]{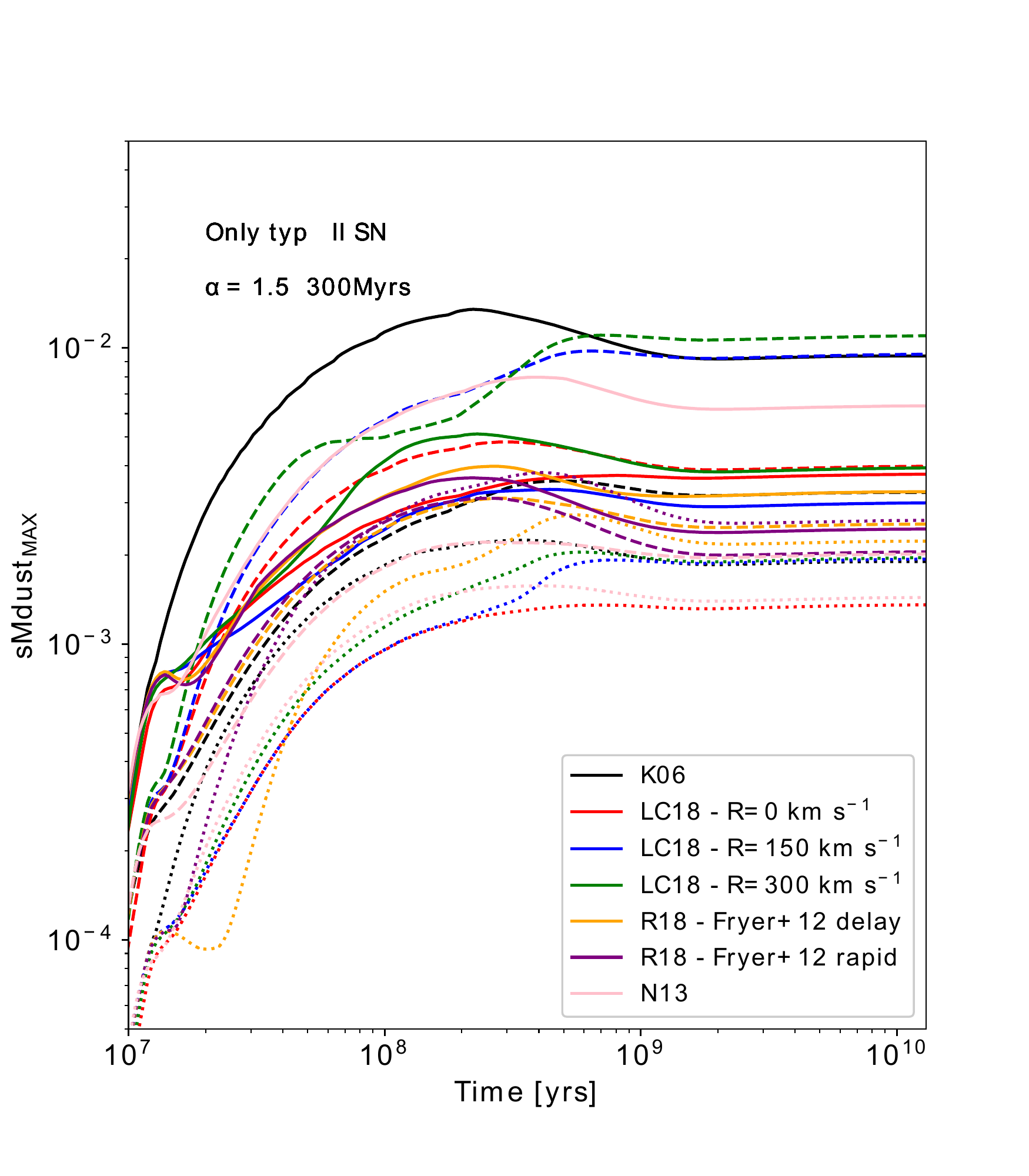}
\caption{Maximum possible sMdust for SNe II computed for different choices of the theoretical metal yields for massive stars listed in the figure. Solid, dashed and dotted lines indicate the upper limits for sMdust obtained for silicate (only pyroxene), carbon and metallic iron dust, respectively.}
         \label{fig:test_yields}
   \end{figure}
   
   \begin{figure}
\centering
\includegraphics[scale=0.5]{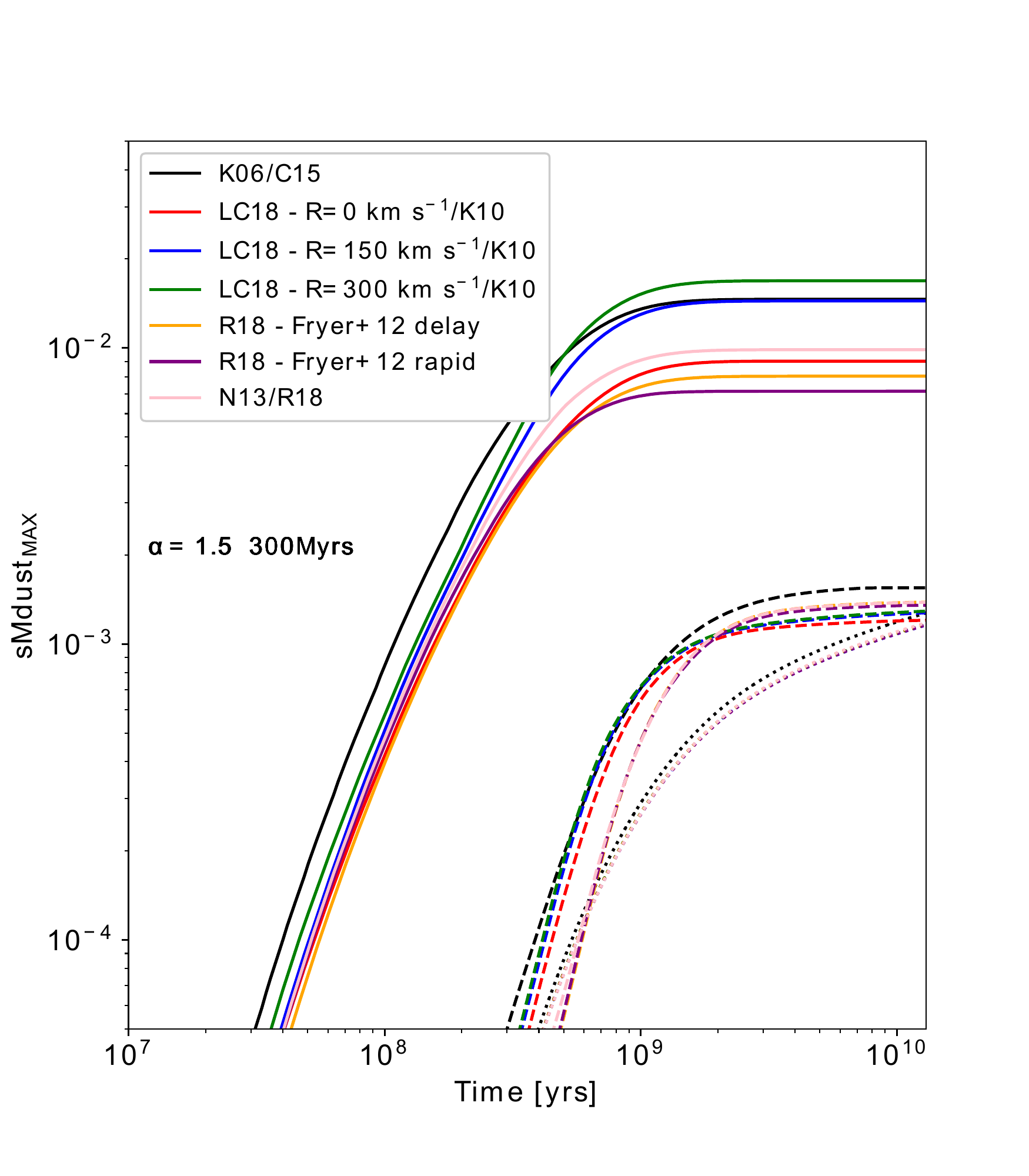}
\caption{Maximum possible sMdust for SNe II (solid lines), TP-AGB stars (dashed lines) and Type Ia SNe (dotted lines) for different choices of the theoretical metal yields for SNe II and TP-AGB stars listed in the figure and provided in Table~\ref{Table:models}.}
         \label{fig:test_yields_AGB_Ia}
   \end{figure}

We here explore the effect on the chemical enrichment by employing different sets of theoretical metal yields taken from the literature, combined with the different time-scales of the delayed SFH and top-heavy IMF. 
We consider 7 different theoretical sets for SNe II, 3 for TP-AGB stars and 2 for Pop III stars (characterised by zero metallicity). For pop III stars we assume the same IMF as for the other stars. 
The metal yields for SNe II by \citet{Kobayashi06} and \citet{Nomoto13} cover a mass range up to $40$~M$_\odot$. \citet{Limongi18} have provided three sets of metal yields up to 120~M$_\odot$ computed for different rotational velocities, R, of massive stars (R=0, 150, 300~km~s$^{-1}$). \citet{Ritter18} have computed two sets of metal yields by employing different formalisms up to 25~M$_\odot$. The properties of the different data sets are summarised in Table~\ref{Table:models}.

Starting from these metal yields we calculate the maximum possible amount of dust produced considering silicates, carbon and iron dust.
For simplicity, we consider the case in which silicate dust is composed entirely by pyroxene (MgSiO$_3$). However, the mass of silicates derived if only olivine (Mg$_2$SiO$_4$) is considered does not change considerably. 
For silicates, the maximum possible mass of dust is estimated at each time-step by evaluating the least abundant of all the elements that form this dust species. The final mass of dust is then obtained by estimating the maximum number of monomer and mass of dust that can be formed from the stoichiometric formula of pyroxene.

A representative example of the results for SNe II Fig.~\ref{fig:test_yields} where the maximum possible sMdust of silicate, carbon and iron dust is plotted a function of time. A top-heavy IMF with $\alpha=1.5$ and $\tau=300$~Myrs have been selected. The trend obtained by employing different data sets is qualitatively similar for different IMF and $\tau$.
The spread in the maximum sMdust spans a factor of about 5 for all the dust species considered. For all the data sets but \citet{Limongi18}, the final mass of silicate dust is larger than carbon by a factor between 1.2 and 3.6. For the three data sets by \citet{Limongi18} the mass of silicate dust is down to one third the mass of carbon dust. The maximum iron mass is around 10\% and 15\% of the total dust mass for \citet{Kobayashi06}, \citet{Limongi18} and for \citet{Nomoto13}, respectively. For the two data sets by \citet{Ritter18} the mass of iron dust is between 30-35\% of the total.
The sets of models which produce the largest maximum mass of dust are from \citet{Kobayashi06} and  \citet{Limongi18} with rotational velocity R=150, 300 km~s$^{-1}$. The three data sets yield very similar values of sMdust from SNe II close to the expected observed peak of dust around 100-200 Myrs.
The most remarkable difference among the dust produced by these three data sets is the predicted chemistry of dust which would be dominated by silicates for \citet{Kobayashi06} and by \citet{Limongi18}. For \citet{Kobayashi06} the maximum amount of silicates produced is three times larger than carbon dust.
The differences between the dust chemistry of the ISM obtained by adopting different theoretical yields might be compared and constrained with future observations.
For these three data sets we also check the amount of oxygen released in the ISM, since oxygen is the most abundant metal and it traces the metallicity of galaxies. The predicted oxygen abundances are compared with observations. We find that the values are always comparable even for different $\tau$ and for the different top-heavy IMF. Furthermore, since reproducing the large mass of observed dust in the Early Universe is challenging, we expect that the comparison with observations would worsen by employing the theoretical data sets yielding a lower amount of metals and dust with respect to \citet{Kobayashi06} and \citet{Limongi18}.  

To test the effect on the yields from TP-AGB stars we consider different combinations of theoretical yields for SNe II and TP-AGB stars. The chemical enrichment from TP-AGB stars is affected by the choice of the SN theoretical yields that change the metallicity of low-mass stars as a function of time and therefore the overall metal and dust masses released in the ISM.
The combinations considered are provided in Table \ref{Table:models}. 
The results are shown in Fig.~\ref{fig:test_yields_AGB_Ia} where we plot the maximum possible sMdust for SNe II, of TP-AGB stars and of SNe Ia.
For all the combinations of yields, we find that the contribution from TP-AGB stars and SNe Ia is negligible, due to the top-heavy IMF. Furthermore, in case a larger amount of dust would be released by TP-AGB stars or SNe Ia, it would start to be relevant only after a few hundreds Myrs, when the dust is expected to start to decrease in the observed galaxies. 
Comparable amount of sMdust could be obtained for TP-AGB stars and SN remnants by assuming drastic low condensation efficiency in SN remnants, that is not our preferred scenario, as discussed in the previous sections. The contribution from TP-AGB stars is more relevant in case the selected IMF is the Chabrier one. However, this scenario is never able to reproduce the observed trends for the galaxies considered in this work.

We additionally tested two different theoretical yields for Pop III stars \citep{Heger10, Nomoto13}. We find that the contribution to the metal enrichment of Pop III stars is negligible with respect to the total.

In summary: the metal yields selected throughout this paper for the systematic calculations are from \citet{Kobayashi06} since it allows for the most favourable conditions for attaining the largest values of sMdust already after 100 Myrs since the beginning of the baryon cycle. The choice of the metal yields for Pop III, TP-AGB stars and Type Ia SNe do not largely affect the results, since these stellar sources provide only a minor contribution to the dust enrichment. We therefore arbitrarily select the theoretical yields by \citet{Heger10} for Pop III stars (at the ages appropriate for the sample), \citet{Cristallo15} for TP-AGB stars and \citet{Iwamoto99} for Type Ia SNe.
\subsubsection{Input quantities for the systematic calculations}
On the basis of the previous tests we are able to define the characteristics required to systematically compute the chemical evolutionary models in order to perform the best fit between model predictions and observations that we here recall:
\begin{itemize}
\item metal yields favouring a fast metal enrichment from SNe II \citep[e.g.][]{Kobayashi06};
\item top-heavy IMF;
\item total mass of baryons $\geq 10$ times the final stellar mass;
\item efficient galactic outflow;
\item condensation fraction for dust produced in SNe II $\geq 25$~\%;
\item no necessity of dust growth in the ISM.
\end{itemize}
The input quantities adopted to compute the grid of models are summarised in Table~\ref{Table:models} (``Systematic calculations'').
\subsubsection{Characterisation of the individual galaxies}

\begin{sidewaystable*}
\centering
\caption{Results of the sample of DGS galaxies and LBGs that have been individually fitted through the models provided in Table~\ref{Table:models}. For each of galaxy analysed we provide the averaged residual, the predicted mass of the different dust components (silicates, carbon and iron) normalised for the final stellar mass, the dust condensation fraction for SNe II, the initial mass of baryons normalised for the final stellar mass, the slope of the top-heavy IMF, the M$_{\rm swept}$ in M$_\odot$, the fraction of the dust in the circumgalactic medium over the total ($f_{\rm dust,out}$), the fraction of gas in the circumgalactic medium over the total ($f_{\rm gas,out}$) and the ratio between the dust removed by the galactic outflow over the total removed and destroyed ($f_{\rm dust,out/(out+SN)}$). The complete table are provided in electronic form.}
\begin{tabular}{llllllllllll}
\hline
name   &   R$^2_{\rm av}$ &  M$_{\rm sil}\times10^{-4}$  &   M$_{\rm car}\times10^{-4}$  &   M$_{\rm ir}\times10^{-4}$   &   f$_{\rm cond}$   &  M$_{\rm gas}$  &   $\alpha$   &  M$_{\rm swept}\times10^3$   & $f_{\rm dust,out}$  &   $f_{\rm gas,out}$  &  $f_{\rm dust,out/(out+SN)}$ \\
\hline
DGS galaxies & & & & & & & & & & & \\
\hline
Haro3   &  4.76   & 21.4   $\pm$  3.6  &  4.8 $\pm$  1.3   &  4.04 $\pm$ 0.72   &  0.44 $\pm$ 0.26   &  36 $\pm$ 14   &   1.24 $\pm$ 0.22   &  4.2 $\pm$ 2.9  &  0.44 $\pm$ 0.26   &  
0.58 $\pm$ 0.29   &  0.66 $\pm$ 0.25 \\
He2-10   &  12.1   & 18.3 $\pm$ 0.8  &  5.84 $\pm$ 0.47   &  3.64 $\pm$ 0.19   &  0.25 $\pm$  0.00   &  20 $\pm$  0     &  1.42 $\pm$ 0.07   &  6.80 $\pm$ 0.23   &  0.21 $\pm$ 0.02   &  0.32 $\pm$ 0.03   &  0.21 $\pm$ 0.03 \\
HS0052+2536   &  5.38   & 14.9 $\pm$ 8.2  &  3.39   $\pm$  1.95   &  2.62   $\pm$  1.50   &  0.28   $\pm$ 0.08   &  36 $\pm$ 21   &   1.29 $\pm$ 0.21   &  4.4   $\pm$  2.9   &  0.25 $\pm$ 0.20   &  0.38 $\pm$ 0.26   &  0.63 $\pm$ 0.26\\
HS0822+3542   &  55.1   & 12.2   $\pm$ 1.5  &  2.41   $\pm$  0.29   &  2.02   $\pm$  0.25   &  0.25   $\pm$  0.00   &  100.0 $\pm$ 0.7    &  1.00 $\pm$  0.00   &  6.2   $\pm$  1.7   &  0.24   $\pm$ 0.06   &  0.41 $\pm$ 0.08   &  0.85 $\pm$ 0.06  \\
... & & & & & & & & & & & \\
\hline
LBGs & & & & & & & & & & & \\
\hline
ID27   &  1.61   & 10   $\pm$  11  &  1.3 $\pm$  1.3   &  1.6$\pm$1.8   &  0.61   $\pm$ 0.32   &  55 $\pm$ 29   &  1.28   $\pm$ 0.21   &  3.9   $\pm$  2.9   &  0.06   $\pm$ 0.07   &  0.12   $\pm$  0.13   &  0.79 $\pm$ 0.21   \\
ID31   &  1.14   & 16 $\pm$ 14  &  2.3 $\pm$ 1.8   &  2.6   $\pm$  2.4   &  0.60 $\pm$ 0.31   &  55   $\pm$ 29     &  1.28   $\pm$ 0.21   &  3.9   $\pm$  2.9   &  0.10 $\pm$ 0.12   &  0.19 $\pm$  0.18   &  0.76 $\pm$ 0.22 \\
lbg10   &  1.96   & 29   $\pm$  24  &  4.1 $\pm$  2.7   &  4.8   $\pm$  4.0   &  0.61   $\pm$  0.32   &  55 $\pm$  29 &  1.28   $\pm$ 0.21   &  3.9 $\pm$  2.9   &  0.20  $\pm$ 0.20    &  0.32 $\pm$ 0.25   &  0.72 $\pm$ 0.23   \\
HZ4   &  2.20   & 36  $\pm$  15  &  6.9   $\pm$  2.9   &  6.2   $\pm$  2.6   &  0.52   $\pm$  0.27   &  56   $\pm$  28   &  1.27   $\pm$ 0.21   &  3.9   $\pm$  2.9   &  0.31 $\pm$ 0.22   &  0.46 $\pm$ 0.27   &  0.67 $\pm$ 0.25  \\
... & & & & & & & & & & & \\
\hline\end{tabular}
\label{Table:results}
\end{sidewaystable*}

Following the procedure described in Section~\ref{sec:method}, we estimate the properties of DGS galaxies and LBGs, such as the chemical composition of their dust, the fraction between the gas and dust ejected in the circumgalactic medium and the fraction of dust destroyed by SN shocks. 
In Figs. from \ref{fig:dust} to \ref{fig:DG} we show the different properties of galaxies derived from the SED fitting or taken from the liteature, i.e. sMdust, metallicity, gas fraction and dust-to-gas ratio, as a function of the sSFR and age together with the corresponding distribution obtained from the fit of individual objects through the models in the chemical evolutionary tracks. 
In Figs.~\ref{fig:Z}-\ref{fig:DG} we also show the metallicity, gas fraction, and dust-to-gas ratio as a function of the sSFR and time for LBGs, for which reliable constraints are not available yet. 
We additionally show the dust-to-gas ratio as a function of the metallicity in Fig. \ref{fig:DG_Z}. For the DGS galaxies such trend is compared with the available estimates, while for LBGs only the values derived from the chemical evolution models are shown.
In Fig. \ref{fig:Mdust_Mstar} we show and the total mass of dust inside the galaxy versus the mass of stars. 

The best fit between the properties of individual galaxies and model predictions are provided in Table~\ref{Table:results}. The complete version of the table is provided in electronic form.

From the performed analysis we derive the following trends:
\begin{itemize}
\begin{figure}
\centering
\includegraphics[scale=0.5]{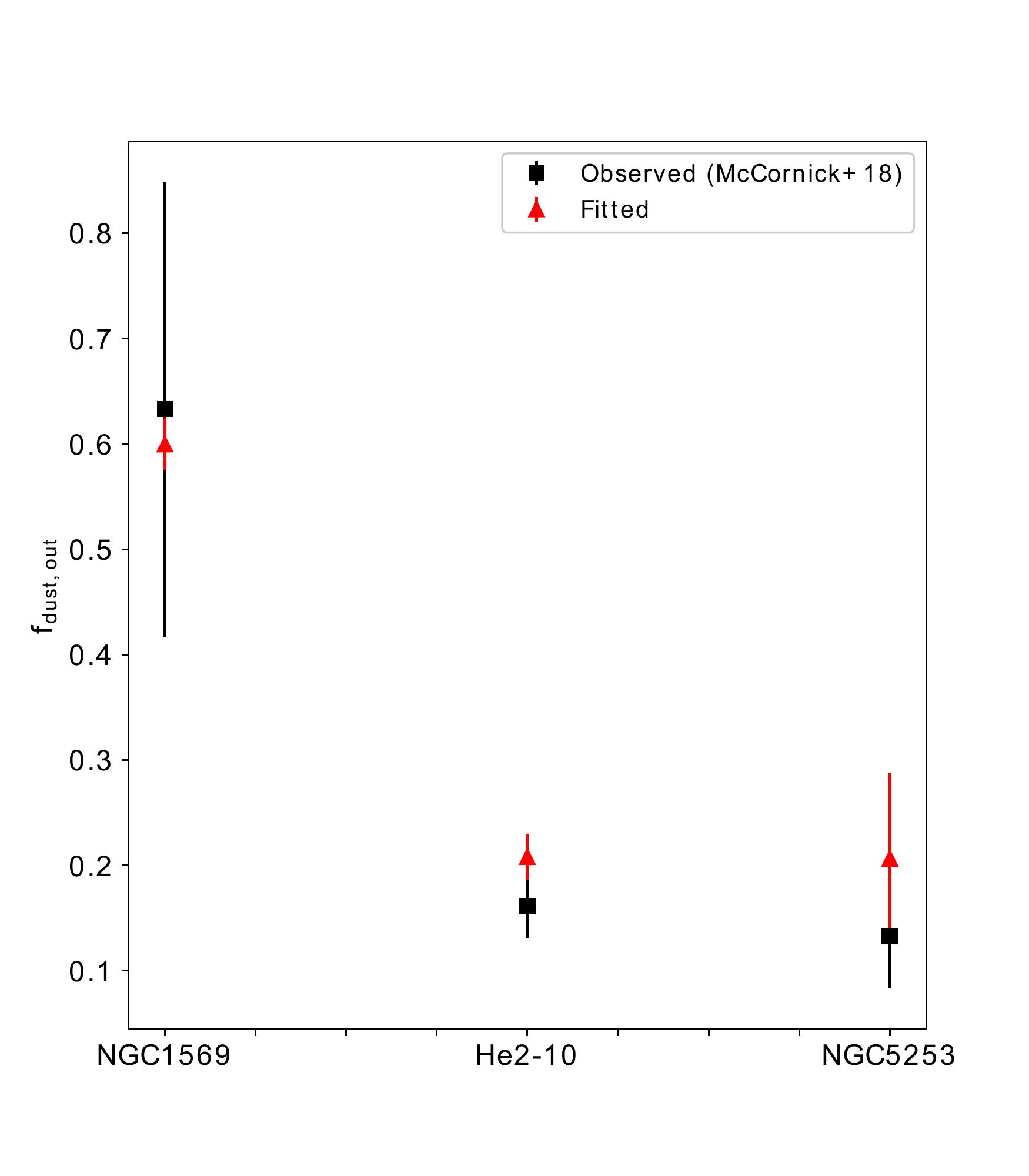}
\caption{Circumgalactic dust mass fraction from \citet{McCormick18} black squares) compared with the predicted value derived from the fit of each galaxy (red triangles).}
\label{Fig:circumgalactic}
   \end{figure}
\item \textbf{Overall properties.} 
The models developed here are in general able to fit the main properties of the observed DGS galaxies and LBGs. 
The stellar production efficiency (defined as the mass of stars over the total mass of baryons initially available) is usually of few \% for both DGS galaxies and LBGs. A strong outflow is required in order to remove the mass of dust and gas present in the galaxy and to reproduce the observation both for DGS galaxies and LBGs. Depending on the specific galaxy under consideration, the amount of gas removed from the ISM ranges between 6 or 12\% for DGS galaxies and LBGs respectively and almost the totality. The exponent of the IMF in Eq. \ref{eq:top-heavy_IMF} is $\alpha \approx 1$ for the majority of the DGS galaxies, while this value is not well constrained for LBGs for which the results are more degenerate. This is not surprising since for LBGs less information are available.
The set of observations for DGS galaxies is reproduced by a condensation fraction of SNe II which is between 25\% and 64\% with an average value of (35$\pm$12)\% considering the entire sample of DGS, while for LBGs a condensation fraction above 40\% is required with an average value of 56$\pm$17\% if all the LBGs are considered. Also in this case, a larger uncertainty on the values of the condensation fraction is found for LBGs.
We also keep in mind that there is some degeneracy between the metal yields provided in the literature and the assumed dust condensation fraction.

The dust-to-gas ratio as a function of the metallicity is fairly well reproduced for DGS galaxies (Fig. \ref{fig:DG_Z}), even though the observed distribution of data points is more tilted than the one we recovered. This discrepancy between our model predictions and observations might be partially due to the large uncertainties affecting the determination of the gas mass from the observations, especially in low-metallicity galaxies. The trend between the total mass of dust and stellar mass is naturally reproduced (Fig. \ref{fig:Mdust_Mstar}).
\item \textbf{Dust chemical composition and condensation fraction}. 
The predicted chemical composition of dust changes as a function of time. This is determined by the IMF and by the condensation fractions assumed for the different species. 
The dust chemical composition is dominated by silicates which are always between 65-80\% of the total mass fraction. Carbon dust mass is instead a factor between 0.6 and 1.6 the mass of iron. We are aware of the fact that such chemical composition of dust is expected to be dependent on the metal yields selected for SNe II.
\item \textbf{Efficiency of the outflow and of SN shocks for dust removal}. We estimate the amount of dust removed from the ISM through galactic winds and destroyed by SN shocks for each of the galaxy considered. Depending on the galaxy, between $\approx$20\% and $\approx$90\% of the dust disappearing from the ISM is removed by galactic outflows rather than destroyed by SN shocks. 

For three of the DGS galaxies in our sample NGC 1569, He2-10, NGC 5253) estimates of circumgalactic dust are available from \citet{McCormick18}. We use such a piece of information for fitting the galaxies against the models, the observed and predicted distributions are shown in Fig.~\ref{Fig:circumgalactic}. 
For the three galaxies considered, our approach provides a satisfactory fit of the observed circumgalactic dust fraction.
\item \textbf{Comparison between DGS galaxies and LBGs.} The metallicity values and gas fractions predicted by our models for LBGs are comparable with the ones derived for DGS, however we predict a faster metal enrichment for the LBGs than for DGS galaxies (lower panel of Fig.~\ref{fig:Z}) and the gas is ejected in a shorter time-scale for LBGs ($\approx 400$~Myrs) than for DGS galaxies ($\approx 1 Gyr$).
From Fig.~\ref{fig:DG} it is possible to notice that the typical dust-to-gas ratio of LBGs is comparable with the upper limit observed for DGS galaxies as consequence of the large condensation fraction required to explain the sMdust of these galaxies.
\end{itemize}

\begin{figure}
\centering
\includegraphics[scale=0.5]{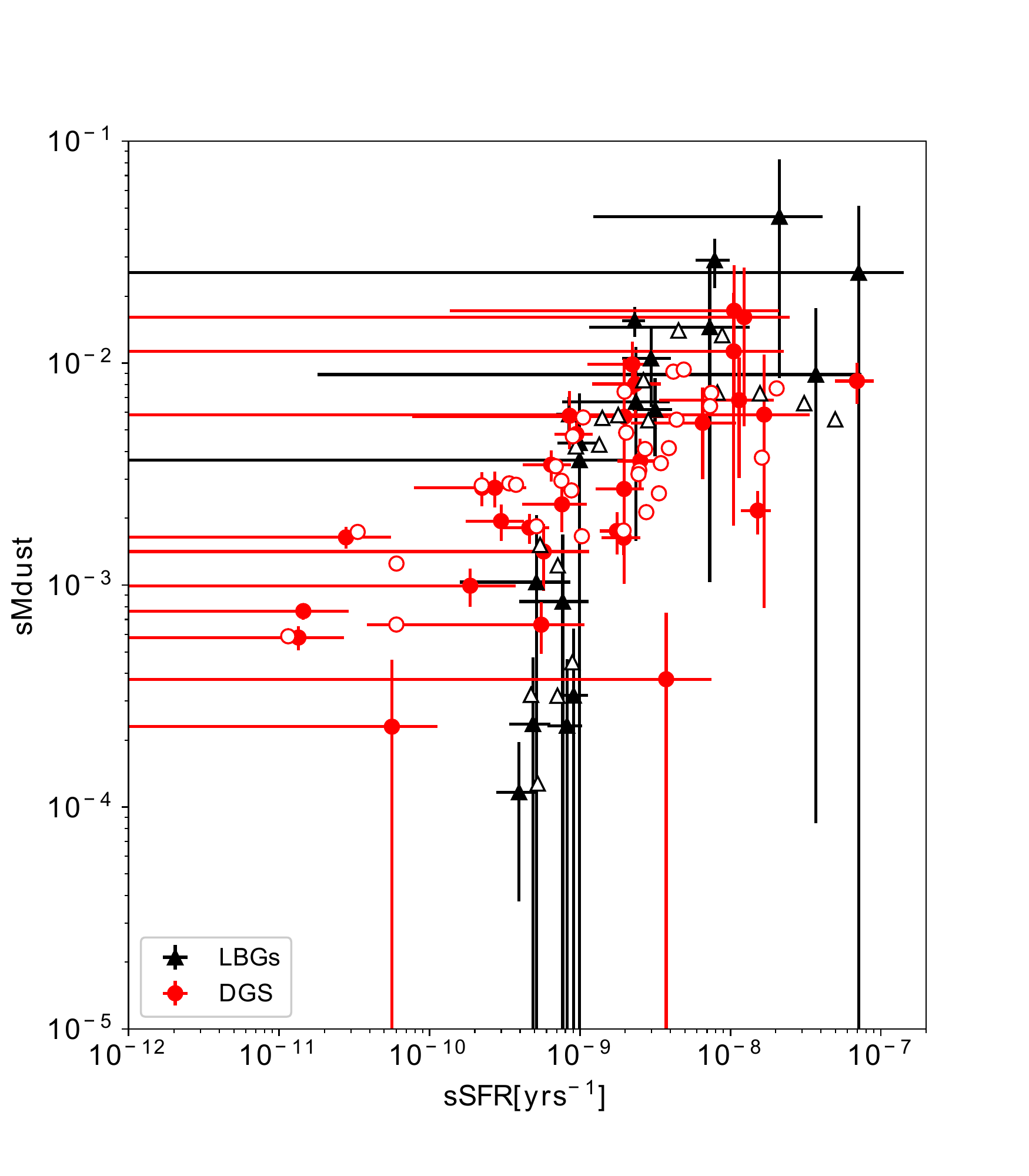}
\includegraphics[scale=0.5]{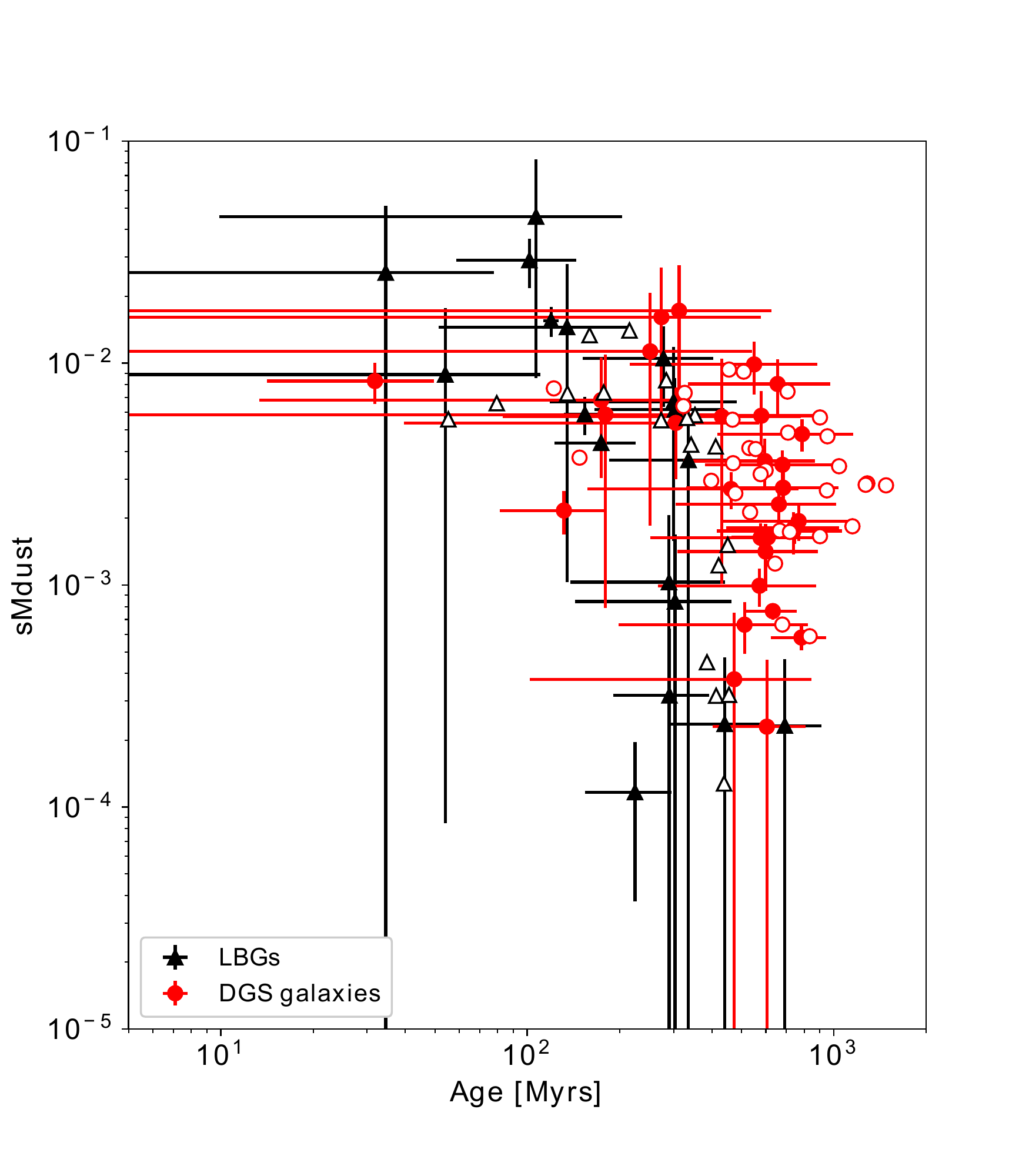}
\caption{Upper panel: the sMdust as a function of the sSFR for DGS galaxies derived from the SED fitting with \textsc{cigale} (full red dots) and LBGs (full black triangles) overplotted with the distribution obtained from the likelihood analysis
of chemical tracks described in Section \ref{sec:method} (empty symbols). 
Lower panel: sMdust as a function of age colour coded as in the upper panel.}
         \label{fig:dust}
   \end{figure}
   
 \begin{figure}
\centering
\includegraphics[scale=0.5]{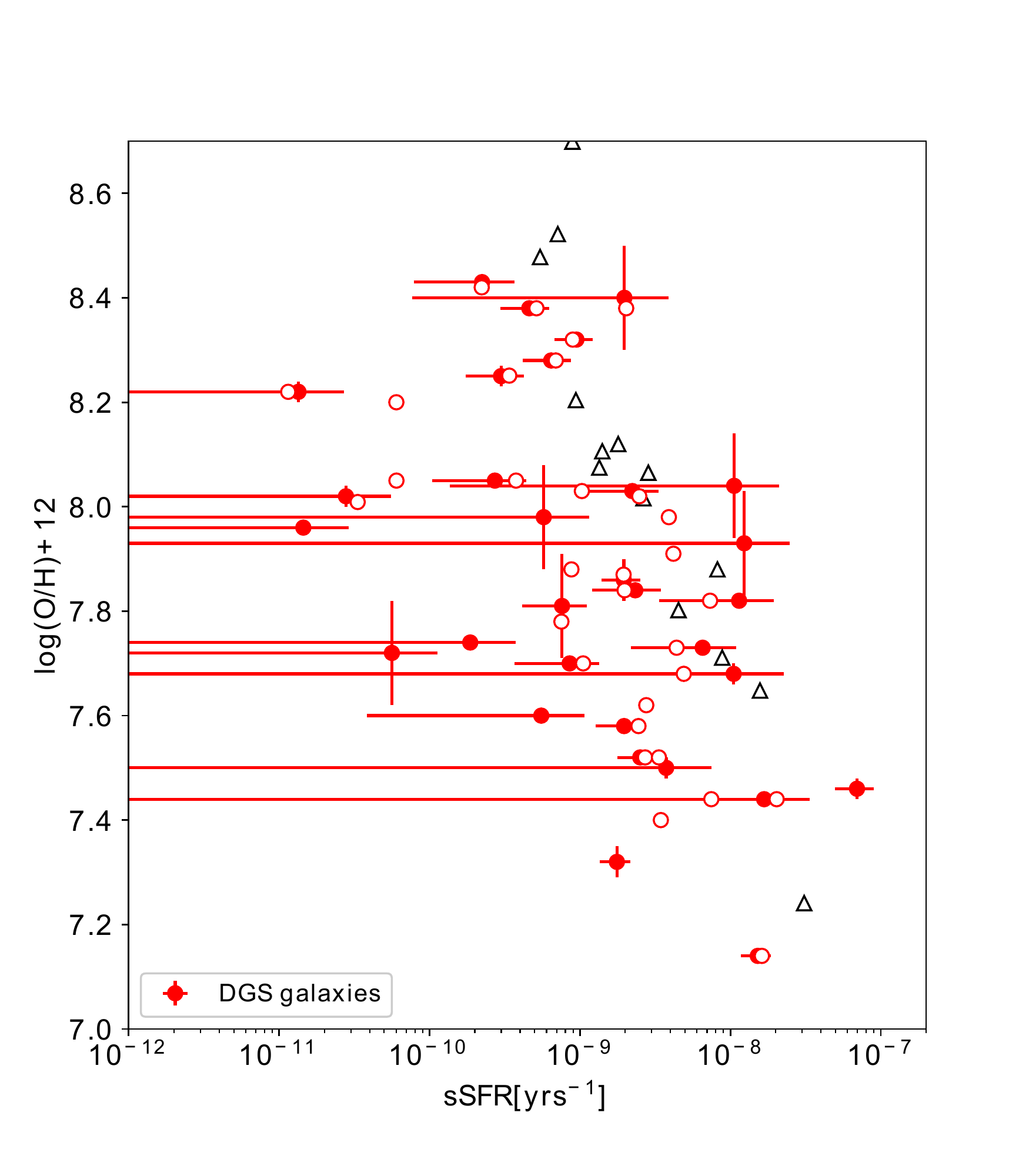}
\includegraphics[scale=0.5]{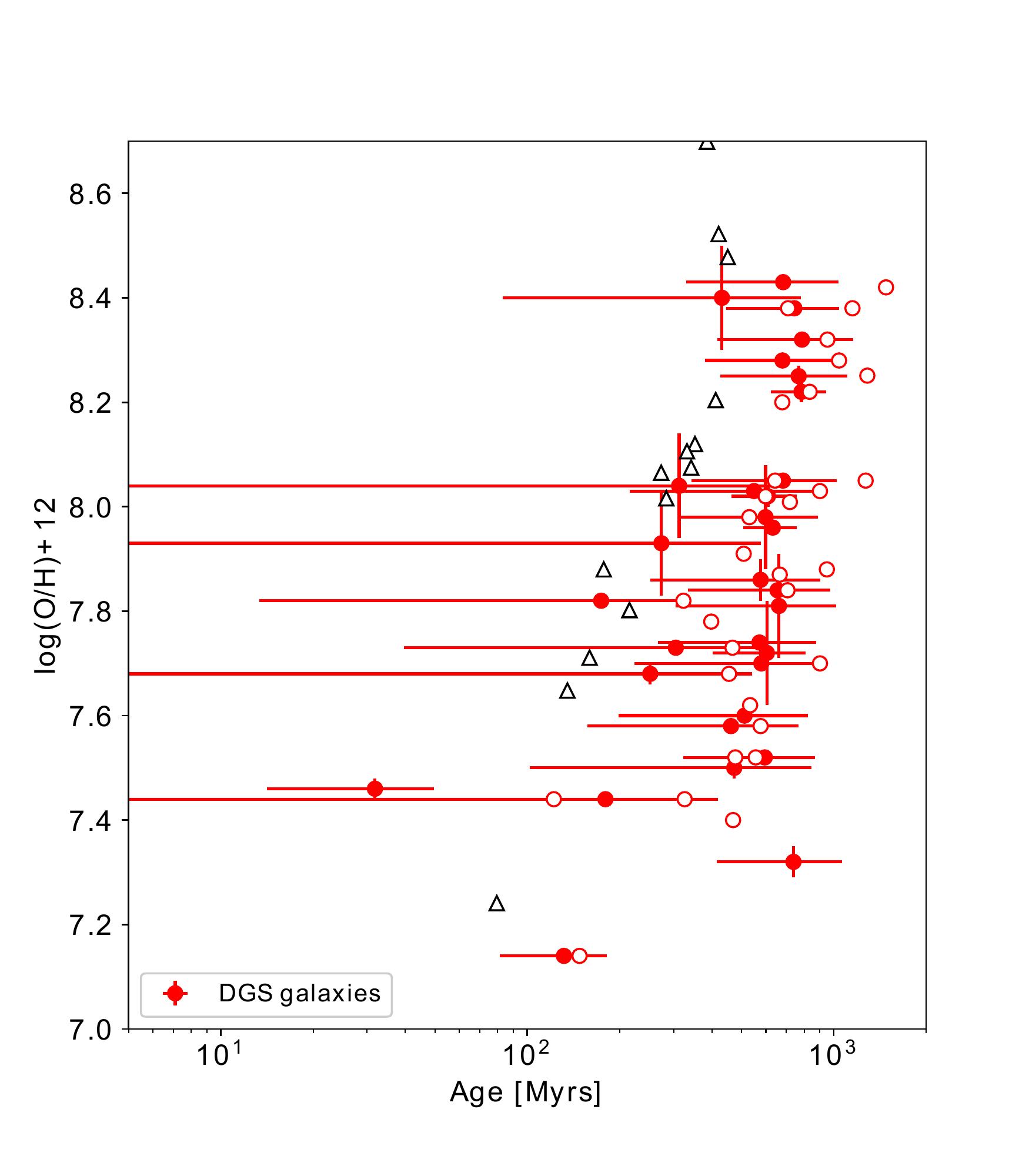}
\caption{Upper panel: the metallicity as a function of the sSFR. Lower panel: the metallicity as a function of age. 
The same colour-coding as in Fig. \ref{fig:dust} is adopted. The predicted distribution for LBGs are here shown for comparison.}
         \label{fig:Z}
   \end{figure}
 
\begin{figure}
\centering
\includegraphics[scale=0.5]{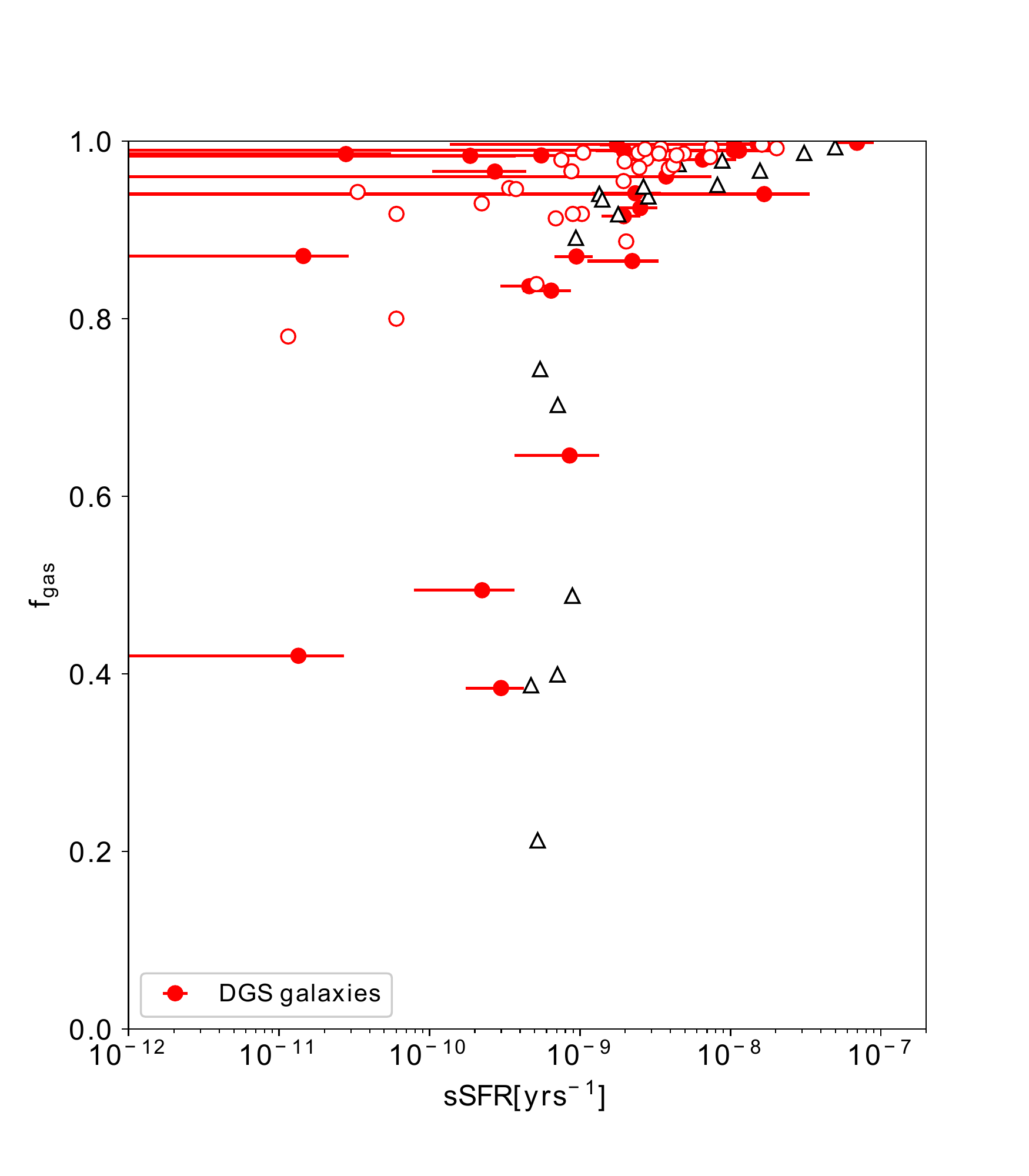}
\includegraphics[scale=0.5]{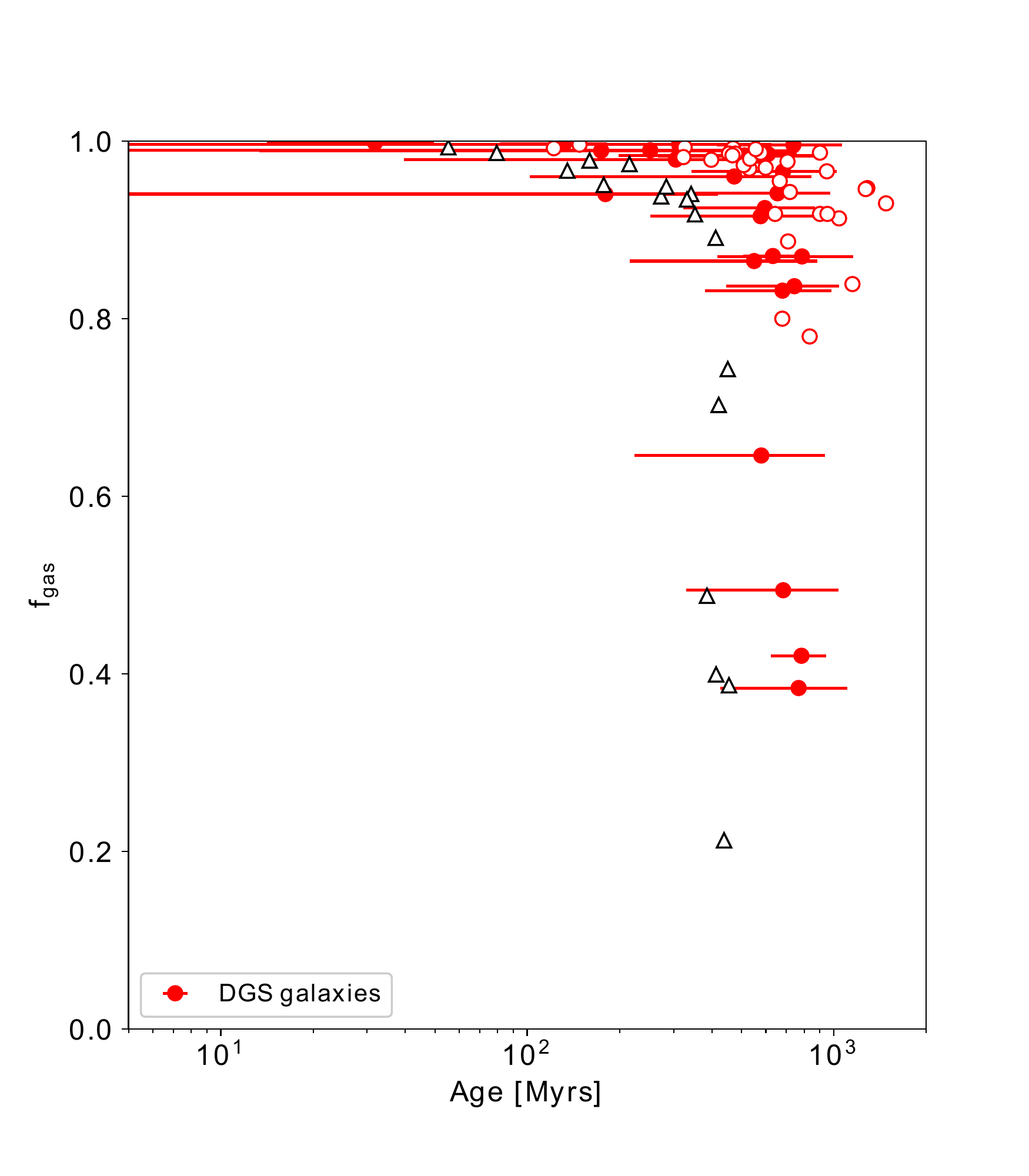}
\caption{Upper panel: the gas fraction as a function of the sSFR. Lower panel: the gas fraction as a function of age. The same colour-coding as in Fig. \ref{fig:dust} is adopted. The predicted distribution for LBGs are here shown for comparison.}
         \label{fig:gas}
   \end{figure}

\begin{figure}
\centering
\includegraphics[scale=0.5]{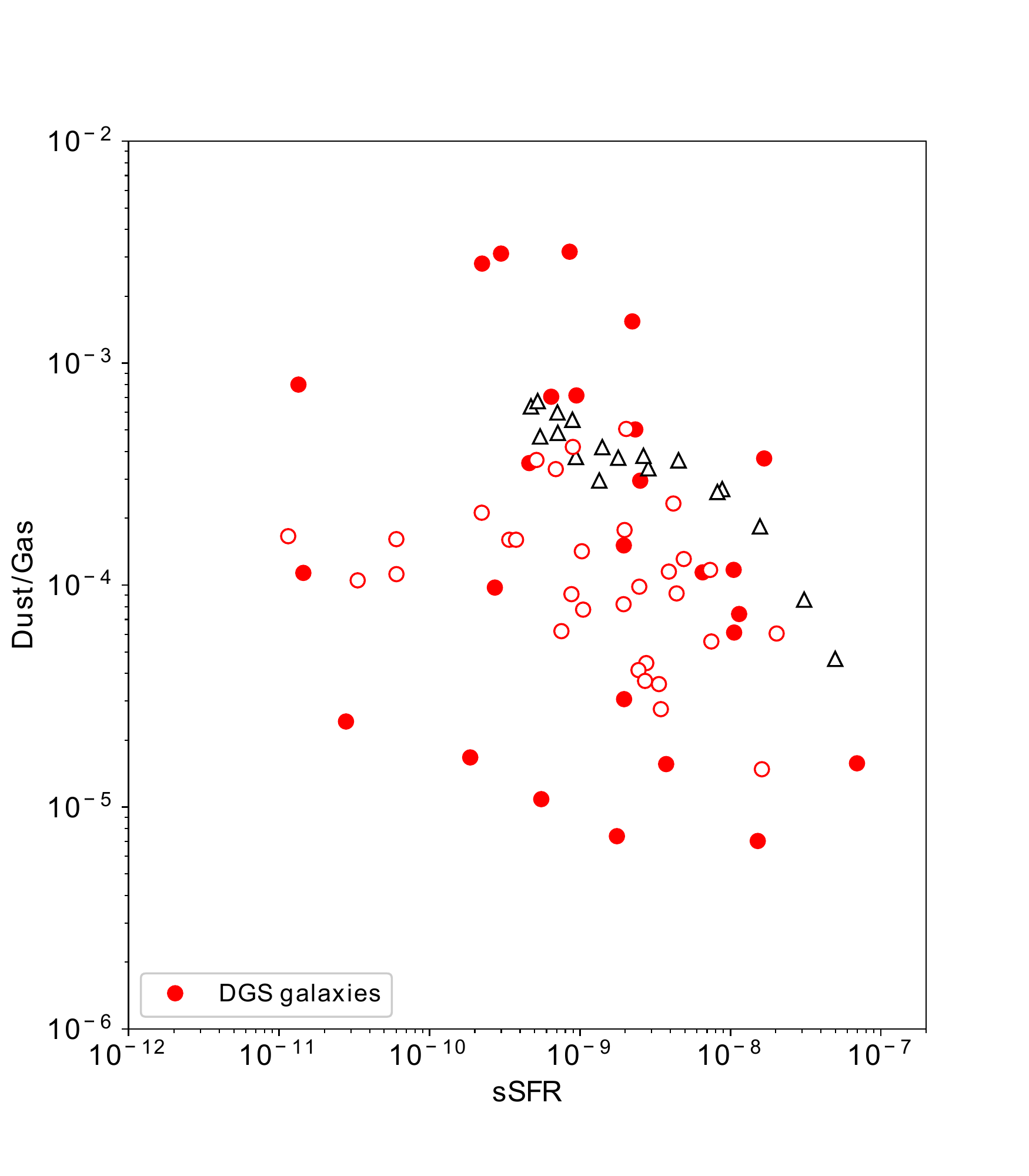}
\includegraphics[scale=0.5]{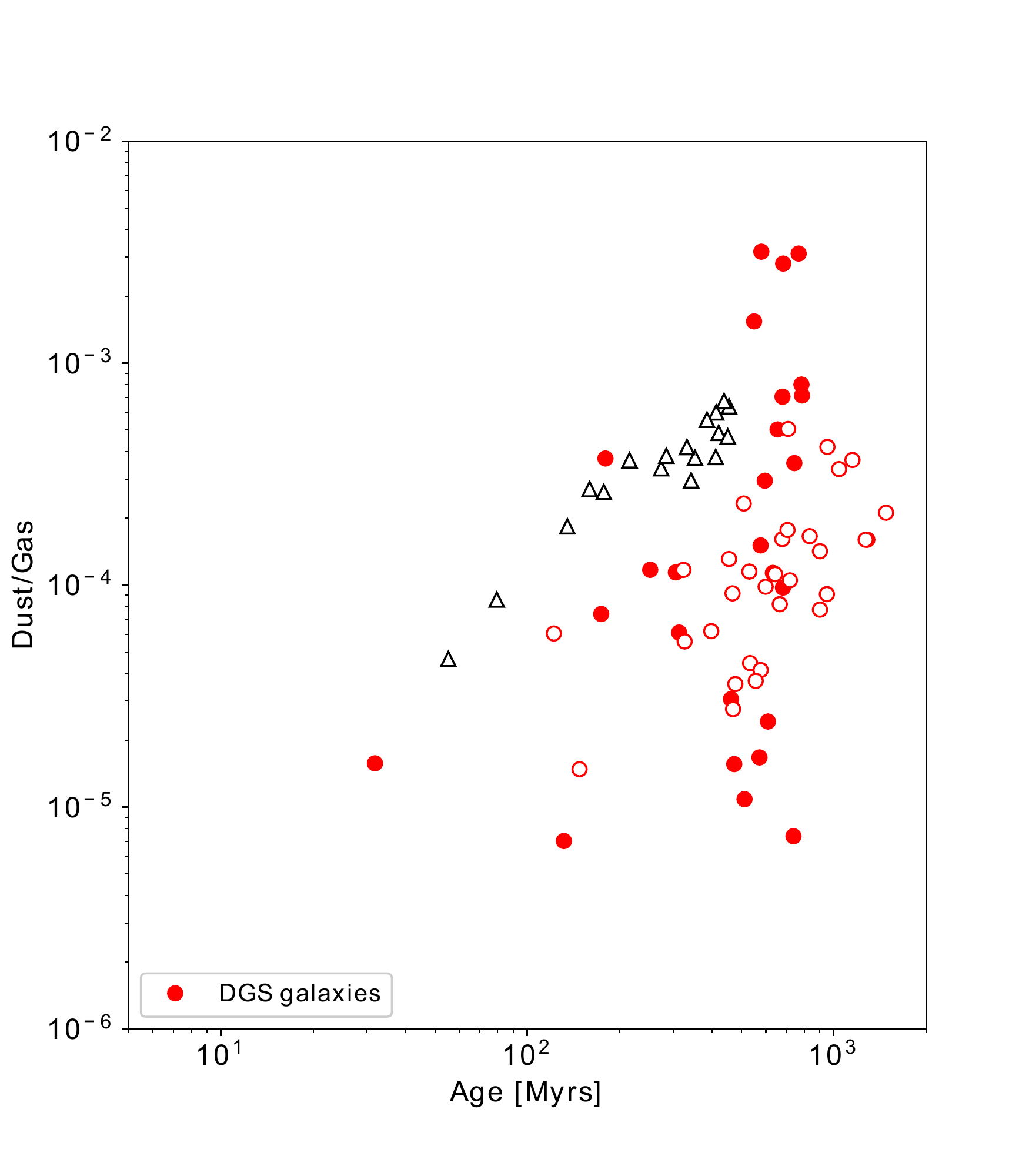}
\caption{Upper panel: the dust-to-gas ratio as a function of the sSFR. Lower panel: the dust-to-gas ratio as a function of age. The same colour-coding as in Fig. \ref{fig:dust} is adopted. The predicted distribution for LBGs are here shown for comparison.}
         \label{fig:DG}
   \end{figure}   
   
   \begin{figure}
\centering
\includegraphics[scale=0.5]{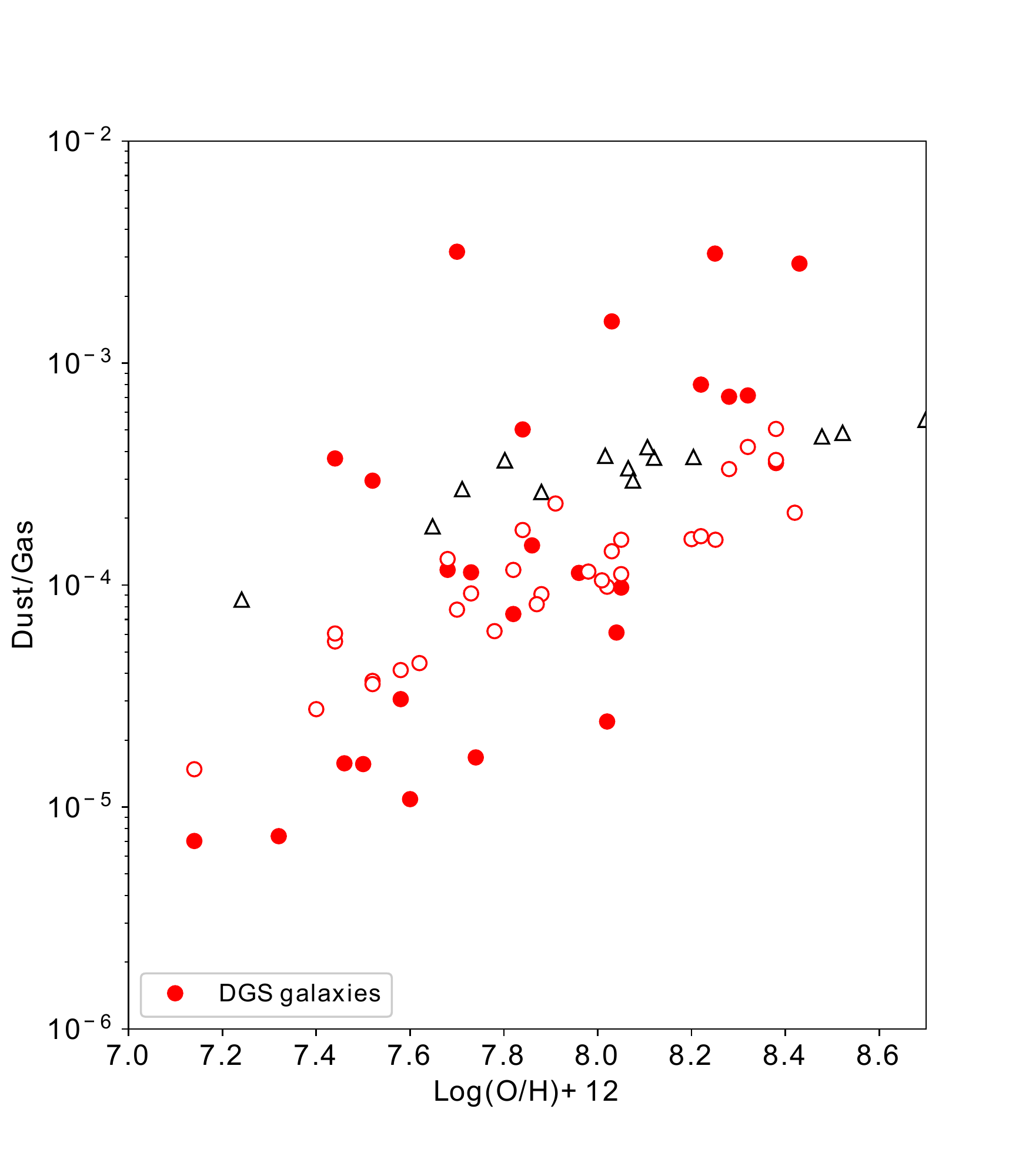}
\caption{The observed and predicted dust-to-gas ratio as a function of the metallicity.
The same colour-coding as in Fig. \ref{fig:dust} is adopted. The predicted distribution for LBGs are here shown for comparison.}
         \label{fig:DG_Z}
   \end{figure}

   \begin{figure}
\centering
\includegraphics[scale=0.5]{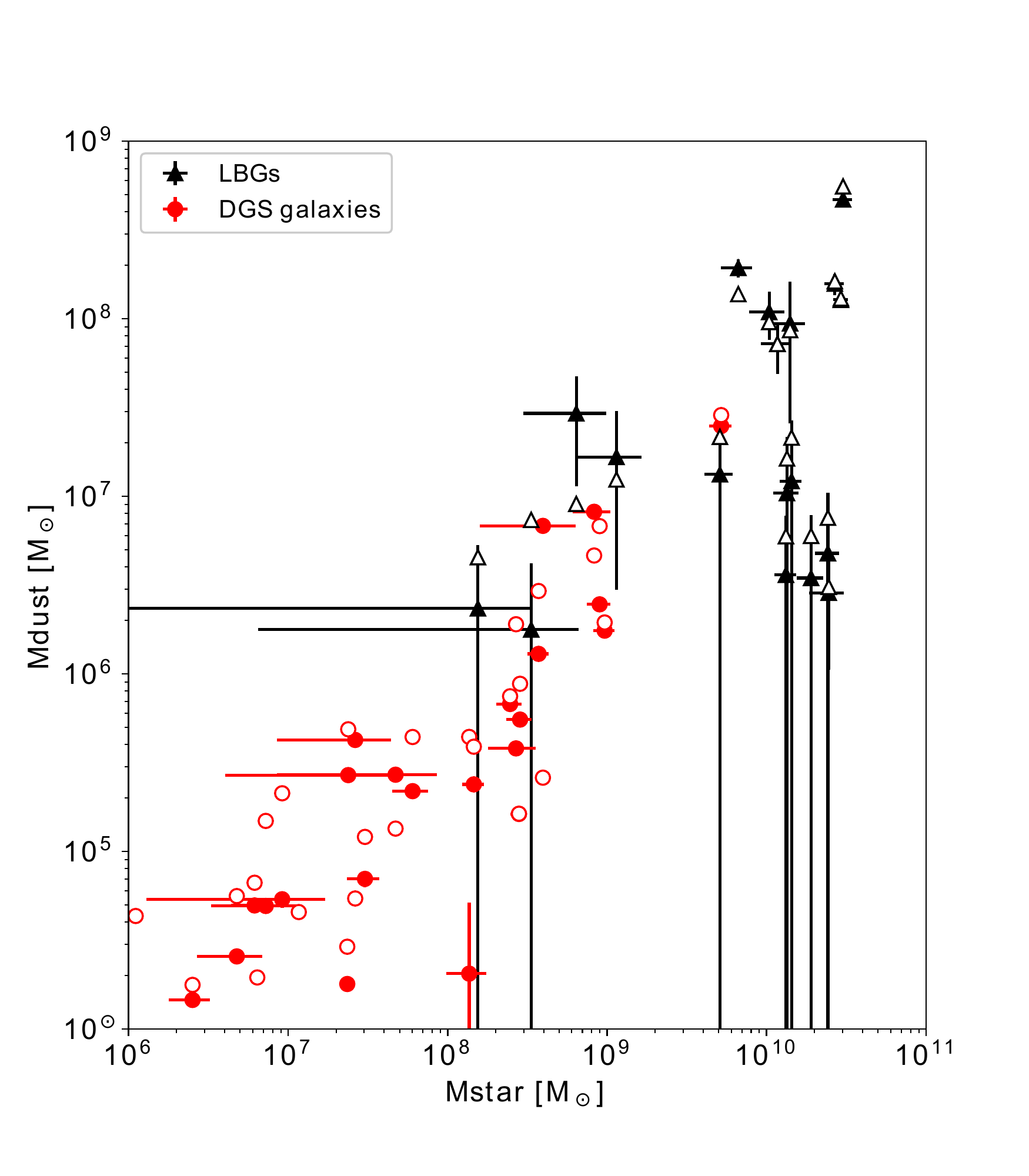}
\caption{The observed and predicted mass of dust as a function of the stellar mass.
The same colour-coding as in Fig. \ref{fig:dust} is adopted. }
         \label{fig:Mdust_Mstar}
   \end{figure}

\section{Discussion}\label{Sec:discussion}
In the present work we show that a top-heavy IMF and high condensation fraction in SN remnants are required in order to have a fast increase of the dust content in the ISM of galaxies within 100 Myrs since the beginning of the baryon cycle. This finding is in agreement with other works in the literature in which the large amount of dust in Quasars at $z>6$ is explained by adopting similar assumptions \citep{Gall11a, Gall11b}. This result is also in agreement with the work from \citet{Liu19} who find that the most of the ALMA observations of galaxies at $z>7$ can be explain by assuming a dust condensation fraction of $\approx 0.5$ for stellar sources. In the local Universe, top-heavy IMF have been adopted to explain some observational properties of ultra compact dwarf galaxies \citep{Dabringhausen12}, ultra faint dwarf galaxies \citep{Geha13, McWilliam2013} and Galactic globular clusters \citep{Marks12}.

On the other hand, in many works in the literature dust growth in the ISM has been considered of fundamental importance in order to explain the amount of dust observed in local and high-redshift galaxies \citep{Asano13, Zhukovska14, Michalowski15, Aoyama17, Ginolfi18, Lesniewska19}. \citet{Popping17} have predicted an efficient dust growth in the ISM for galaxies with stellar masses $\log (M_{*}/M_{\odot}) > 7$, while dust enrichment from stellar sources appears to be dominant for galaxies characterised by lower masses.
\citet{Mancini15} and \citet{Graziani19} have found a similar trend but for larger stellar masses of $\log (M_{*}/M_{\odot}) > 9$ and $8.5 < \log (M_{*}/M_{\odot}) < 9.5$, respectively. The dependence between the efficiency of grain growth in the ISM and the stellar mass of the galaxies found in these works is ascribed to the presence of more metals in more massive galaxies. \citet{Schneider16} have also pointed out the dependence of grain growth efficiency with the density of the gas in the galaxy. 
\citet{daCunha10} use the model from \citet{Calura08} for dwarf galaxies which includes grain growth in the ISM, and in which a Salpeter IMF is adopted for these galaxies. However, the galaxies at larger sMdust and sSFR are not reproduced by their models (see Fig. 10, left panel of \citet{daCunha10}). This result is not in contradiction with our analysis, since we need to assume a top-heavy IMF in order to reproduce the larger values of sMdust and this option has not been explored in their analysis.

In the analysis performed by \citet{RR15} the models developed by \citet{Asano13} are employed.
As shown in Fig.~11 of \citet{RR15}, the sMdust of galaxies with sSFR between $10^{-9}$ and $10^{-8}$ is not reproduced.
Following the analysis by \citet{RR15}, \citet{Devis17, Devis19} reinterpreted the properties of the same sample of galaxies by studying their chemical evolution under different scenarios (e.g. with and without dust growth in the ISM). They have obtained their best fit models and the properties of DGS derived by \citet{RR15} by employing a Chabrier IMF and assuming that a) a star formation history characterised by several bursts as proposed by \citet{Zhukovska14}, b) dust can grow in the ISM but at lower rates with less efficient star formation, and c) a stronger inflow and outflow in DGS galaxies than in less star-forming galaxies. 
Despite the ability of such a model to reproduce the overall observed trends of different samples of local galaxies, the DGS characterised by the largest sSFR remain challenging to explain (see the top-left panel of Fig.~C1 in \citet{Devis17}). 
The tension between this kind of models and observations is even more critical in our analysis since our SED fitting yields values of the sMdust at earlier epochs (larger sSFR) that are even larger than the ones obtained by \citet{RR15} and employed by \citet{Devis17, Devis19}, as discussed in Section~\ref{Sec:prop}. Such a difference implies that larger values of sMdust should be achieved in DGS galaxies earlier than in the aforementioned works. In our work, this effect is obtained by adopting a top-heavy IMF which increases the amount of metals available to form dust already at early epochs coupled with a condensation fraction $>25$\% for SNe II. 
In such a framework, dust growth in the ISM is not needed to produce the total amount of dust observed. Moreover, we show that this process would become efficient when the galactic outflow is expected to start to reduce the amount of gas and dust in the galaxies. Therefore, even if present, the outflow would cancel the effect of dust growth in the ISM.

The trend between the dust-to-gas ratio and the metallicity is well reproduced for DGS galaxies by \citet{Zhukovska14} who has included dust growth in the ISM and very low condensation fraction for silicates in her calculations ($f_{\rm sil}=10^{-3}$). The same observed trend is not well reproduced by assuming a more efficient condensation in SN remnants and no dust growth in the ISM, similarly to what we derive in Fig. \ref{fig:DG_Z}. Contrary to our work and to the analysis by \citet{Devis17, Devis19} a closed-box model is employed in \citet{Zhukovska14} together with a variety of SFH for the galaxies. However, the comparison with the observations is limited to the trend between the dust-to-gas and metallicity. In our analysis, the low condensation fraction from SNe II required to reproduce this trend does not reproduce the large value of sMdust at the beginning of the cycle. The discrepancy between our model predictions and observations might be partially due to the large uncertanties in the determination of the mass of gas in low metallicity galaxies. 

\citet{Ginolfi18} have also proposed dust growth in the ISM of dwarf galaxies to explain the observed amount of dust. However, the investigation is limited to a Salpeter IMF, and the possibility of producing a larger amount of dust by employing a top-heavy IMF is not explored in their work.

From the point of view of the micro physics of grain accretion, \citet{Ferrara16} has indicated that dust accretion in the diffuse ISM is difficult to obtain due to the low gas densities that imply a low accretion rate, even taking into account that the energy barrier between SiO molecules and silicate grain surface has been experimentally found to be zero even at low temperatures \citep{Rouille15}. 
One way of increasing the depletion of the gas elements in the ISM and the size of dust grains is by forming icy mantles around the bulk of the grain in dense molecular clouds, where the density is higher than in the diffuse medium. The icy mantle formed around dust grains can prevent the process of grain accretion by creating an energy barrier between molecules in the gas phase and the bulk of the grain. Icy mantles can however also partly evaporate when stars are formed. Such a process might be particularly efficient if the SFR is relatively high as it is the case for DGS galaxies and LBGs. 
\citet{Ceccarelli18} have further developed the idea proposed by \citet{Ferrara16} by studying how the chemistry of the icy mantles changes as a function of different parameters, including the metallicity. The aim of such a work, is to estimate the probability for Si-bearing species to encounter in the icy mantle and to form clusters that can build up silicate grains. The authors have found a very low probability for cluster formation due to this process, especially at the low metallicities that characterise the galaxies considered in this paper.
Conversely, the results of a recent experiment suggests that silicate and carbon dust may be formed on the grain surface following the evaporation of icy mantles \citep{Rouille20}.
Other laboratory experiments and theoretical works suggest that a non-negligible amount of dust can be formed in low-temperature dusty plasma \citep{Bleecker06,Hollenstein00}. However, under which conditions such a process is relevant (e.g. intensity of the radiation field providing different ionisation fractions for the elements) has only been investigated for the typical conditions of the Milky Way \citep{Zhukovska16}.

The inability of models to reproduce the largest sMdust observed may suggest that dust condensation in SN remnants might be the preferred scenario to reproduce the observations discussed in this paper.

However, how much dust is produced and destroyed in the SN remnants is still very debated. A large amount of dust ($0.1-0.5$~M$_\odot$) is formed \textit{in situ} in SN remnants \citep{Matsuura11}, but it is still unclear how much dust is destroyed by the reverse shock or reformed. \citet{Matsuura19} have interpreted the increase in the 31.5~$\mu$m photometry in SN1987A obtained from SOFIA FORCAST 11.1, 19.7 in 2016 with respect to 10 years earlier as a possible indication of dust re-condensation in the forward shock. 
Indeed, the mass of dust needed to explain the emission of this dust component at a temperature of $\sim 85$~K is more than 10 times larger than the one estimated 10 years before. This might indicate that dust could be reformed after being destroyed by shocks. Such an interpretation should be confirmed by the future time of the observed spectra. Indeed, another possible explanation for the observed emission is that the decrease in the density of the expanding ejecta has allowed X-rays to efficiently heat the freshly formed dust.
The re-formation of dust in the post-shocked regions has also found theoretical confirmation \citep{Sarangi18}.
From the observational point of view \citet{Gall14} have found that large dust grains are rapidly formed in SN remnants. Such grains are large enough to be able to survive to the reverse shock. 
From the theoretical side, different authors have studied the formation and dust survival in SN ejecta. On one hand, \citet{Bianchi07, Bocchio16} have estimated that only between 2-20\% of the dust mass observed in SN remnants will survive to the reverse shock, and that the mass of dust observed around SN1987A has not been yet destroyed by the passage of the reverse shock. On the other hand, different calculations suggest that a lager fraction (42-98\%) of the newly produced dust can survive to the reverse shock in clumpy ejecta \citep{Biscaro16}. The survival of dust grains can also depend on the their size, on the energetic of the explosion, on the gas density of the ISM and on the thickness of the hydrogen envelope \citep{Nozawa07}.

In our work we also show that another condition to reproduce the metallicity trend, is related to the initial gas content in the galaxies that has to be at least more than 10 times the final mass of stars. In this context, galactic outflows are needed to efficiently remove gas and dust from the galaxy.
Such a result is observationally supported for LBGs in which outflows are commonly observed \citep{Shapley03, Pettini02, Gallerani18}, while star formation is rather inefficient.
As far as DGS galaxies are concerned, circumgalactic dust has been detected in different galaxies included in our sample \citep[e.g. NGC 1569, NGC 5253, He 2-10][]{McCormick18, Suzuki18}. 
On the theoretical point of view, inflows and outflows are needed to explain the metallicity and gas content of galaxies at redshift of 3.4 \citep{Troncoso14}.

\section{Conclusions}\label{Sec:conclusions}
In this paper we develop and revise the existing prescriptions for dust evolution in galaxies in order to interpret the observations of DGS galaxies in the local Universe and of LBGs at high-redshift.

Different investigations in the literature show the inadequacy of the current models for reproducing the largest values of the sMdust of mass observed at the largest sSFR for DGS galaxies. The inability of the current models for reproducing the observations, is even more severe in our investigation, in which the sMdust and sSFR are even larger than the previous estimates in the literature \citep{RR15}.
This indicates that the state-of-the-art framework adopted for interpreting the dust content in these galaxies need to be revised. In particular, the dust prescriptions that assume low condensation efficiency for SN and dust growth in the ISM are not able to reproduce the largest sMdust at early epochs estimated for DGS galaxies and LBGs.
For LBGs we develop a new description for metal and dust evolution that is in part based on the results obtained for DGS that are their nearby counterparts.

For both samples of galaxies, we reproduce the observations by adopting a top-heavy IMF which favour the fast enrichment of the ISM through SNe II and dust condensation fraction $> 25$\% for DGS galaxies and $>40$\% for LBGs. In this context, the dust enrichment from TP-AGB stars and from Type Ia SNe plays a minor role, since a top-heavy IMF is not favourable for the formation of massive stars.
Galactic outflows are essential to reproduce the decline of the sMdust (and of the gas fraction for DGS galaxies) as a function of sSFR. Dust destruction by SN usually plays a secondary role with respect to dust removal from the galactic outflow. 
A typical star formation efficiency (the mass of gas converted into stars) of few per cent is instead required to reproduce the trends with the metallicity.

Despite grain growth is often considered to be a fundamental process to explain the amount of dust observed, we find that the contribution of such a process starts to be relevant when the outflow start to remove dust from the ISM, and therefore its presence would be masked by the efficient removal of dust from the galaxies.
\begin{acknowledgements}
AN acknowledges the support of the Centre National
d’Etudes Spatiale (CNES) through a post-doctoral fellowship.
PT work was supported by the Programme National “Physique et
Chimie du Milieu Interstellaire” (PCMI) of CNRS/INSU with INC/INP co-funded by CEA and CNES.
BC was supported by the ERC Consolidator Grant (Hungary) funding scheme (Project RADIOSTAR, G.A. n. 724560).
HH thanks the Ministry of Science and Technology for support through grant MOST 107-2923-M-001-003-MY3 and MOST 108- 2112-M-001-007-MY3, and the Academia Sinica for Investigator Award AS-IA-109-M02. 
We are grateful to the anonymous referee for carefully reading the manuscript and for helping in improving it.
\end{acknowledgements}

\bibliographystyle{aa}
\bibliography{nanni}

\begin{thebibliography}{94}
\expandafter\ifx\csname natexlab\endcsname\relax\def\natexlab#1{#1}\fi

\bibitem[{{Ag{\'u}ndez} \& {Wakelam}(2013)}]{Agundez13}
{Ag{\'u}ndez}, M. \& {Wakelam}, V. 2013, Chemical Reviews, 113, 8710

\bibitem[{{Aoyama} {et~al.}(2017){Aoyama}, {Hou}, {Shimizu}, {Hirashita},
  {Todoroki}, {Choi}, \& {Nagamine}}]{Aoyama17}
{Aoyama}, S., {Hou}, K.-C., {Shimizu}, I., {et~al.} 2017, \mnras, 466, 105

\bibitem[{{Asano} {et~al.}(2013){Asano}, {Takeuchi}, {Hirashita}, \&
  {Inoue}}]{Asano13}
{Asano}, R.~S., {Takeuchi}, T.~T., {Hirashita}, H., \& {Inoue}, A.~K. 2013,
  EPS, 65, 213

\bibitem[{{Bianchi} \& {Schneider}(2007)}]{Bianchi07}
{Bianchi}, S. \& {Schneider}, R. 2007, \mnras, 378, 973

\bibitem[{{Biscaro} \& {Cherchneff}(2016)}]{Biscaro16}
{Biscaro}, C. \& {Cherchneff}, I. 2016, \aap, 589, A132

\bibitem[{Bleecker {et~al.}(2006)Bleecker, Bogaerts, \& Goedheer}]{Bleecker06}
Bleecker, K., Bogaerts, A., \& Goedheer, W. 2006, Physical review. E,
  Statistical, nonlinear, and soft matter physics, 73, 026405

\bibitem[{{Bocchio} {et~al.}(2016){Bocchio}, {Marassi}, {Schneider}, {Bianchi},
  {Limongi}, \& {Chieffi}}]{Bocchio16}
{Bocchio}, M., {Marassi}, S., {Schneider}, R., {et~al.} 2016, \aap, 587, A157

\bibitem[{{Boquien} {et~al.}(2019){Boquien}, {Burgarella}, {Roehlly}, {Buat},
  {Ciesla}, {Corre}, {Inoue}, \& {Salas}}]{Boquien19}
{Boquien}, M., {Burgarella}, D., {Roehlly}, Y., {et~al.} 2019, \aap, 622, A103

\bibitem[{{Bruzual} \& {Charlot}(2003)}]{BC03}
{Bruzual}, G. \& {Charlot}, S. 2003, \mnras, 344, 1000

\bibitem[{{Burgarella} {et~al.}(2020){Burgarella}, {Nanni}, {Hirashita},
  {Theule}, {Inoue}, \& {Takeuchi}}]{Burgarella20}
{Burgarella}, D., {Nanni}, A., {Hirashita}, H., {et~al.} 2020, arXiv e-prints,
  arXiv:2002.01858

\bibitem[{{Cai} {et~al.}(2020){Cai}, {Zotti}, \& {Bonato}}]{Cai20}
{Cai}, Z.-Y., {Zotti}, G.~D., \& {Bonato}, M. 2020, \apj, 891, 74

\bibitem[{{Calura} {et~al.}(2014){Calura}, {Gilli}, {Vignali}, {Pozzi},
  {Pipino}, \& {Matteucci}}]{Calura14}
{Calura}, F., {Gilli}, R., {Vignali}, C., {et~al.} 2014, \mnras, 438, 2765

\bibitem[{{Calura} {et~al.}(2008){Calura}, {Pipino}, \& {Matteucci}}]{Calura08}
{Calura}, F., {Pipino}, A., \& {Matteucci}, F. 2008, \aap, 479, 669

\bibitem[{{Ceccarelli} {et~al.}(2018){Ceccarelli}, {Viti}, {Balucani}, \&
  {Taquet}}]{Ceccarelli18}
{Ceccarelli}, C., {Viti}, S., {Balucani}, N., \& {Taquet}, V. 2018, \mnras,
  476, 1371

\bibitem[{{Chabrier}(2003)}]{Chabrier03}
{Chabrier}, G. 2003, \pasp, 115, 763

\bibitem[{{C{\^o}t{\'e}} {et~al.}(2017){C{\^o}t{\'e}}, {O'Shea}, {Ritter},
  {Herwig}, \& {Venn}}]{Cote17}
{C{\^o}t{\'e}}, B., {O'Shea}, W., {Ritter}, C., {Herwig}, F., \& {Venn}, K.~A.
  2017, \apj, 835, 128

\bibitem[{{Cristallo} {et~al.}(2015){Cristallo}, {Straniero}, {Piersanti}, \&
  {Gobrecht}}]{Cristallo15}
{Cristallo}, S., {Straniero}, O., {Piersanti}, L., \& {Gobrecht}, D. 2015,
  \apjs, 219, 40

\bibitem[{{da Cunha} {et~al.}(2010){da Cunha}, {Eminian}, {Charlot}, \&
  {Blaizot}}]{daCunha10}
{da Cunha}, E., {Eminian}, C., {Charlot}, S., \& {Blaizot}, J. 2010, \mnras,
  403, 1894

\bibitem[{{Dabringhausen} {et~al.}(2012){Dabringhausen}, {Kroupa},
  {Pflamm-Altenburg}, \& {Mieske}}]{Dabringhausen12}
{Dabringhausen}, J., {Kroupa}, P., {Pflamm-Altenburg}, J., \& {Mieske}, S.
  2012, \apj, 747, 72

\bibitem[{{De Vis} {et~al.}(2017){De Vis}, {Gomez}, {Schofield}, {Maddox},
  {Dunne}, {Baes}, {Cigan}, {Clark}, {Gomez}, {Lara-L{\'o}pez}, \&
  {Owers}}]{Devis17}
{De Vis}, P., {Gomez}, H.~L., {Schofield}, S.~P., {et~al.} 2017, \mnras, 471,
  1743

\bibitem[{{De Vis} {et~al.}(2019){De Vis}, {Jones}, {Viaene}, {Casasola},
  {Clark}, {Baes}, {Bianchi}, {Cassara}, {Davies}, {De Looze}, {Galametz},
  {Galliano}, {Lianou}, {Madden}, {Manilla-Robles}, {Mosenkov}, {Nersesian},
  {Roychowdhury}, {Xilouris}, \& {Ysard}}]{Devis19}
{De Vis}, P., {Jones}, A., {Viaene}, S., {et~al.} 2019, \aap, 623, A5

\bibitem[{{Dwek}(1998)}]{Dwek98}
{Dwek}, E. 1998, \apj, 501, 643

\bibitem[{{Dwek} {et~al.}(2007){Dwek}, {Galliano}, \& {Jones}}]{Dwek07}
{Dwek}, E., {Galliano}, F., \& {Jones}, A.~P. 2007, \apj, 662, 927

\bibitem[{{Ferrara} {et~al.}(2016){Ferrara}, {Viti}, \&
  {Ceccarelli}}]{Ferrara16}
{Ferrara}, A., {Viti}, S., \& {Ceccarelli}, C. 2016, \mnras, 463, L112

\bibitem[{{Gail} \& {Sedlmayr}(1999)}]{Gail99}
{Gail}, H.~P. \& {Sedlmayr}, E. 1999, \aap, 347, 594

\bibitem[{{Gall} {et~al.}(2011{\natexlab{a}}){Gall}, {Andersen}, \&
  {Hjorth}}]{Gall11a}
{Gall}, C., {Andersen}, A.~C., \& {Hjorth}, J. 2011{\natexlab{a}}, \aap, 528,
  A13

\bibitem[{{Gall} {et~al.}(2011{\natexlab{b}}){Gall}, {Andersen}, \&
  {Hjorth}}]{Gall11b}
{Gall}, C., {Andersen}, A.~C., \& {Hjorth}, J. 2011{\natexlab{b}}, \aap, 528,
  A14

\bibitem[{{Gall} {et~al.}(2014){Gall}, {Hjorth}, {Watson}, {Dwek}, {Maund},
  {Fox}, {Leloudas}, {Malesani}, \& {Day-Jones}}]{Gall14}
{Gall}, C., {Hjorth}, J., {Watson}, D., {et~al.} 2014, \nat, 511, 326

\bibitem[{{Gallerani} {et~al.}(2018){Gallerani}, {Pallottini}, {Feruglio},
  {Ferrara}, {Maiolino}, {Vallini}, {Riechers}, \& {Pavesi}}]{Gallerani18}
{Gallerani}, S., {Pallottini}, A., {Feruglio}, C., {et~al.} 2018, \mnras, 473,
  1909

\bibitem[{{Geha} {et~al.}(2013){Geha}, {Brown}, {Tumlinson}, {Kalirai},
  {Simon}, {Kirby}, {Vand enBerg}, {Mu{\~n}oz}, {Avila}, {Guhathakurta}, \&
  {Ferguson}}]{Geha13}
{Geha}, M., {Brown}, T.~M., {Tumlinson}, J., {et~al.} 2013, \apj, 771, 29

\bibitem[{{Ginolfi} {et~al.}(2018){Ginolfi}, {Graziani}, {Schneider},
  {Marassi}, {Valiante}, {Dell'Agli}, {Ventura}, \& {Hunt}}]{Ginolfi18}
{Ginolfi}, M., {Graziani}, L., {Schneider}, R., {et~al.} 2018, \mnras, 473,
  4538

\bibitem[{{Gioannini} {et~al.}(2017){Gioannini}, {Matteucci}, \&
  {Calura}}]{Gioannini17}
{Gioannini}, L., {Matteucci}, F., \& {Calura}, F. 2017, \mnras, 471, 4615

\bibitem[{{Gomez} {et~al.}(2012){Gomez}, {Clark}, {Nozawa}, {Krause}, {Gomez},
  {Matsuura}, {Barlow}, {Besel}, {Dunne}, {Gear}, {Hargrave}, {Henning},
  {Ivison}, {Sibthorpe}, {Swinyard}, \& {Wesson}}]{Gomez12}
{Gomez}, H.~L., {Clark}, C.~J.~R., {Nozawa}, T., {et~al.} 2012, \mnras, 420,
  3557

\bibitem[{{Graziani} {et~al.}(2019){Graziani}, {Schneider}, {Ginolfi}, {Hunt},
  {Maio}, {Glatzle}, \& {Ciardi}}]{Graziani19}
{Graziani}, L., {Schneider}, R., {Ginolfi}, M., {et~al.} 2019, arXiv e-prints,
  arXiv:1909.07388

\bibitem[{{Heger} \& {Woosley}(2010)}]{Heger10}
{Heger}, A. \& {Woosley}, S.~E. 2010, \apj, 724, 341

\bibitem[{{Herwig}(2005)}]{Herwig05}
{Herwig}, F. 2005, \araa, 43, 435

\bibitem[{{Hirashita}(1999)}]{Hirashita99}
{Hirashita}, H. 1999, \apj, 522, 220

\bibitem[{{Hirashita}(2000)}]{Hirashita00}
{Hirashita}, H. 2000, \pasj, 52, 585

\bibitem[{{Hirashita}(2012)}]{Hirashita12}
{Hirashita}, H. 2012, \mnras, 422, 1263

\bibitem[{{Hirashita} \& {Kuo}(2011)}]{Hirashita11}
{Hirashita}, H. \& {Kuo}, T.-M. 2011, \mnras, 416, 1340

\bibitem[{{H{\"o}fner} \& {Olofsson}(2018)}]{Hofner18}
{H{\"o}fner}, S. \& {Olofsson}, H. 2018, \aapr, 26, 1

\bibitem[{Hollenstein(2000)}]{Hollenstein00}
Hollenstein, C. 2000, Plasma Physics and Controlled Fusion, 42, R93

\bibitem[{{Iwamoto} {et~al.}(1999){Iwamoto}, {Brachwitz}, {Nomoto},
  {Kishimoto}, {Umeda}, {Hix}, \& {Thielemann}}]{Iwamoto99}
{Iwamoto}, K., {Brachwitz}, F., {Nomoto}, K., {et~al.} 1999, \apjs, 125, 439

\bibitem[{{Karakas}(2010)}]{Karakas10}
{Karakas}, A.~I. 2010, \mnras, 403, 1413

\bibitem[{{Kobayashi} {et~al.}(2006){Kobayashi}, {Umeda}, {Nomoto}, {Tominaga},
  \& {Ohkubo}}]{Kobayashi06}
{Kobayashi}, C., {Umeda}, H., {Nomoto}, K., {Tominaga}, N., \& {Ohkubo}, T.
  2006, \apj, 653, 1145

\bibitem[{{Le{\'s}niewska} \& {Micha{\l}owski}(2019)}]{Lesniewska19}
{Le{\'s}niewska}, A. \& {Micha{\l}owski}, M.~J. 2019, \aap, 624, L13

\bibitem[{{Limongi} \& {Chieffi}(2018)}]{Limongi18}
{Limongi}, M. \& {Chieffi}, A. 2018, \apjs, 237, 13

\bibitem[{{Lisenfeld} \& {Ferrara}(1998)}]{Lisenfeld98}
{Lisenfeld}, U. \& {Ferrara}, A. 1998, \apj, 496, 145

\bibitem[{{Liu} \& {Hirashita}(2019)}]{Liu19}
{Liu}, H.-M. \& {Hirashita}, H. 2019, \mnras, 490, 540

\bibitem[{{Lodders}(2010)}]{Lodders10}
{Lodders}, K. 2010, Astrophysics and Space Science Proceedings, 16, 379

\bibitem[{{Madden} {et~al.}(2013){Madden}, {R{\'e}my-Ruyer}, {Galametz},
  {Cormier}, {Lebouteiller}, {Galliano}, {Hony}, {Bendo}, {Smith}, {Pohlen},
  {Roussel}, {Sauvage}, {Wu}, {Sturm}, {Poglitsch}, {Contursi}, {Doublier},
  {Baes}, {Barlow}, {Boselli}, {Boquien}, {Carlson}, {Ciesla}, {Cooray},
  {Cortese}, {de Looze}, {Irwin}, {Isaak}, {Kamenetzky}, {Karczewski}, {Lu},
  {MacHattie}, {O'Halloran}, {Parkin}, {Rangwala}, {Schirm}, {Schulz},
  {Spinoglio}, {Vaccari}, {Wilson}, \& {Wozniak}}]{Madden13}
{Madden}, S.~C., {R{\'e}my-Ruyer}, A., {Galametz}, M., {et~al.} 2013, \pasp,
  125, 600

\bibitem[{{Mancini} {et~al.}(2015){Mancini}, {Schneider}, {Graziani},
  {Valiante}, {Dayal}, {Maio}, {Ciardi}, \& {Hunt}}]{Mancini15}
{Mancini}, M., {Schneider}, R., {Graziani}, L., {et~al.} 2015, \mnras, 451, L70

\bibitem[{{Marassi} {et~al.}(2019){Marassi}, {Schneider}, {Limongi}, {Chieffi},
  {Graziani}, \& {Bianchi}}]{Marassi19}
{Marassi}, S., {Schneider}, R., {Limongi}, M., {et~al.} 2019, \mnras, 484, 2587

\bibitem[{{Marks} {et~al.}(2012){Marks}, {Kroupa}, {Dabringhausen}, \&
  {Pawlowski}}]{Marks12}
{Marks}, M., {Kroupa}, P., {Dabringhausen}, J., \& {Pawlowski}, M.~S. 2012,
  \mnras, 422, 2246

\bibitem[{{Mathis} {et~al.}(1977){Mathis}, {Rumpl}, \& {Nordsieck}}]{Mathis77}
{Mathis}, J.~S., {Rumpl}, W., \& {Nordsieck}, K.~H. 1977, \apj, 217, 425

\bibitem[{{Matsuura} {et~al.}(2019){Matsuura}, {De Buizer}, {Arendt}, {Dwek},
  {Barlow}, {Bevan}, {Cigan}, {Gomez}, {Rho}, {Wesson}, {Bouchet}, {Danziger},
  \& {Meixner}}]{Matsuura19}
{Matsuura}, M., {De Buizer}, J.~M., {Arendt}, R.~G., {et~al.} 2019, \mnras,
  482, 1715

\bibitem[{{Matsuura} {et~al.}(2011){Matsuura}, {Dwek}, {Meixner}, {Otsuka},
  {Babler}, {Barlow}, {Roman-Duval}, {Engelbracht}, {Sandstrom},
  {Laki{\'c}evi{\'c}}, {van Loon}, {Sonneborn}, {Clayton}, {Long}, {Lundqvist},
  {Nozawa}, {Gordon}, {Hony}, {Panuzzo}, {Okumura}, {Misselt}, {Montiel}, \&
  {Sauvage}}]{Matsuura11}
{Matsuura}, M., {Dwek}, E., {Meixner}, M., {et~al.} 2011, Science, 333, 1258

\bibitem[{{Mattsson}(2020)}]{Mattsson20}
{Mattsson}, L. 2020, \mnras, 491, 4334

\bibitem[{{Mattsson} {et~al.}(2019){Mattsson}, {Fynbo}, \&
  {Villarroel}}]{Mattsson19}
{Mattsson}, L., {Fynbo}, J.~P.~U., \& {Villarroel}, B. 2019, \mnras, 490, 5788

\bibitem[{{McCormick} {et~al.}(2018){McCormick}, {Veilleux}, {Mel{\'e}ndez},
  {Martin}, {Bland -Hawthorn}, {Cecil}, {Heitsch}, {M{\"u}ller}, {Rupke}, \&
  {Engelbracht}}]{McCormick18}
{McCormick}, A., {Veilleux}, S., {Mel{\'e}ndez}, M., {et~al.} 2018, \mnras,
  477, 699

\bibitem[{{McKee}(1989)}]{McKee89}
{McKee}, C. 1989, in IAU Symposium, Vol. 135, Interstellar Dust, ed. L.~J.
  {Allamandola} \& A.~G.~G.~M. {Tielens}, 431

\bibitem[{{McKinnon} {et~al.}(2018){McKinnon}, {Vogelsberger}, {Torrey},
  {Marinacci}, \& {Kannan}}]{McKinnon18}
{McKinnon}, R., {Vogelsberger}, M., {Torrey}, P., {Marinacci}, F., \& {Kannan},
  R. 2018, \mnras, 478, 2851

\bibitem[{{McWilliam} {et~al.}(2013){McWilliam}, {Wallerstein}, \&
  {Mottini}}]{McWilliam2013}
{McWilliam}, A., {Wallerstein}, G., \& {Mottini}, M. 2013, \apj, 778, 149

\bibitem[{{Micha{\l}owski}(2015)}]{Michalowski15}
{Micha{\l}owski}, M.~J. 2015, \aap, 577, A80

\bibitem[{{Murray} {et~al.}(2005){Murray}, {Quataert}, \&
  {Thompson}}]{Murray05}
{Murray}, N., {Quataert}, E., \& {Thompson}, T.~A. 2005, \apj, 618, 569

\bibitem[{{Nanni} {et~al.}(2013){Nanni}, {Bressan}, {Marigo}, \&
  {Girardi}}]{Nanni13}
{Nanni}, A., {Bressan}, A., {Marigo}, P., \& {Girardi}, L. 2013, \mnras, 434,
  2390

\bibitem[{{Nanni} {et~al.}(2014){Nanni}, {Bressan}, {Marigo}, \&
  {Girardi}}]{Nanni14}
{Nanni}, A., {Bressan}, A., {Marigo}, P., \& {Girardi}, L. 2014, \mnras
  [\eprint[arXiv]{1312.0875}]

\bibitem[{{Nomoto} {et~al.}(2013){Nomoto}, {Kobayashi}, \&
  {Tominaga}}]{Nomoto13}
{Nomoto}, K., {Kobayashi}, C., \& {Tominaga}, N. 2013, \araa, 51, 457

\bibitem[{{Nozawa} {et~al.}(2006){Nozawa}, {Kozasa}, \& {Habe}}]{Nozawa06}
{Nozawa}, T., {Kozasa}, T., \& {Habe}, A. 2006, \apj, 648, 435

\bibitem[{{Nozawa} {et~al.}(2007){Nozawa}, {Kozasa}, {Habe}, {Dwek}, {Umeda},
  {Tominaga}, {Maeda}, \& {Nomoto}}]{Nozawa07}
{Nozawa}, T., {Kozasa}, T., {Habe}, A., {et~al.} 2007, \apj, 666, 955

\bibitem[{{Nozawa} {et~al.}(2003){Nozawa}, {Kozasa}, {Umeda}, {Maeda}, \&
  {Nomoto}}]{Nozawa03}
{Nozawa}, T., {Kozasa}, T., {Umeda}, H., {Maeda}, K., \& {Nomoto}, K. 2003,
  \apj, 598, 785

\bibitem[{{Nozawa} {et~al.}(2011){Nozawa}, {Maeda}, {Kozasa}, {Tanaka},
  {Nomoto}, \& {Umeda}}]{Nozawa11}
{Nozawa}, T., {Maeda}, K., {Kozasa}, T., {et~al.} 2011, \apj, 736, 45

\bibitem[{{Pettini} {et~al.}(2002){Pettini}, {Rix}, {Steidel}, {Hunt},
  {Shapley}, \& {Adelberger}}]{Pettini02}
{Pettini}, M., {Rix}, S.~A., {Steidel}, C.~C., {et~al.} 2002, \apss, 281, 461

\bibitem[{{Pforr} {et~al.}(2012){Pforr}, {Maraston}, \& {Tonini}}]{Pforr12}
{Pforr}, J., {Maraston}, C., \& {Tonini}, C. 2012, \mnras, 422, 3285

\bibitem[{{Popping} {et~al.}(2017){Popping}, {Somerville}, \&
  {Galametz}}]{Popping17}
{Popping}, G., {Somerville}, R.~S., \& {Galametz}, M. 2017, \mnras, 471, 3152

\bibitem[{{R{\'e}my-Ruyer} {et~al.}(2014){R{\'e}my-Ruyer}, {Madden},
  {Galliano}, {Galametz}, {Takeuchi}, {Asano}, {Zhukovska}, {Lebouteiller},
  {Cormier}, {Jones}, {Bocchio}, {Baes}, {Bendo}, {Boquien}, {Boselli},
  {DeLooze}, {Doublier-Pritchard}, {Hughes}, {Karczewski}, \&
  {Spinoglio}}]{RR14}
{R{\'e}my-Ruyer}, A., {Madden}, S.~C., {Galliano}, F., {et~al.} 2014, \aap,
  563, A31

\bibitem[{{R{\'e}my-Ruyer} {et~al.}(2013){R{\'e}my-Ruyer}, {Madden},
  {Galliano}, {Hony}, {Sauvage}, {Bendo}, {Roussel}, {Pohlen}, {Smith},
  {Galametz}, {Cormier}, {Lebouteiller}, {Wu}, {Baes}, {Barlow}, {Boquien},
  {Boselli}, {Ciesla}, {De Looze}, {Karczewski}, {Panuzzo}, {Spinoglio},
  {Vaccari}, \& {Wilson}}]{RR13}
{R{\'e}my-Ruyer}, A., {Madden}, S.~C., {Galliano}, F., {et~al.} 2013, \aap,
  557, A95

\bibitem[{{R{\'e}my-Ruyer} {et~al.}(2015){R{\'e}my-Ruyer}, {Madden},
  {Galliano}, {Lebouteiller}, {Baes}, {Bendo}, {Boselli}, {Ciesla}, {Cormier},
  {Cooray}, {Cortese}, {De Looze}, {Doublier-Pritchard}, {Galametz}, {Jones},
  {Karczewski}, {Lu}, \& {Spinoglio}}]{RR15}
{R{\'e}my-Ruyer}, A., {Madden}, S.~C., {Galliano}, F., {et~al.} 2015, \aap,
  582, A121

\bibitem[{{Ritter} {et~al.}(2018){Ritter}, {Herwig}, {Jones}, {Pignatari},
  {Fryer}, \& {Hirschi}}]{Ritter18}
{Ritter}, C., {Herwig}, F., {Jones}, S., {et~al.} 2018, \mnras, 480, 538

\bibitem[{{Rouill{\'e}} {et~al.}(2020){Rouill{\'e}}, {J{\"a}ger}, \&
  {Henning}}]{Rouille20}
{Rouill{\'e}}, G., {J{\"a}ger}, C., \& {Henning}, T. 2020, \apj, 892, 96

\bibitem[{{Rouill{\'e}} {et~al.}(2015){Rouill{\'e}}, {J{\"a}ger},
  {Krasnokutski}, {Krebsz}, \& {Henning}}]{Rouille15}
{Rouill{\'e}}, G., {J{\"a}ger}, C., {Krasnokutski}, S.~A., {Krebsz}, M., \&
  {Henning}, T. 2015, arXiv e-prints, arXiv:1502.00388

\bibitem[{{Rowlands} {et~al.}(2014){Rowlands}, {Gomez}, {Dunne},
  {Arag{\'o}n-Salamanca}, {Dye}, {Maddox}, {da Cunha}, \& {van der
  Werf}}]{Rowlands14}
{Rowlands}, K., {Gomez}, H.~L., {Dunne}, L., {et~al.} 2014, \mnras, 441, 1040

\bibitem[{{Sarangi} {et~al.}(2018){Sarangi}, {Dwek}, \& {Arendt}}]{Sarangi18}
{Sarangi}, A., {Dwek}, E., \& {Arendt}, R.~G. 2018, \apj, 859, 66

\bibitem[{{Schneider} {et~al.}(2016){Schneider}, {Hunt}, \&
  {Valiante}}]{Schneider16}
{Schneider}, R., {Hunt}, L., \& {Valiante}, R. 2016, \mnras, 457, 1842

\bibitem[{{Shapley} {et~al.}(2003){Shapley}, {Steidel}, {Pettini}, \&
  {Adelberger}}]{Shapley03}
{Shapley}, A.~E., {Steidel}, C.~C., {Pettini}, M., \& {Adelberger}, K.~L. 2003,
  \apj, 588, 65

\bibitem[{{Spyromilio} {et~al.}(2001){Spyromilio}, {Leibundgut}, \&
  {Gilmozzi}}]{Spyromilio01}
{Spyromilio}, J., {Leibundgut}, B., \& {Gilmozzi}, R. 2001, \aap, 376, 188

\bibitem[{{Suzuki} {et~al.}(2018){Suzuki}, {Kaneda}, {Onaka}, {Yamagishi},
  {Ishihara}, {Kokusho}, \& {Tsuchikawa}}]{Suzuki18}
{Suzuki}, T., {Kaneda}, H., {Onaka}, T., {et~al.} 2018, \mnras, 477, 3065

\bibitem[{{Troncoso} {et~al.}(2014){Troncoso}, {Maiolino}, {Sommariva},
  {Cresci}, {Mannucci}, {Marconi}, {Meneghetti}, {Grazian}, {Cimatti},
  {Fontana}, {Nagao}, \& {Pentericci}}]{Troncoso14}
{Troncoso}, P., {Maiolino}, R., {Sommariva}, V., {et~al.} 2014, \aap, 563, A58

\bibitem[{{Valiante} {et~al.}(2014){Valiante}, {Schneider}, {Salvadori}, \&
  {Gallerani}}]{Valiante14}
{Valiante}, R., {Schneider}, R., {Salvadori}, S., \& {Gallerani}, S. 2014,
  \mnras, 444, 2442

\bibitem[{{Ventura} {et~al.}(2012){Ventura}, {Criscienzo}, {Schneider},
  {Carini}, {Valiante}, {D'Antona}, {Gallerani}, {Maiolino}, \&
  {Tornamb{\'e}}}]{Ventura12}
{Ventura}, P., {Criscienzo}, M.~D., {Schneider}, R., {et~al.} 2012, \mnras,
  424, 2345

\bibitem[{{Wang} {et~al.}(2017){Wang}, {Hirashita}, \& {Hou}}]{Wang17}
{Wang}, W.-C., {Hirashita}, H., \& {Hou}, K.-C. 2017, \mnras, 465, 3475

\bibitem[{{Ysard} {et~al.}(2019){Ysard}, {Koehler}, {Jimenez-Serra}, {Jones},
  \& {Verstraete}}]{Ysard19}
{Ysard}, N., {Koehler}, M., {Jimenez-Serra}, I., {Jones}, A.~P., \&
  {Verstraete}, L. 2019, \aap, 631, A88

\bibitem[{{Zhukovska}(2014)}]{Zhukovska14}
{Zhukovska}, S. 2014, \aap, 562, A76

\bibitem[{{Zhukovska} {et~al.}(2016){Zhukovska}, {Dobbs}, {Jenkins}, \&
  {Klessen}}]{Zhukovska16}
{Zhukovska}, S., {Dobbs}, C., {Jenkins}, E.~B., \& {Klessen}, R.~S. 2016, \apj,
  831, 147

\end{thebibliography}
\appendix
\section{Properties of DGS galaxies and best-fitting models from \textsc{cigale}}
\begin{sidewaystable*}
\caption{Properties of DGS galaxies and associated errors from the fitting with \textsc{cigale} \citep{Burgarella20} (of which the $\chi^2$ is provided), and from the literature \citep{RR13, RR14}. From \citet{RR13, RR14} we select $Z=\log(O/H)+12$, the mass of atomic and molecular hydrogen (M$_{\rm HI}$, M$_{\rm H2, Z}$) and mean atomic weight of the galaxy ($\mu_{\rm gal}$) \citep[see Table A.1 of][and references therein]{RR14}. All the masses are given in units of M$_\odot$, the sSFR is in yrs$^{-1}$, the age in yrs and the $\tau$ in Myrs.}
\label{Table:gal_prop}
 \begin{tabular}{lrrrrrrrrrrrrrrrrrrrr}
%\begin{tabular}{|l|r|r|r|r|r|r|r|r|r|r|r|r|r|l|l|r|}
\hline
id & $\chi^2$ & sSFR  & sSFR$_{\rm err}$ & sMdust & sMdust$_{\rm err}$ & age & age$_{\rm err}$ &
  $\tau$ & $\tau_{\rm err}$  & Mstar  & Mstar$_{\rm err}$ & $Z$ & $Z_{\rm err}$ &
M$_{\rm HI}$ & M$_{\rm H2, Z}$  & $\mu_{\rm gal}$ \\
\hline
  Haro3 & 1.2 & 6.4E-10 & 2.3E-10 & 0.003484 & 5.54E-4 & 680 & 301 & 181 & 112 & 3.72E8 & 5.6E7 & 8.28 & 0.01 & 1.13E9 & 2.01E8 & 1.38\\
  He2-10 & 2.6 & 2.2E-10 & 1.5E-10 & 0.00274 & 4.8E-4 & 683 & 353 & 129 & 92 & 8.98E8 & 1.51E8 & 8.43 & 0.01 & 3.10E8 & 3.21E8 & 1.39\\
  HS0052+2536 & 3.5 & 1.059E-8 & 1.046E-8 & 0.0172 & 0.0103 & 313 & 314 & 438 & 364 & 3.96E8 & 2.37E8 & 8.04 & 0.1 & $\leq$4.82E10 & $\leq$3.26E10 & 1.38\\
  HS0822+3542 & 4.8 & 1.76E-9 & 4.1E-10 & 0.00175 & 3.77E-4 & 738 & 323 & 488 & 357 & 6.33E5 & 8.88E4 & 7.32 & 0.03 & 5.75E7 & 5.21E7 & 1.37\\
  HS1236+3937 & 0.5 & 6.0E-11 & 6.0E-11 & 2.3E-4 & 2.3E-4 & 605 & 203 & 63 & 56 & 1.37E8 & 3.8E7 & 7.72 & 0.1 & - & - & 1.37\\
  HS1304+3529 & 3.8 & 1.236E-8 & 1.236E-8 & 0.0161 & 1.09E-2 & 274 & 305 & 403 & 356 & 2.63E7 & 1.78E7 & 7.93 & 0.1 & - & - & 1.38\\
  HS1319+3224 & 0.7 & 7.6E-10 & 3.5E-10 & 2.31E-3 & 5.76E-4 & 662 & 357 & 222 & 198 & 3.03E7 & 6.90E6 & 7.81 & 0.1 & - & - & 1.37\\
  HS1330+3651 & 4.7 & 5.7E-10 & 5.7E-10 & 1.42E-3 & 4.74E-4 & 599 & 290 & 117 & 186 & 2.68E8 & 8.9E7 & 7.98 & 0.1 & - & - & 1.38\\
  HS1442+4250 & 1.3 & 5.5E-10 & 5.1E-10 & 6.63E-4 & 1.73E-4 & 511 & 313 & 158 & 213 & 1.17E7 & 2.7E6 & 7.6 & 0.01 & 3.10E8 & 2.1E8 & 1.37\\
  HS2352+2733 & 2.6 & 1.97E-9 & 1.9E-9 & 5.73E-3 & 0.00472 & 431 & 348 & 323 & 329 & 4.72E7 & 3.86E7 & 8.4 & 0.1 & - & - & 1.38\\
  IZw18 & 3.1 & 1.52E-8 & 3.42E-9 & 2.17E-3 & 4.82E-4 & 132 & 50 & 415 & 361 & 1.11E6  & 2.11E5  & 7.14 & 0.01 & 1.00E8 & 1.49E8 & 1.37\\
  Mrk1450 & 3.4 & 2.33E-9 & 1.13E-9 & 8.05E-3 & 0.00234 & 655 & 320 & 495 & 344 & 6.16E6  & 1.76E6 & 7.84 & 0.01 & 4.30E7 & 2.91E7 & 1.37\\
  Mrk153 & 2.0 & 1.96E-9 & 5.7E-10 & 0.00163 & 2.69E-4 & 577 & 325 & 397 & 368 & 1.46E8 & 2.29E7 & 7.86 & 0.04 & $\leq$6.81E8 & $\leq$4.61E8 & 1.37\\
  Mrk209 & 1.8 & 1.9E-10 & 1.9E-10 & 9.93E-4 & 1.95E-4 & 572 & 305 & 95 & 94 & 6.41E6  & 1.22E6  & 7.74 & 0.01 & 2.76E7 & 2.5E8 & 1.37\\
  Mrk930 & 1.5 & 2.23E-9 & 1.11E-9 & 9.87E-3 & 2.65E-3 & 549 & 334 & 390 & 354 & 8.29E8 & 2.19E8 & 8.03 & 0.01 & 3.19E9 & 6.57E8 & 1.38\\
  NGC1140 & 4.6 & 4.6E-10 & 1.6E-10 & 1.82E-3 & 2.83E-4 & 743 & 298 & 173 & 89 & 9.65E8 & 1.43E8 & 8.38 & 0.01 & 3.50E9 & 8.02E7 & 1.38\\
  NGC1569 & 1.2 & 3.0E-11 & 3.0E-11 & 0.00164 & 1.84E-4 & 609 & 146 & 63 & 32 & 3.93E6  & 3.94E5  & 8.02 & 0.02 &  1.77E8 & 1.54E7 & 1.38\\
  NGC5253 & 4.3 & 3.0E-10 & 1.2E-10 & 0.00194 & 3.58E-4 & 768 & 341 & 156 & 89 & 2.85E8 & 5.05E7 & 8.25 & 0.02 & 1.06E8 & 2.26E7 & 1.38\\
  NGC625 & 4.9 & 1.0E-11 & 1.0E-11 & 5.8E-4 & 7.3E-5 & 784 & 162 & 54 & 31 & 2.8E8 & 3.23E7 & 8.22 & 0.02 & 1.10E8 & 3.72E7 & 1.38\\
  NGC6822 & 4.2 & 1.0E-11 & 1.0E-11 & 7.63E-4 & 6.7E-5 & 632 & 125 & 52 & 30 & 2.35E7 & 1.66E6 & 7.96 & 0.01 &  1.04E8 & 1.05E7 & 1.38\\
  Pox186 & 3.5 & 8.5E-10 & 4.9E-10 & 0.00579 & 1.69E-3 & 580 & 356 & 230 & 253 & 2.52E6 & 7.24E5 & 7.7 & 0.01 & $\leq$2.00E6 & $\leq$1.35E6 & 1.37\\
  SBS1159+545 & 1.6 & 1.677E-8 & 1.677E-8 & 0.00586 & 0.00507 & 180 & 240 & 365 & 353 & 9.18E6 & 7.88E6 & 7.44 & 0.01 & $\leq$6.30E7 & $\leq$4.26E7 & 1.37\\
  SBS1211+540 & 1.3 & 1.97E-9 & 6.9E-10 & 2.71E-3 & 5.09E-4 & 462 & 305 & 310 & 350 & 1.45E6 & 2.51E5 & 7.58 & 0.01 & 5.60E7 & 3.79E7 & 1.37\\
  SBS1249+493 & 0.8 & 1.051E-8 & 1.214E-8 & 0.011295 & 0.00944 & 252 & 288 & 366 & 352 & 2.38E7 & 1.98E7 & 7.68 & 0.02 &  1.00E9 & 6.77E8 & 1.37\\
  SBS1533+574 & 2.2 & 2.7E-10 & 1.7E-10 & 2.74E-3 & 5.06E-4 & 682 & 339 & 135 & 93 & 2.46E8 & 4.38E7 & 8.05 & 0.01 & 3.00E9 & 2.03E9 & 1.38\\
  Tol1214-277 & 0.9 & 2.52E-9 & 7.5E-10 & 0.00362 & 9.56E-4 & 595 & 273 & 510 & 365 & 6.03E7 & 1.52E7 & 7.52 & 0.01 & $\leq$3.22E8 & $\leq$2.18E8 & 1.37\\
  UGC4483 & 4.7 & 6.944E-8 & 1.998E-8 & 8.30E-3 & 0.00175 & 32 & 18 & 346 & 349 & 1.23E5 & 2.12E4 & 7.46 & 0.02 & 2.52E7 & 2.23E7 & 1.37\\
  UGCA20 & 0.4 & 3.74E-9 & 3.74E-9 & 3.76E-4 & 3.76E-4 & 473 & 371 & 243 & 311 & 1.16E6 & 7.59E5 & 7.5 & 0.02 & 6.92E6 & 1.35E7 & 1.37\\
  UM133 & 2.5 & 1.143E-8 & 8.06E-9 & 6.81E-3 & 3.77E-3 & 174 & 161 & 348 & 354 & 7.24E6 & 3.96E6 & 7.82 & 0.01 & 2.15E8 & 2.71E8 & 1.37\\
  UM448 & 2.2 & 9.5E-10 & 2.7E-10 & 4.79E-3 & 7.89E-4 & 787 & 370 & 283 & 182 & 2.15E9 & 8.15E8 & 8.32 & 0.01 & 6.00E9 & 1.92E10 & 1.38\\
  UM461 & 1.3 & 6.54E-9 & 4.37E-9 & 5.38E-3 & 2.37E-3 & 305 & 265 & 380 & 356 & 4.77E6 & 2.09E6 & 7.73 & 0.01 & 7.31E7  & 9.09E7 & 1.37\\
\hline\end{tabular}
%\hline
%\end{tabular}
\end{sidewaystable*}
%\end{center}
%\end{table*}

%\begin{adjustbox}{angle=90}

%\end{adjustbox}
\begin{figure}
\centering
\includegraphics[scale=0.5]{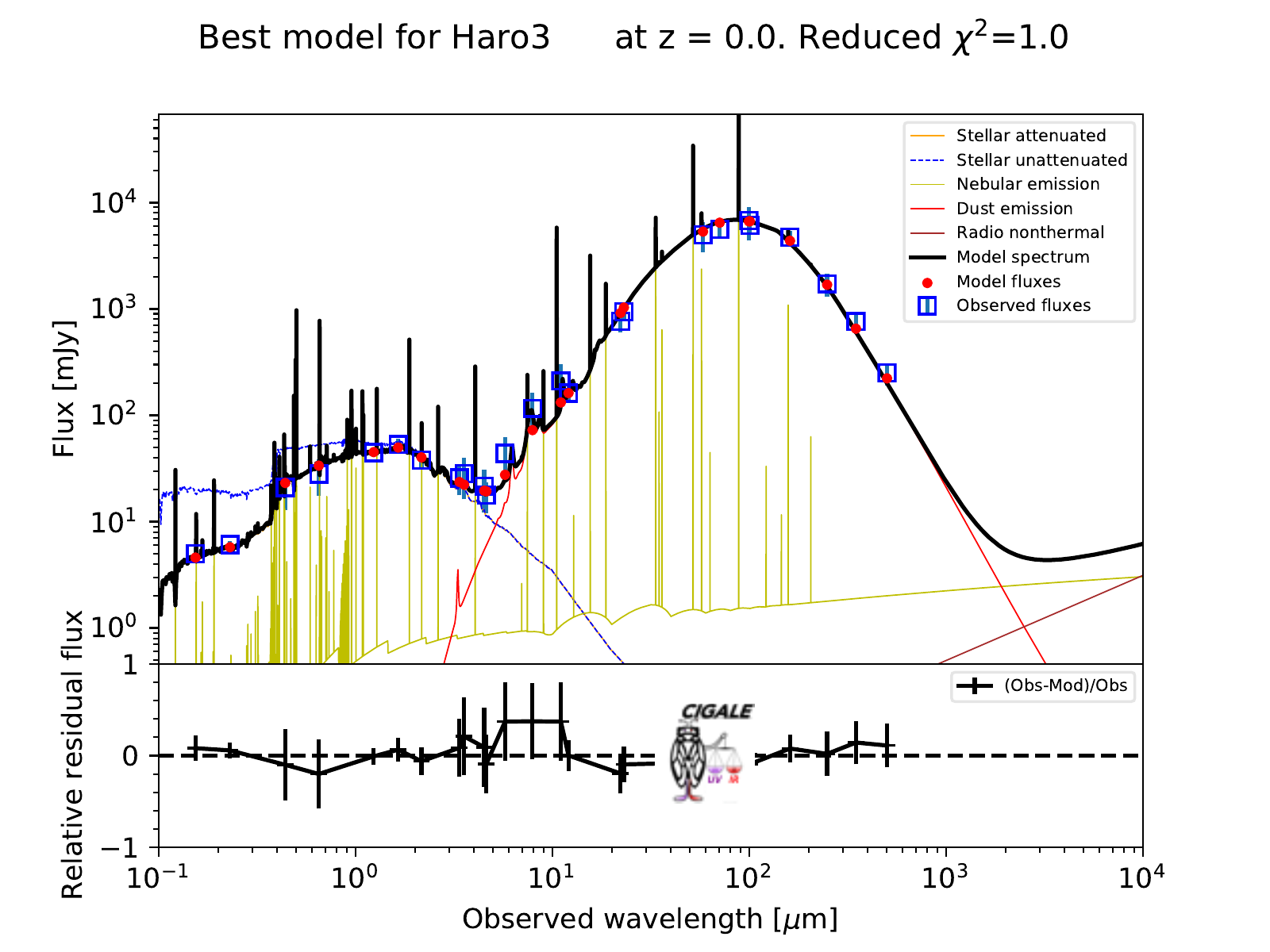}
%\caption{Haro3}
         \label{Fig:Haro3}
   \end{figure}

   \begin{figure}
\centering
\includegraphics[scale=0.5]{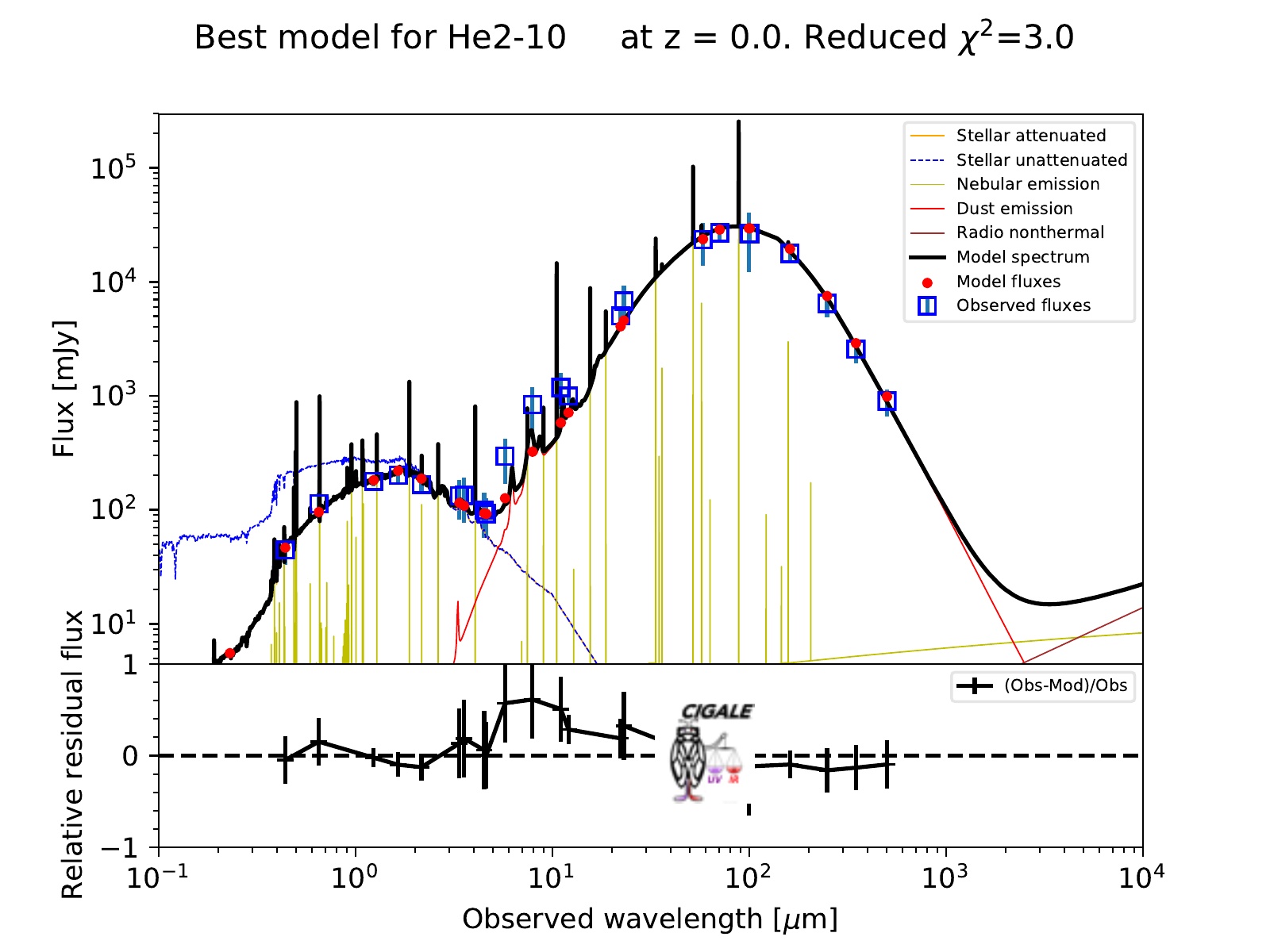}
%\caption{}
         \label{Fig:Haro2-10}
   \end{figure}

   \begin{figure}
\centering
\includegraphics[scale=0.5]{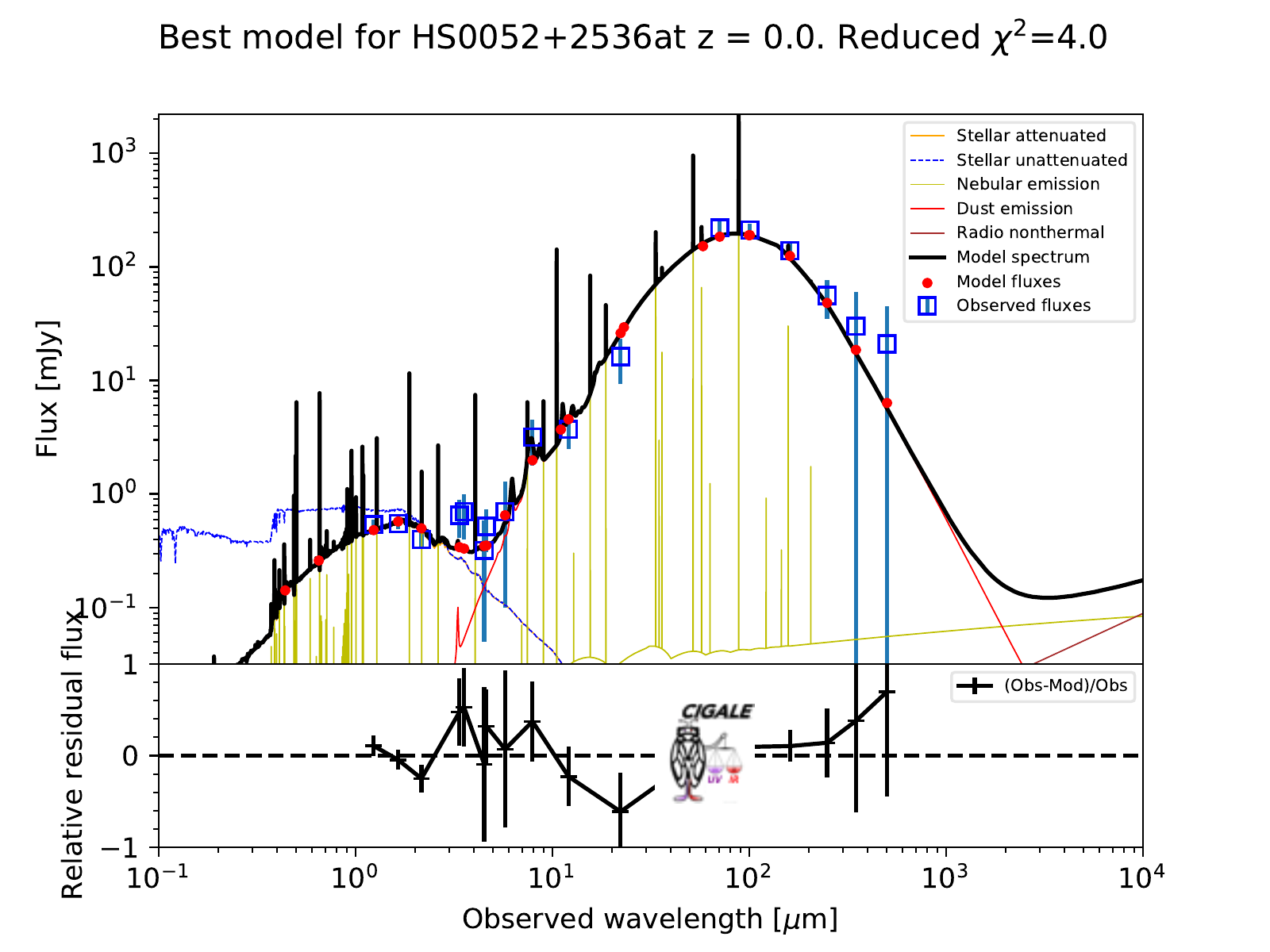}
%\caption{}
         \label{Fig:HS0052+2536}
   \end{figure}

      \begin{figure}
\centering
\includegraphics[scale=0.5]{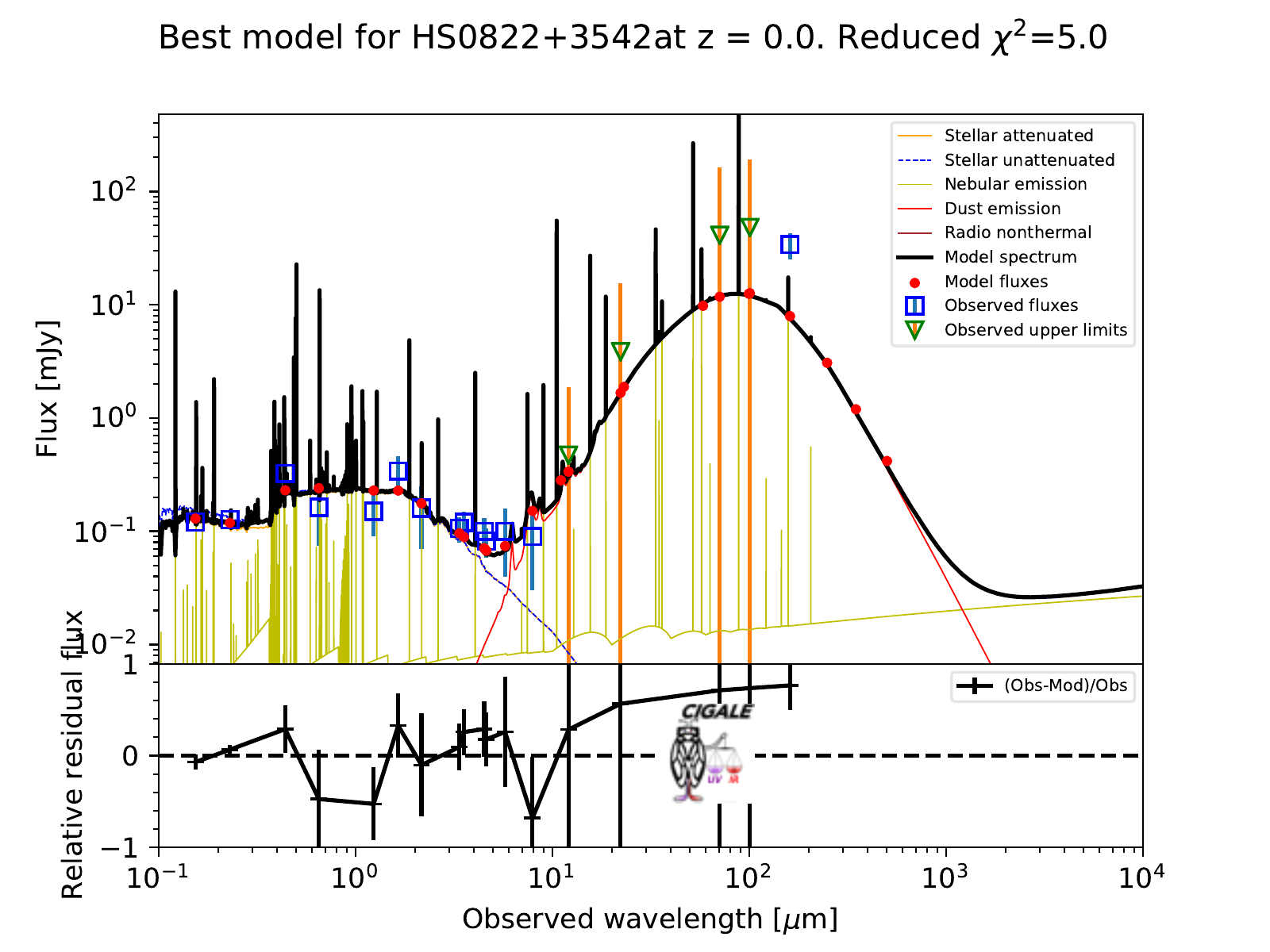}
%\caption{}
         \label{Fig:HS0822+3542}
   \end{figure}
   
    \begin{figure}
\centering
\includegraphics[scale=0.5]{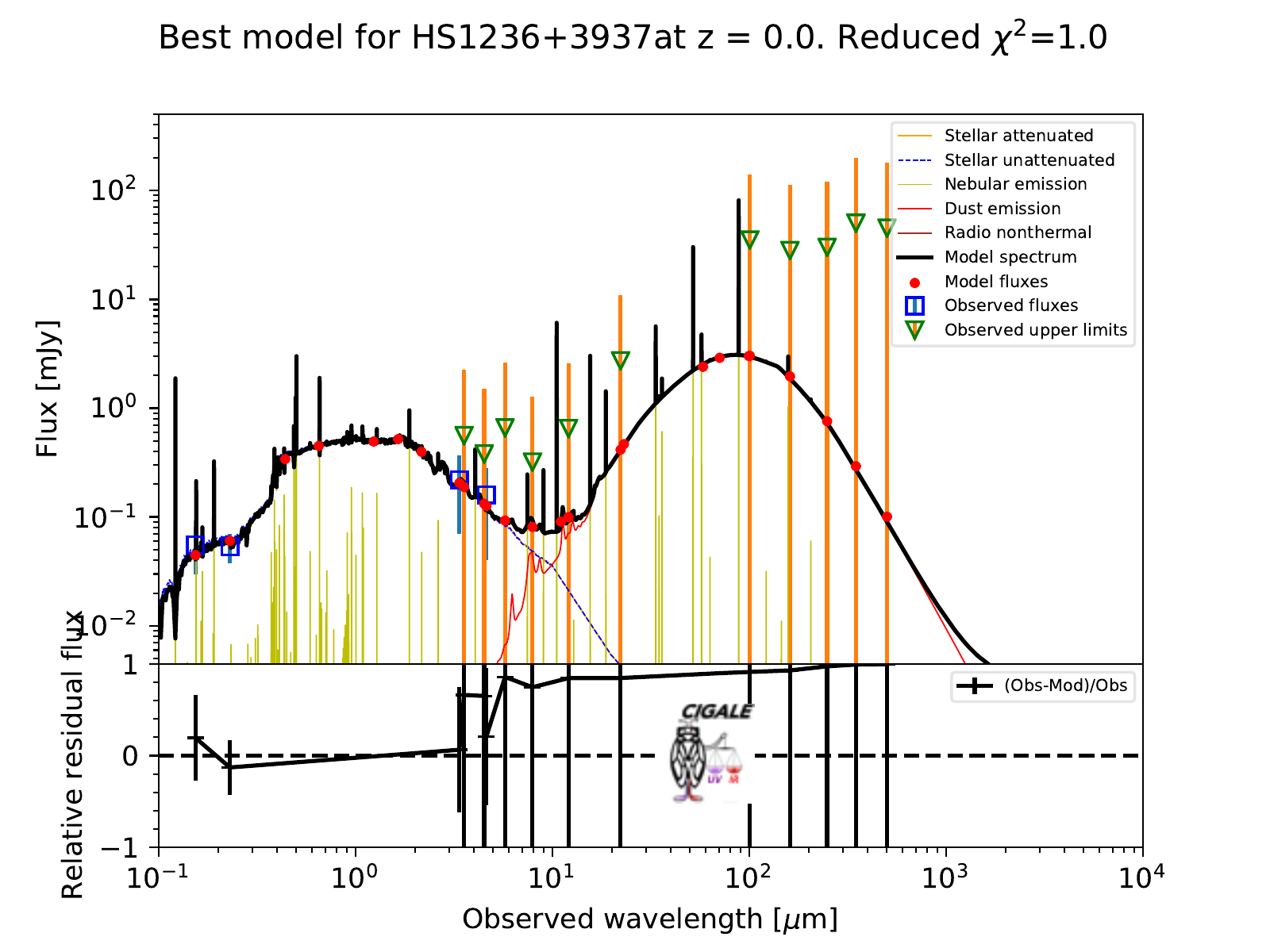}
%\caption{}
         \label{Fig:HS1236+3937}
   \end{figure}
   
    \begin{figure}
\centering
\includegraphics[scale=0.5]{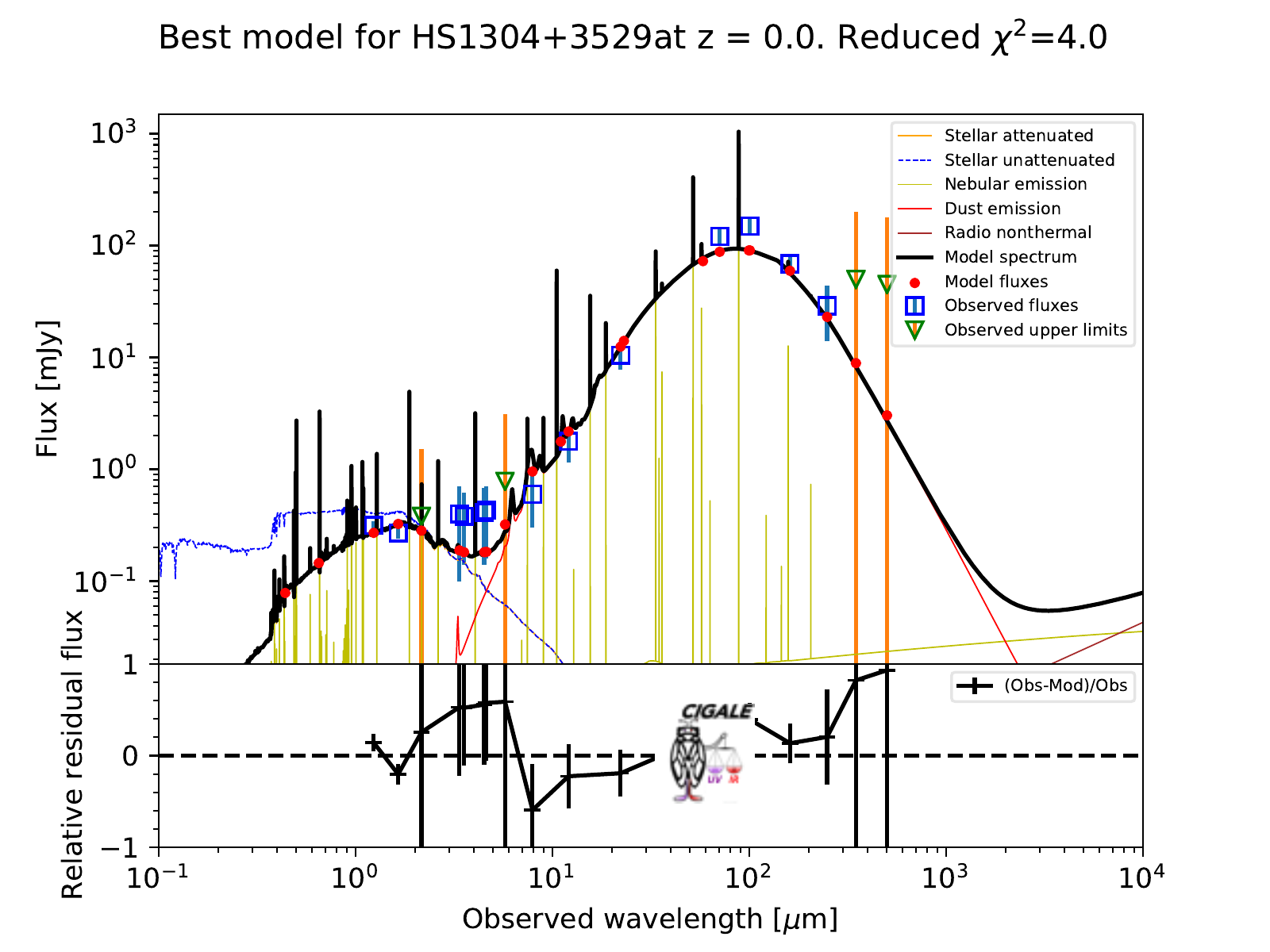}
%%\caption{}
         \label{Fig:HS1304+3529}
   \end{figure}

    \begin{figure}
\centering
\includegraphics[scale=0.5]{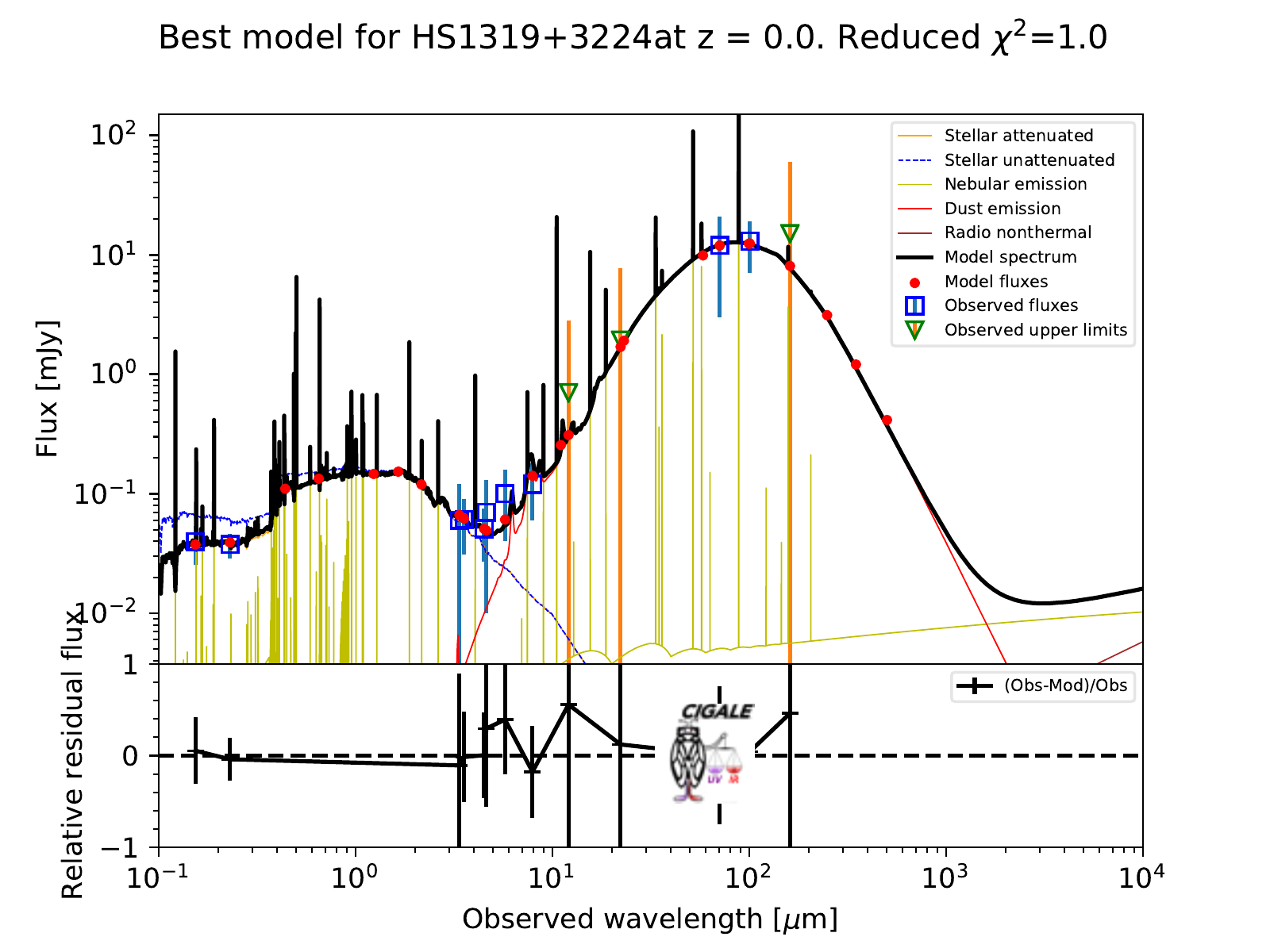}
%\caption{}
         \label{Fig:HS1319+3224}
   \end{figure}

    \begin{figure}
\centering
\includegraphics[scale=0.5]{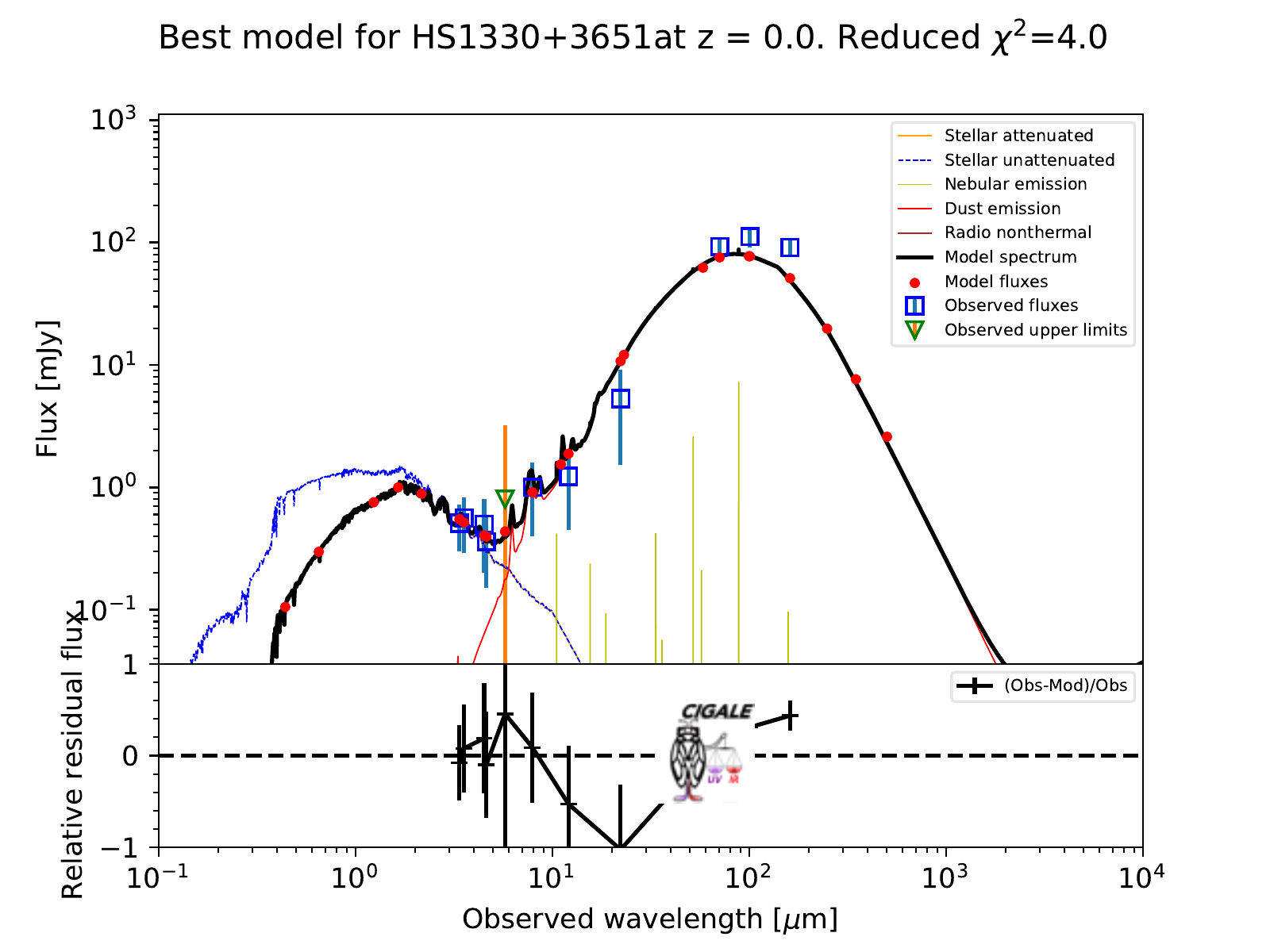}
%\caption{}
         \label{Fig:HS1330+3651}
   \end{figure}

    \begin{figure}
\centering
\includegraphics[scale=0.5]{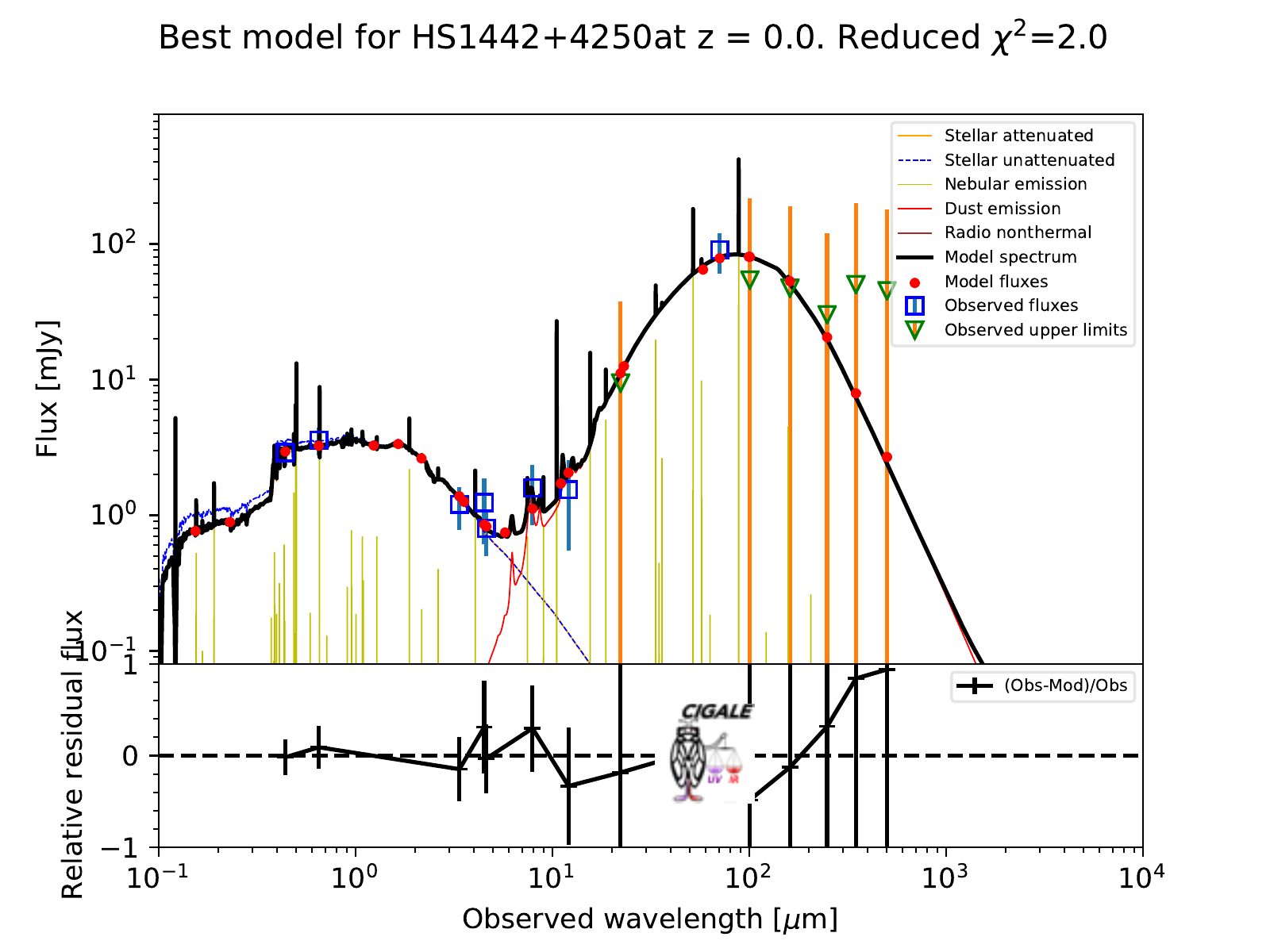}
%\caption{}
         \label{Fig:HS1442+4250}
   \end{figure}
    \begin{figure}
\centering
\includegraphics[scale=0.5]{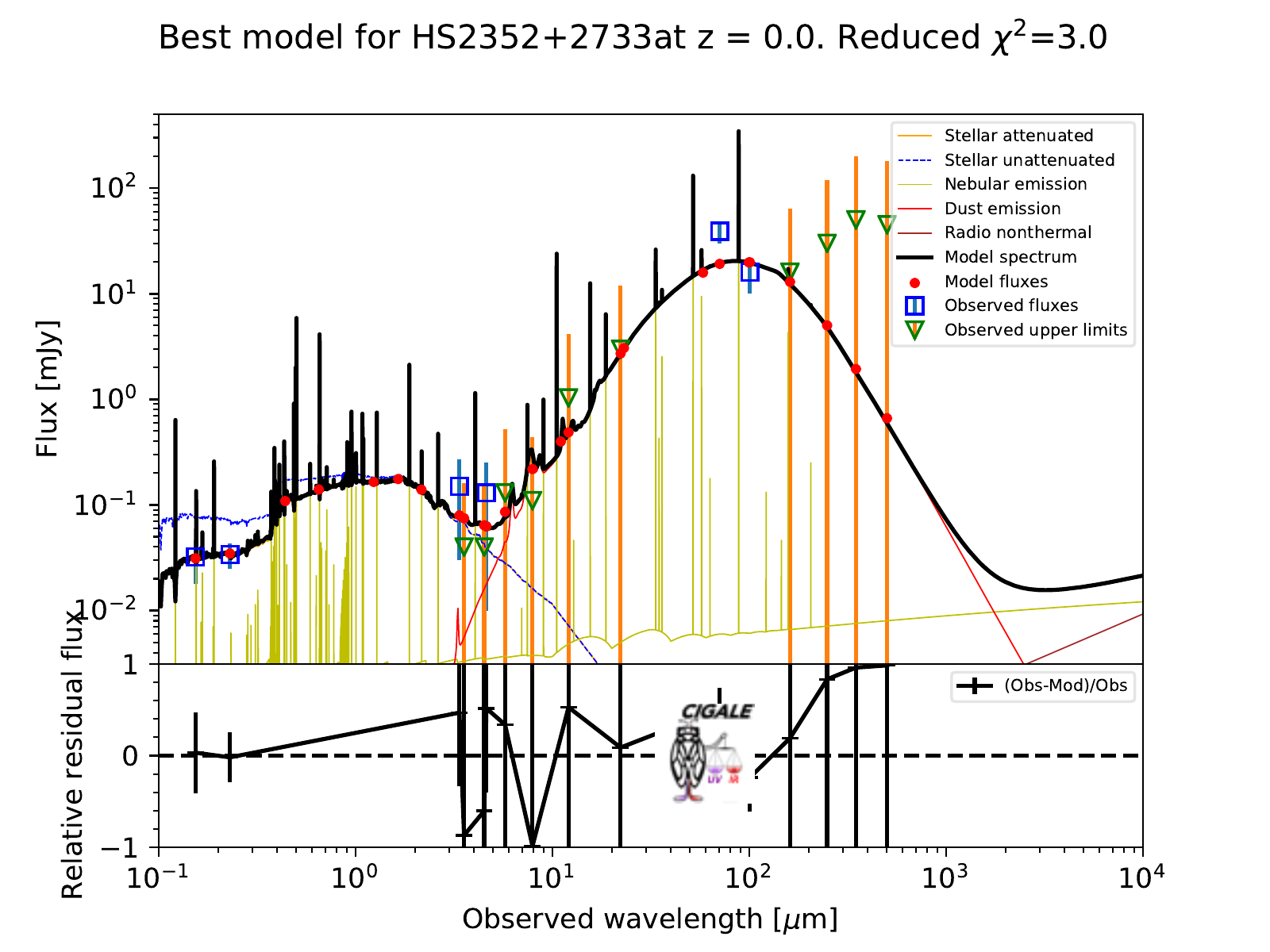}
%\caption{}
         \label{Fig:HS2352+2733}
   \end{figure}

    \begin{figure}
\centering
\includegraphics[scale=0.5]{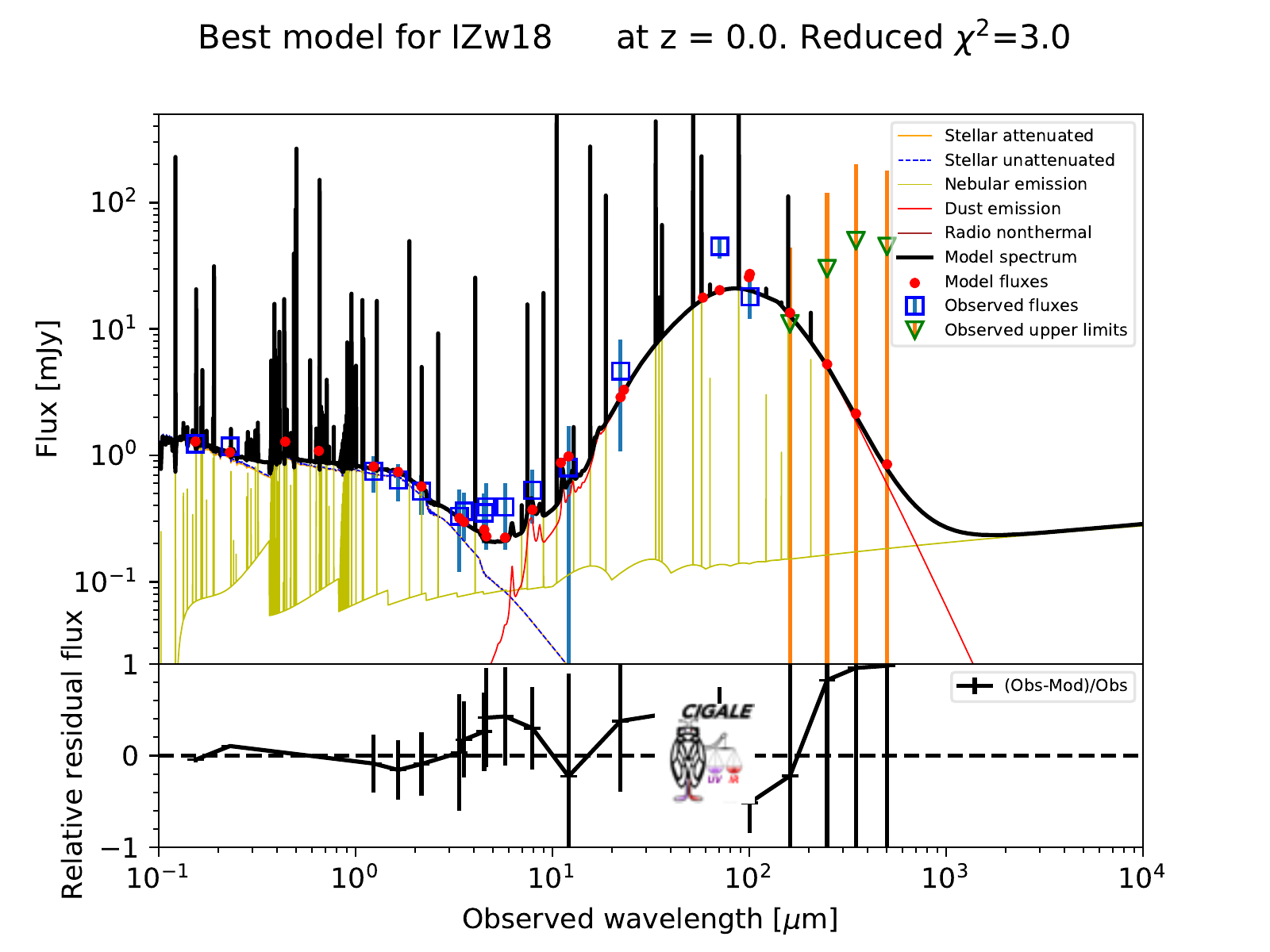}
%\caption{}
         \label{Fig:IZw18}
   \end{figure}   
  
      \begin{figure}
\centering
\includegraphics[scale=0.5]{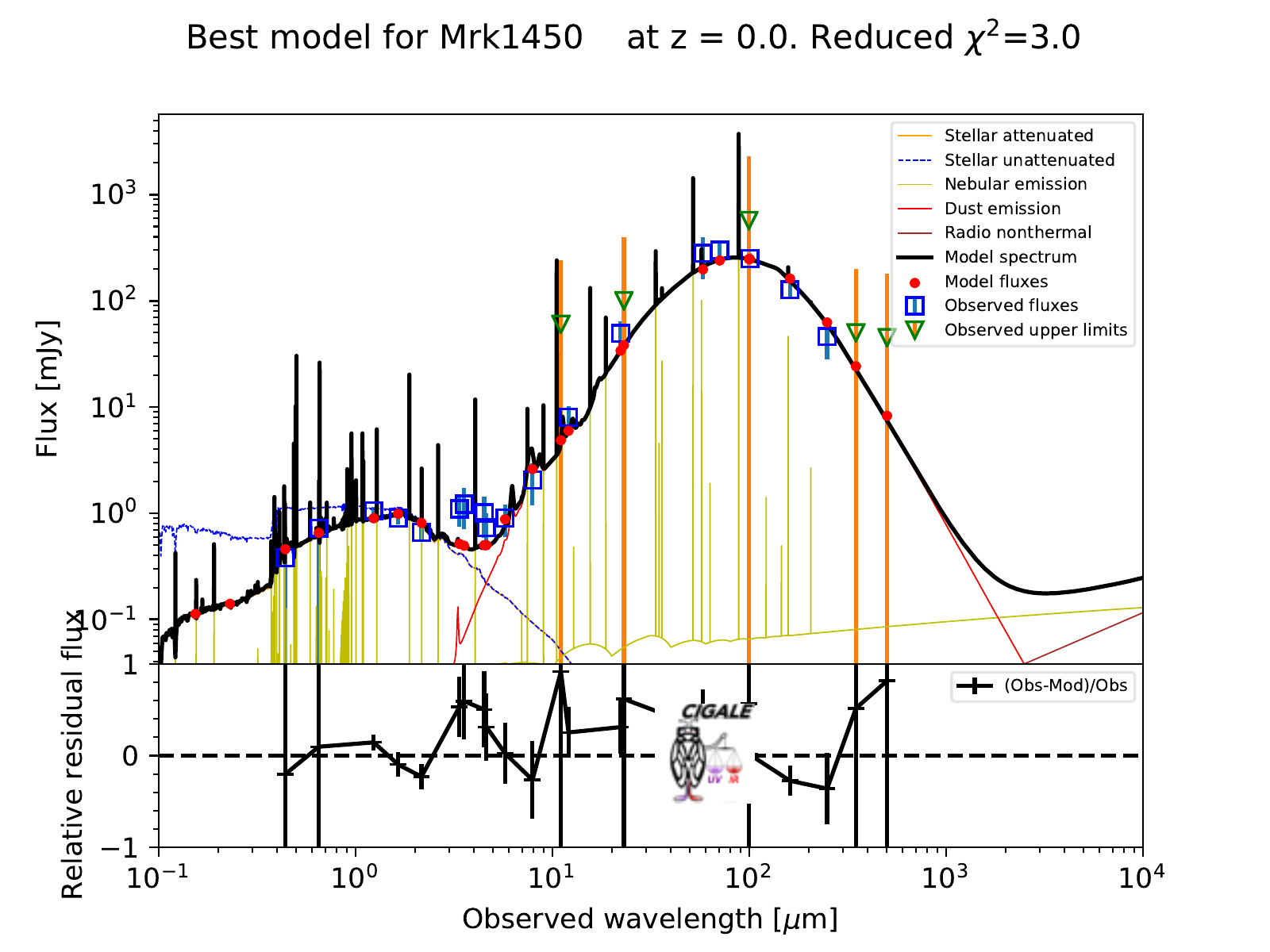}
%\caption{}
         \label{Fig:Mrk1450}
   \end{figure} 
   
       \begin{figure}
\centering
\includegraphics[scale=0.5]{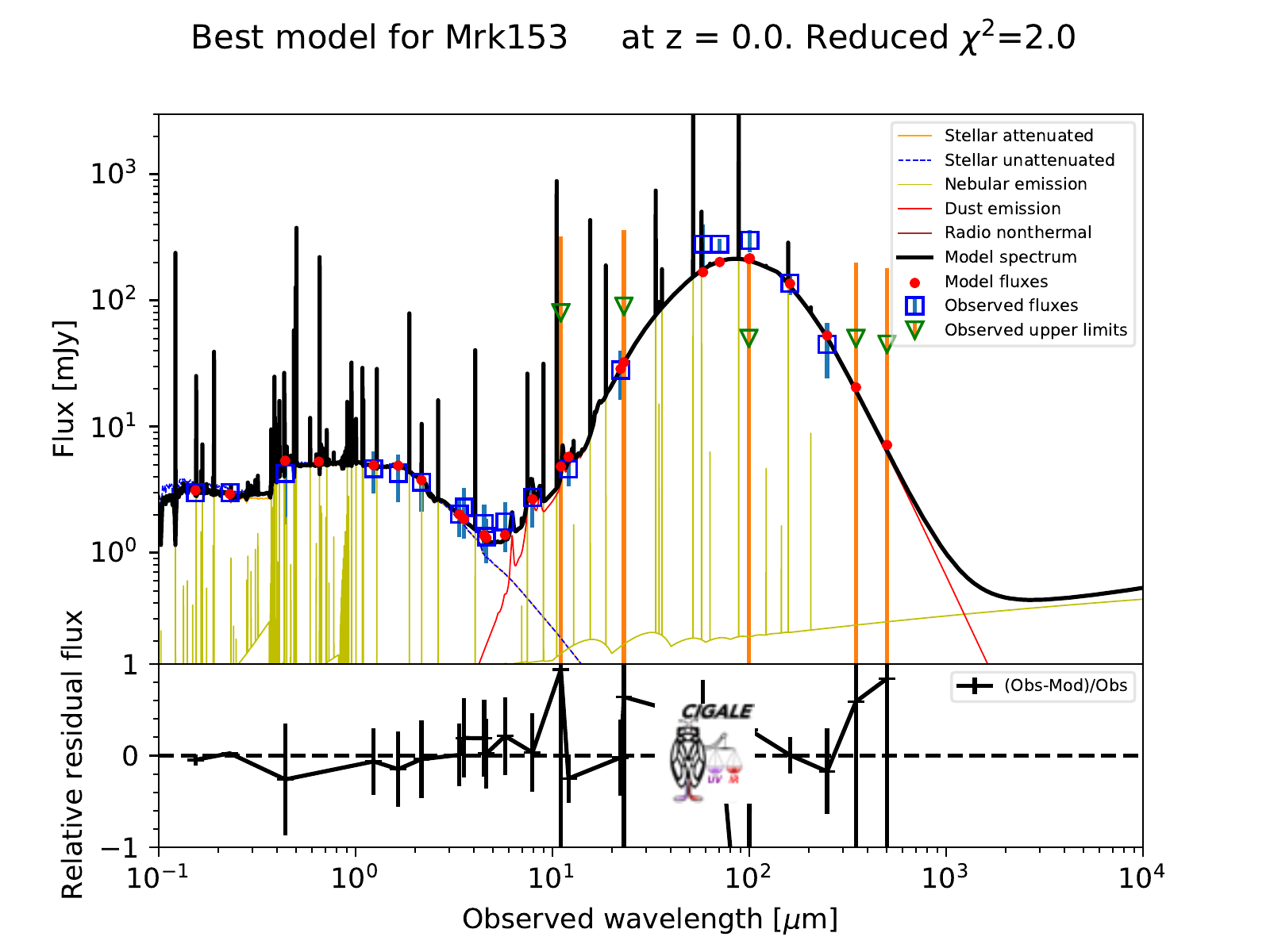}
%\caption{}
         \label{Fig:Mrk153}
   \end{figure} 
   
   \begin{figure}
\centering
\includegraphics[scale=0.5]{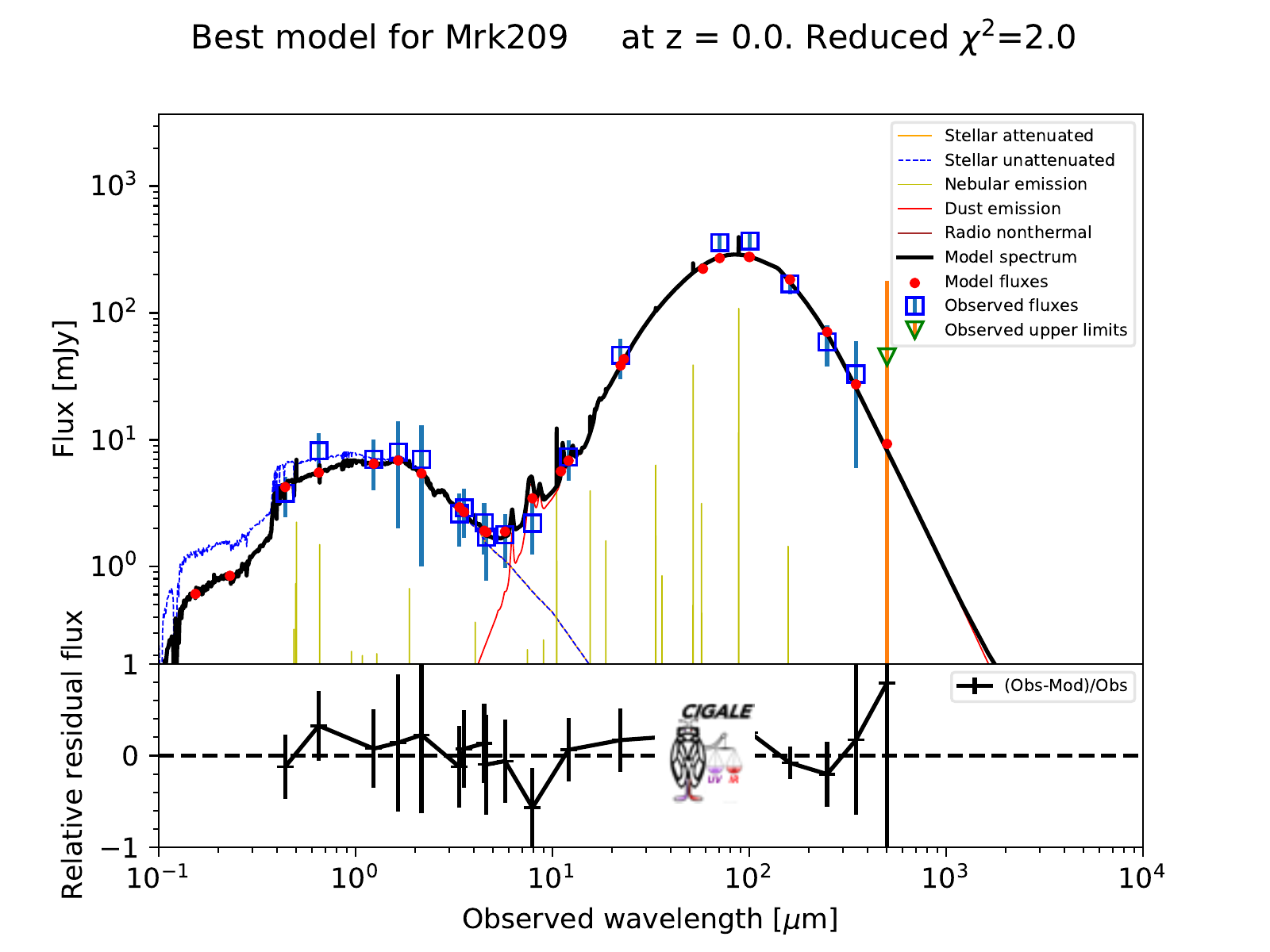}
%\caption{}
         \label{Fig:Mrk209}
   \end{figure} 
   
      \begin{figure}
\centering
\includegraphics[scale=0.5]{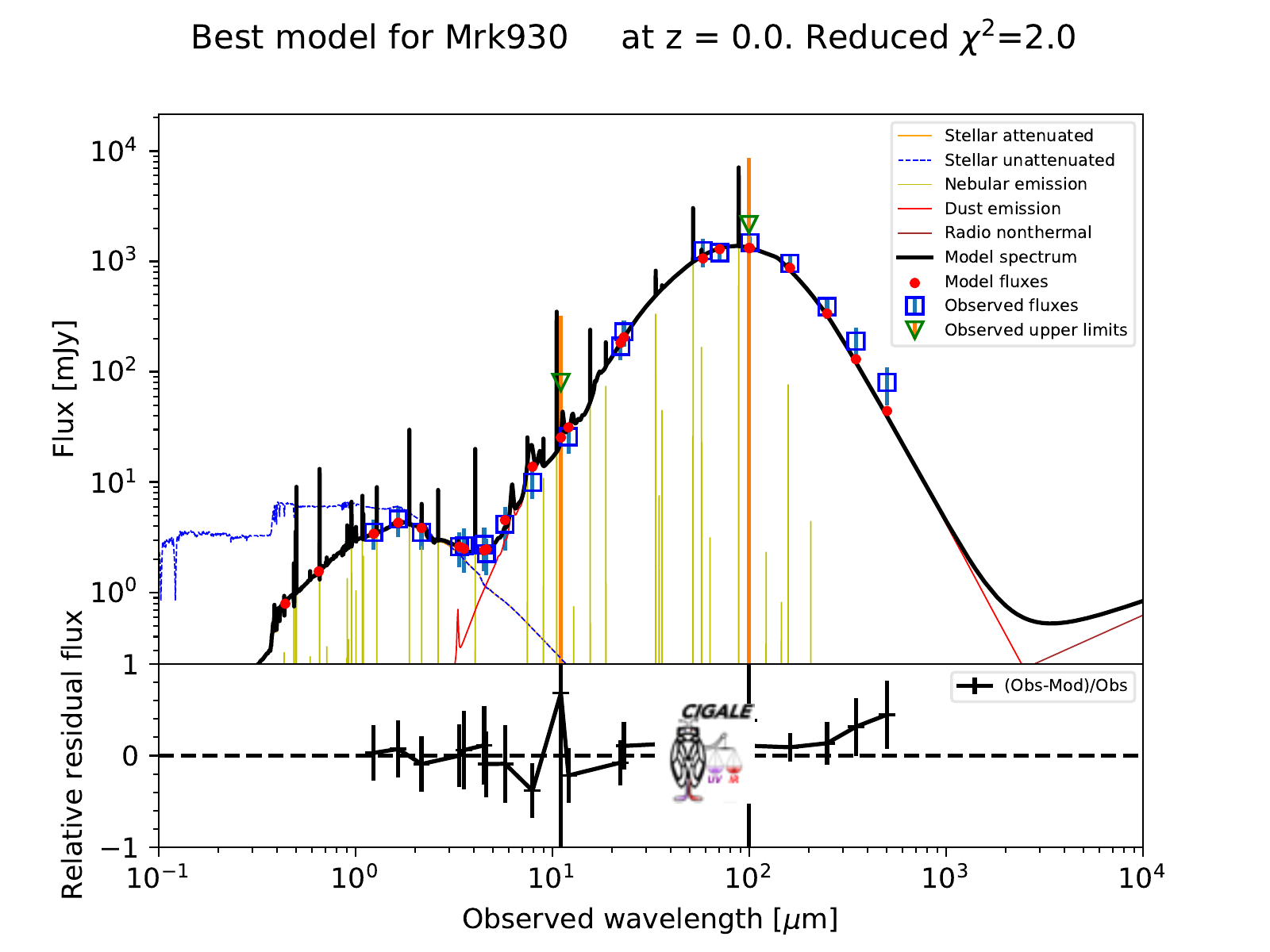}
%\caption{}
         \label{Fig:Mrk930}
   \end{figure} 
   
      \begin{figure}
\centering
\includegraphics[scale=0.5]{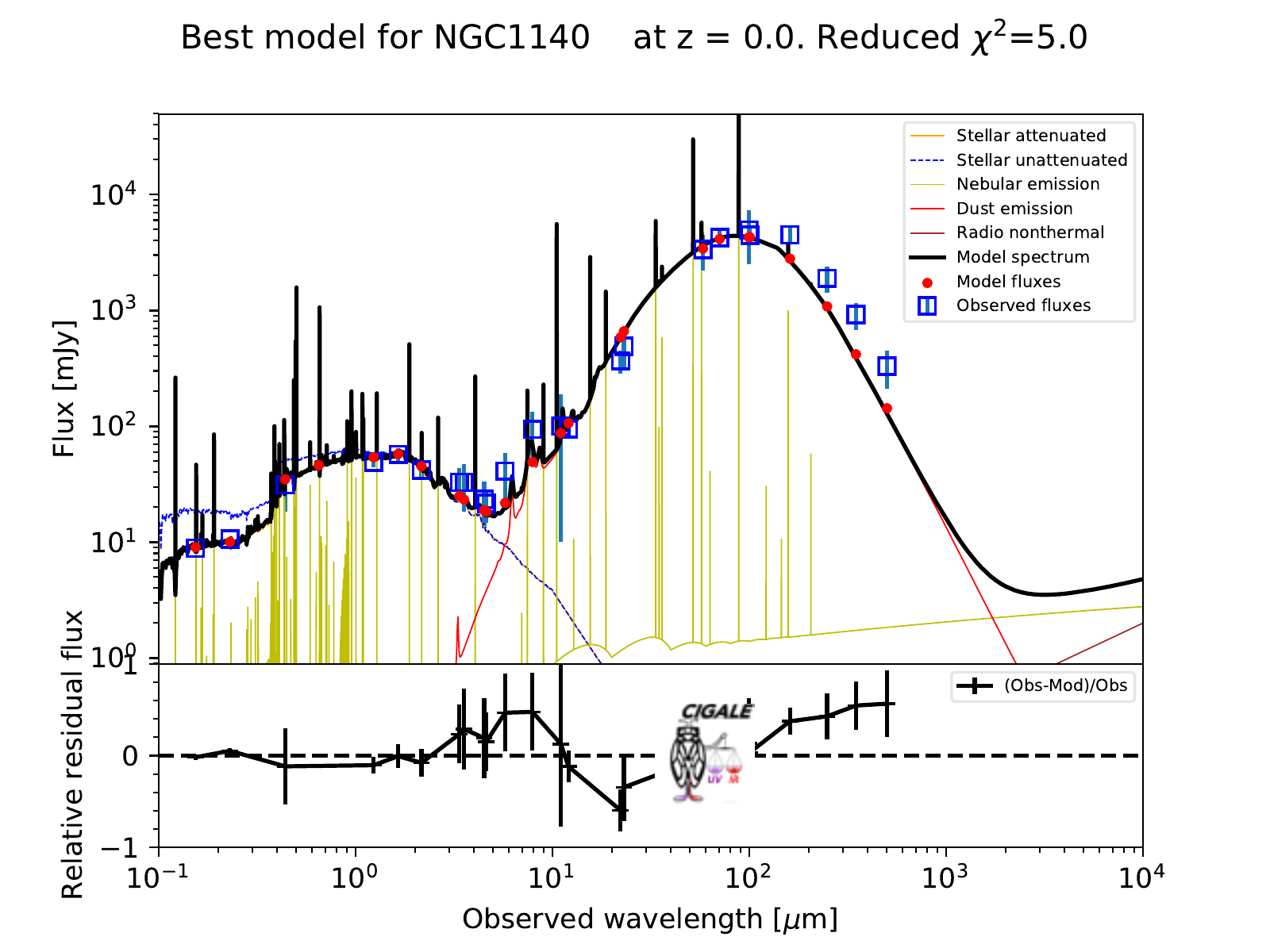}
%\caption{}
         \label{Fig:NGC1140}
   \end{figure} 
   
      \begin{figure}
\centering
\includegraphics[scale=0.5]{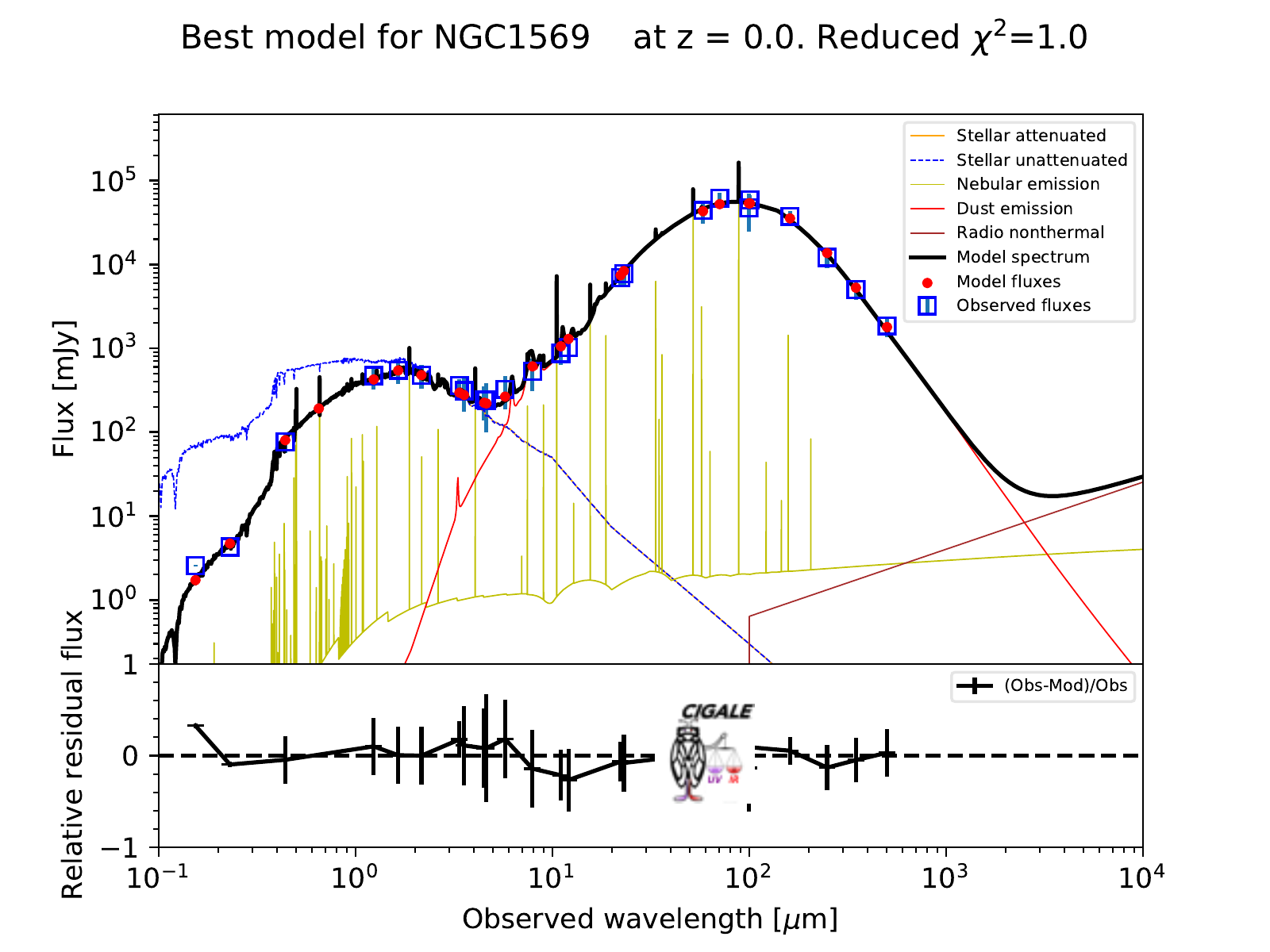}
%\caption{}
         \label{Fig:NGC1569}
   \end{figure} 
   
      \begin{figure}
\centering
\includegraphics[scale=0.5]{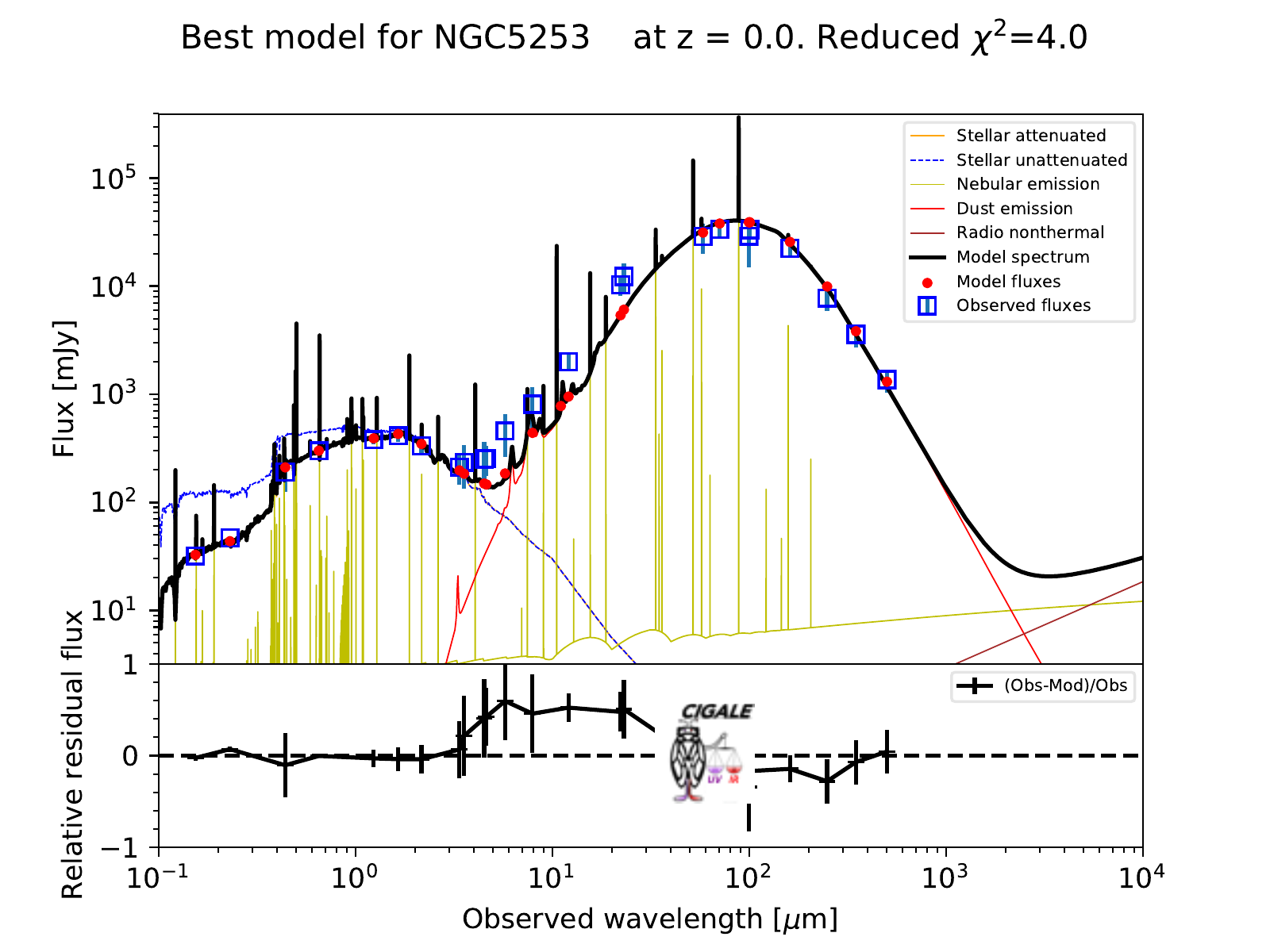}
%\caption{}
         \label{Fig:NGC5253}
   \end{figure} 
   
      \begin{figure}
\centering
\includegraphics[scale=0.5]{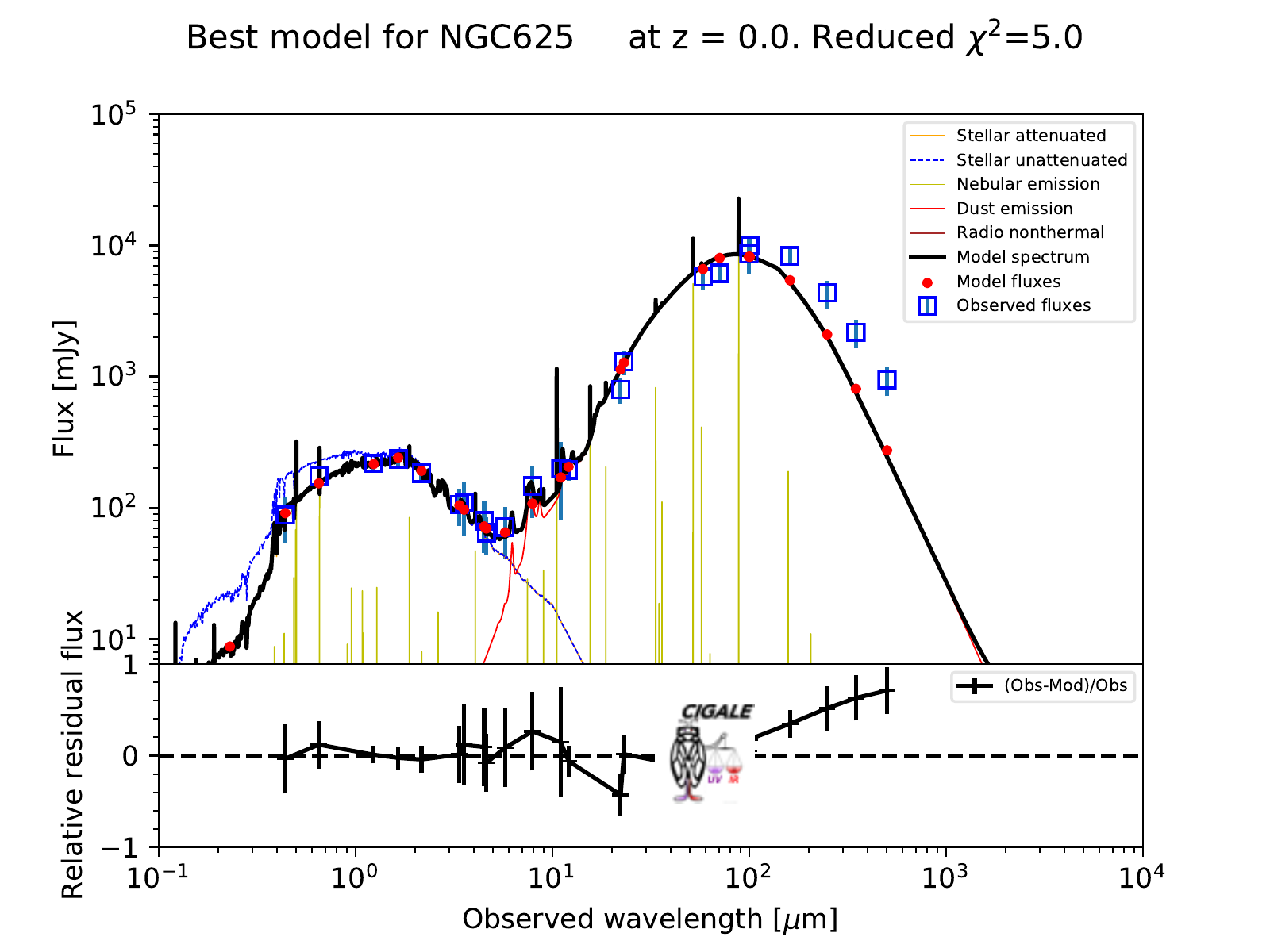}
%\caption{}
         \label{Fig:NGC625}
   \end{figure} 
   
      \begin{figure}
\centering
\includegraphics[scale=0.5]{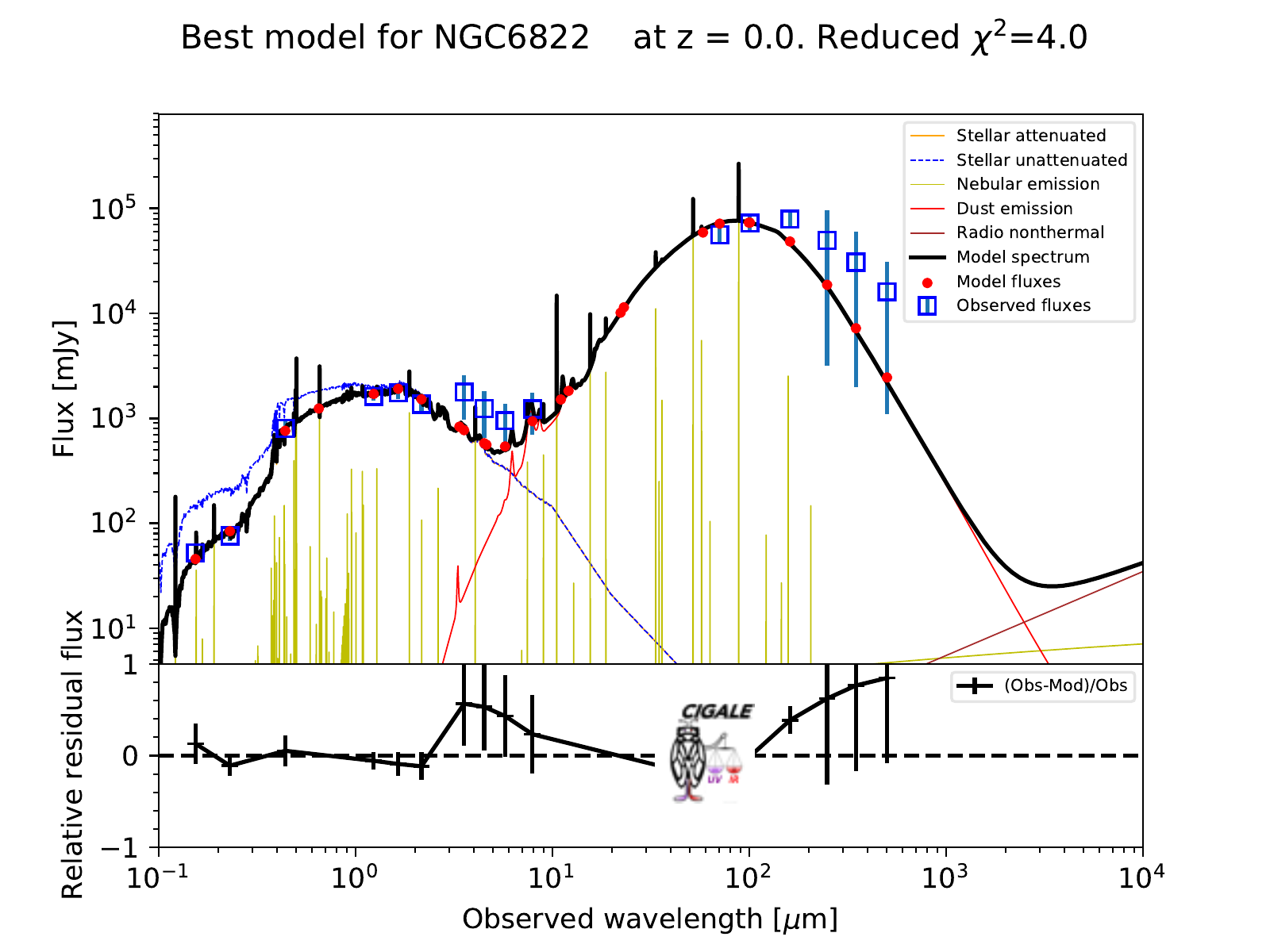}
%\caption{}
         \label{Fig:NGC6822}
   \end{figure} 
   
      \begin{figure}
\centering
\includegraphics[scale=0.5]{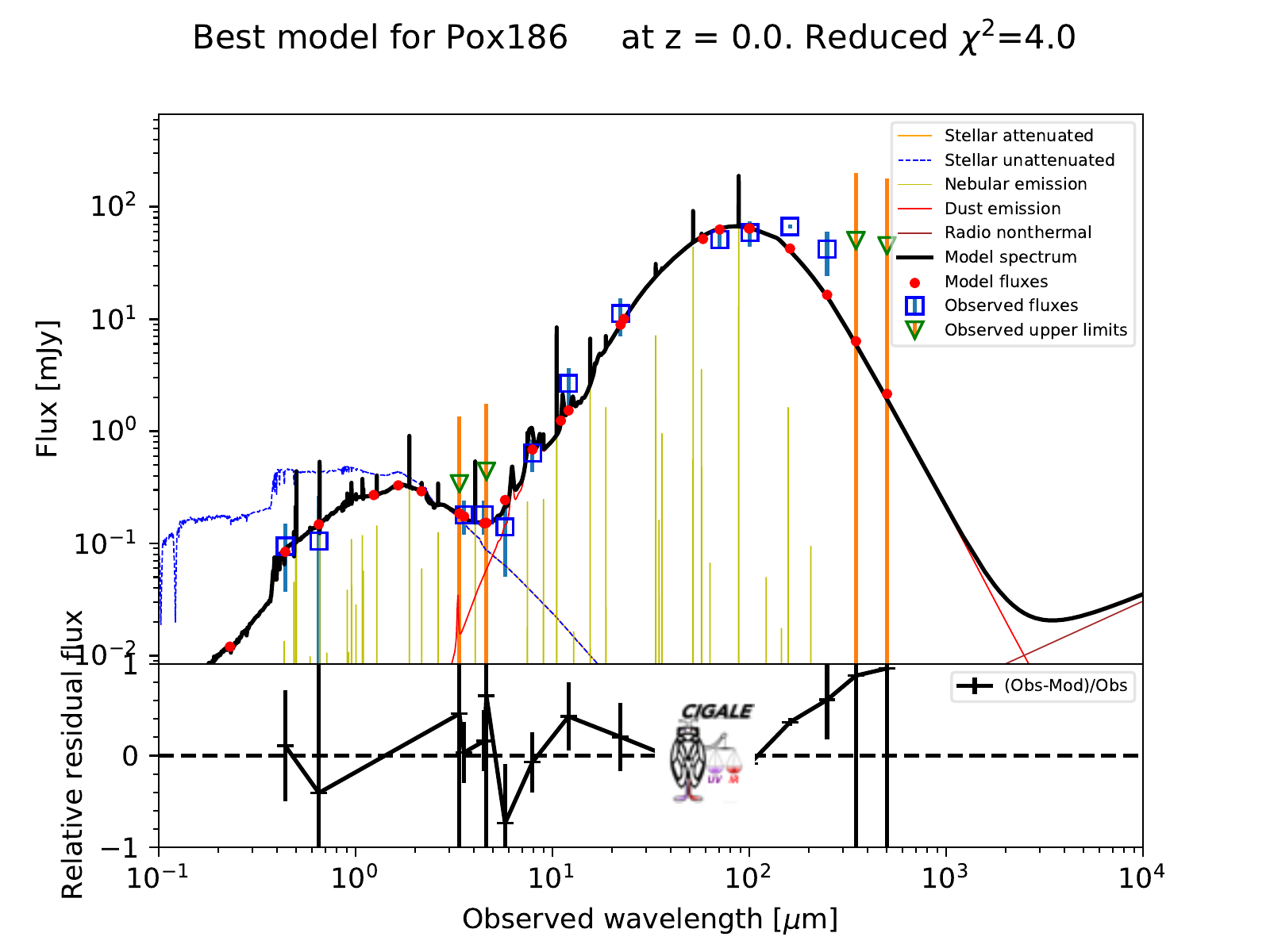}
%\caption{}
         \label{Fig:Pox186}
   \end{figure} 
   
      \begin{figure}
\centering
\includegraphics[scale=0.5]{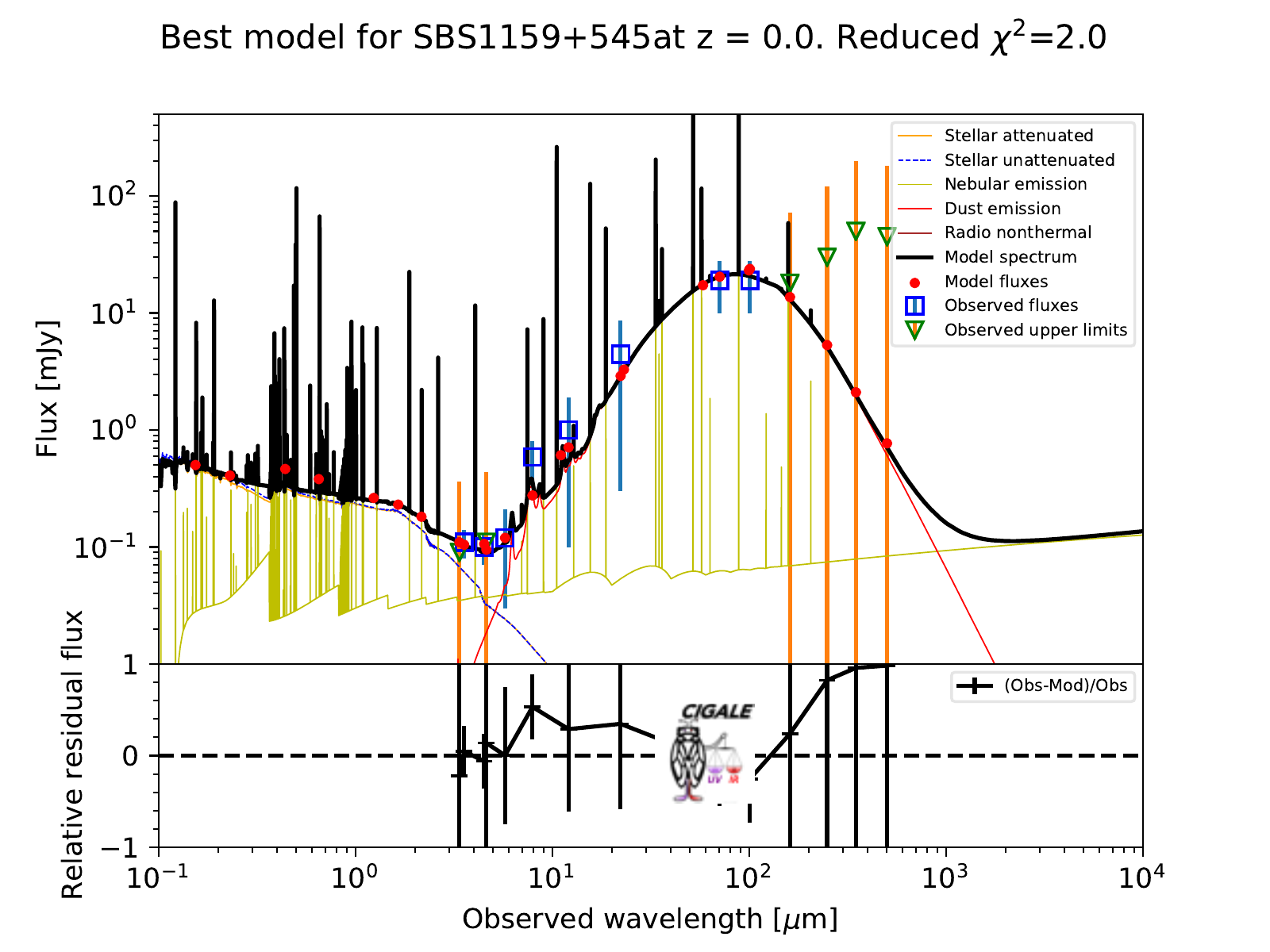}
%\caption{}
         \label{Fig:SBS1159+545}
   \end{figure} 

      \begin{figure}
\centering
\includegraphics[scale=0.5]{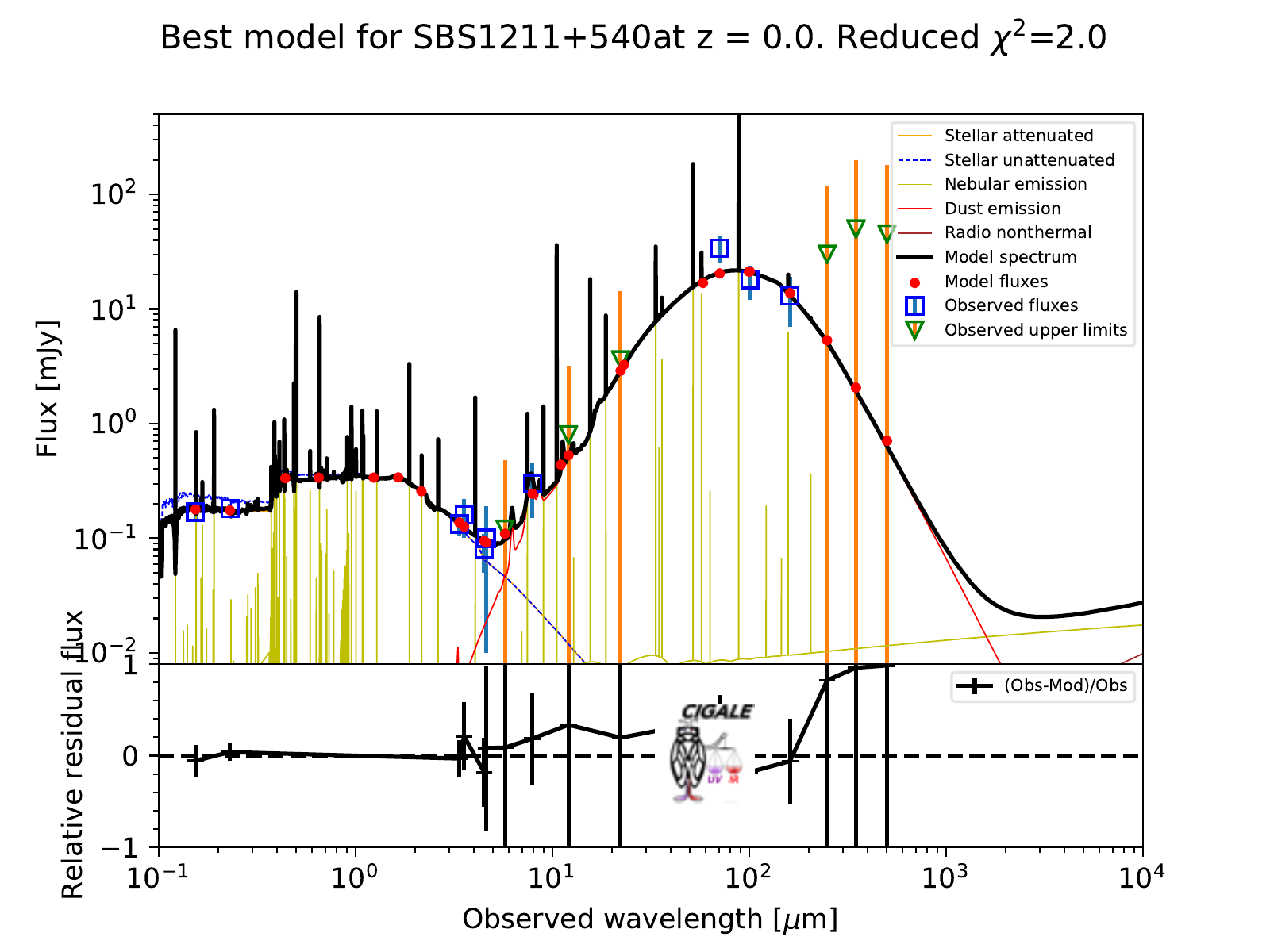}
%\caption{}
         \label{Fig:SBS1211+540}
   \end{figure} 

      \begin{figure}
\centering
\includegraphics[scale=0.5]{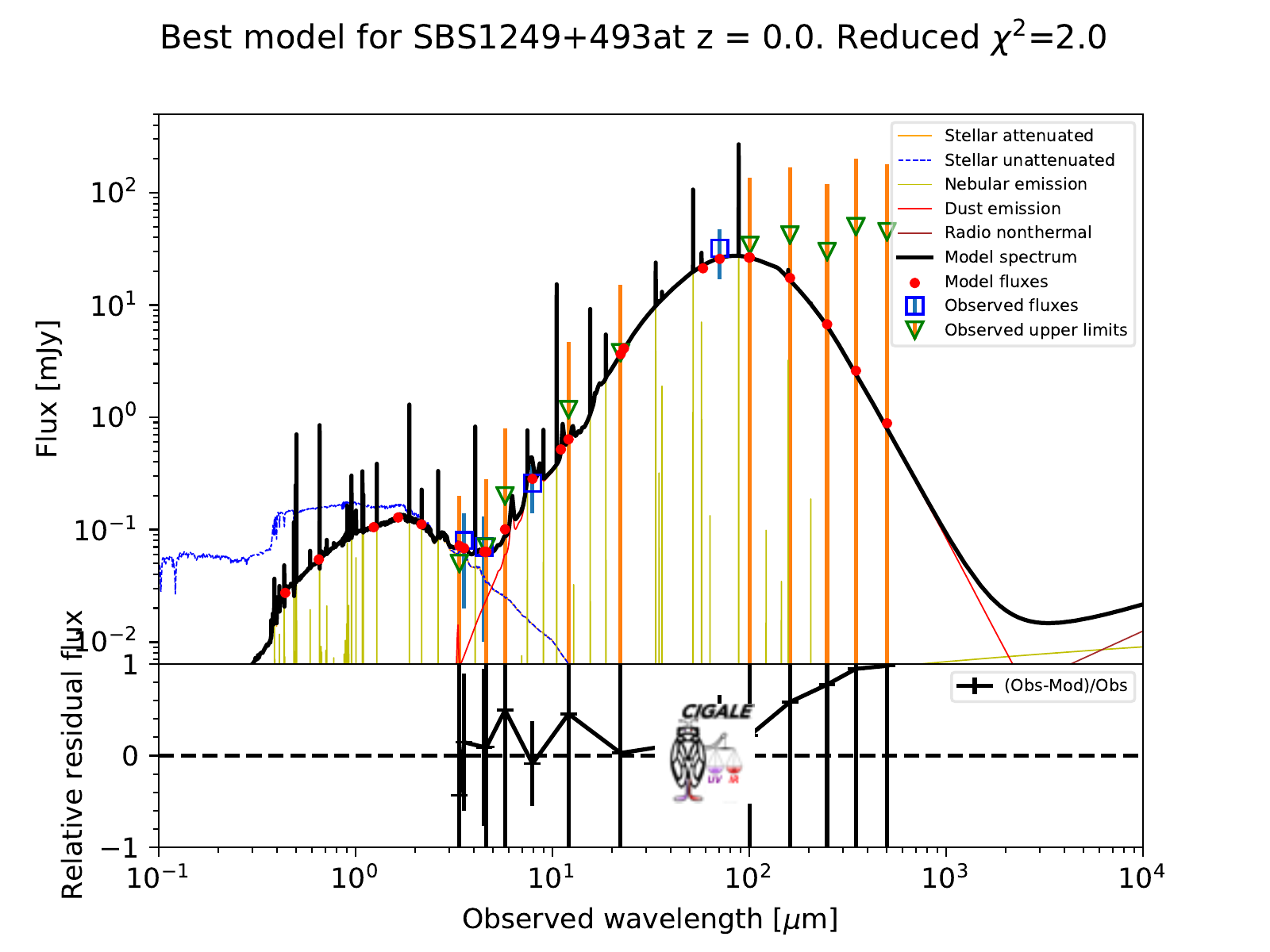}
%\caption{}
         \label{Fig:SBS1249+493}
   \end{figure} 
   
      \begin{figure}
\centering
\includegraphics[scale=0.5]{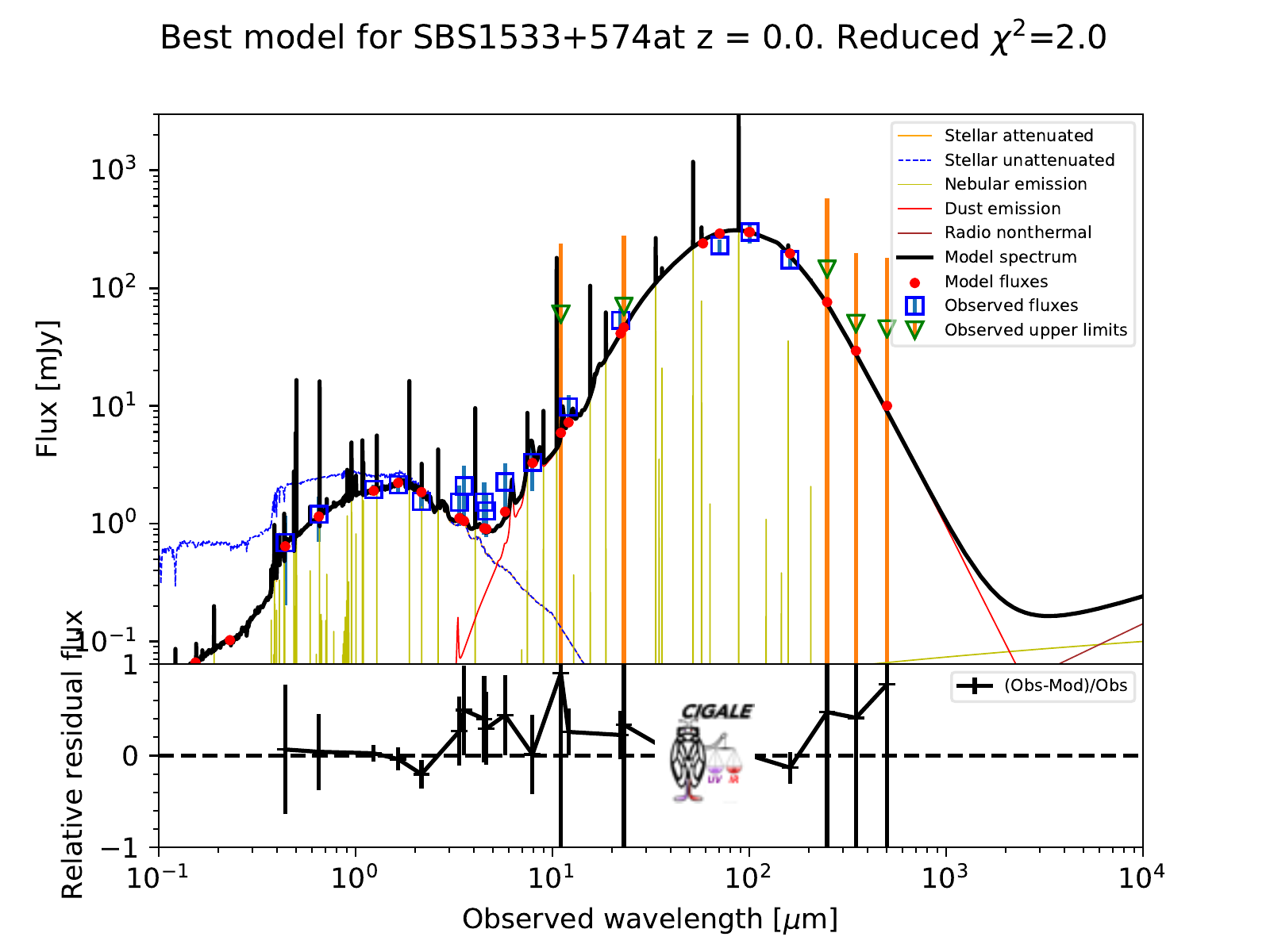}
%\caption{}
         \label{Fig:SBS1533+574}
   \end{figure} 

      \begin{figure}
\centering
\includegraphics[scale=0.5]{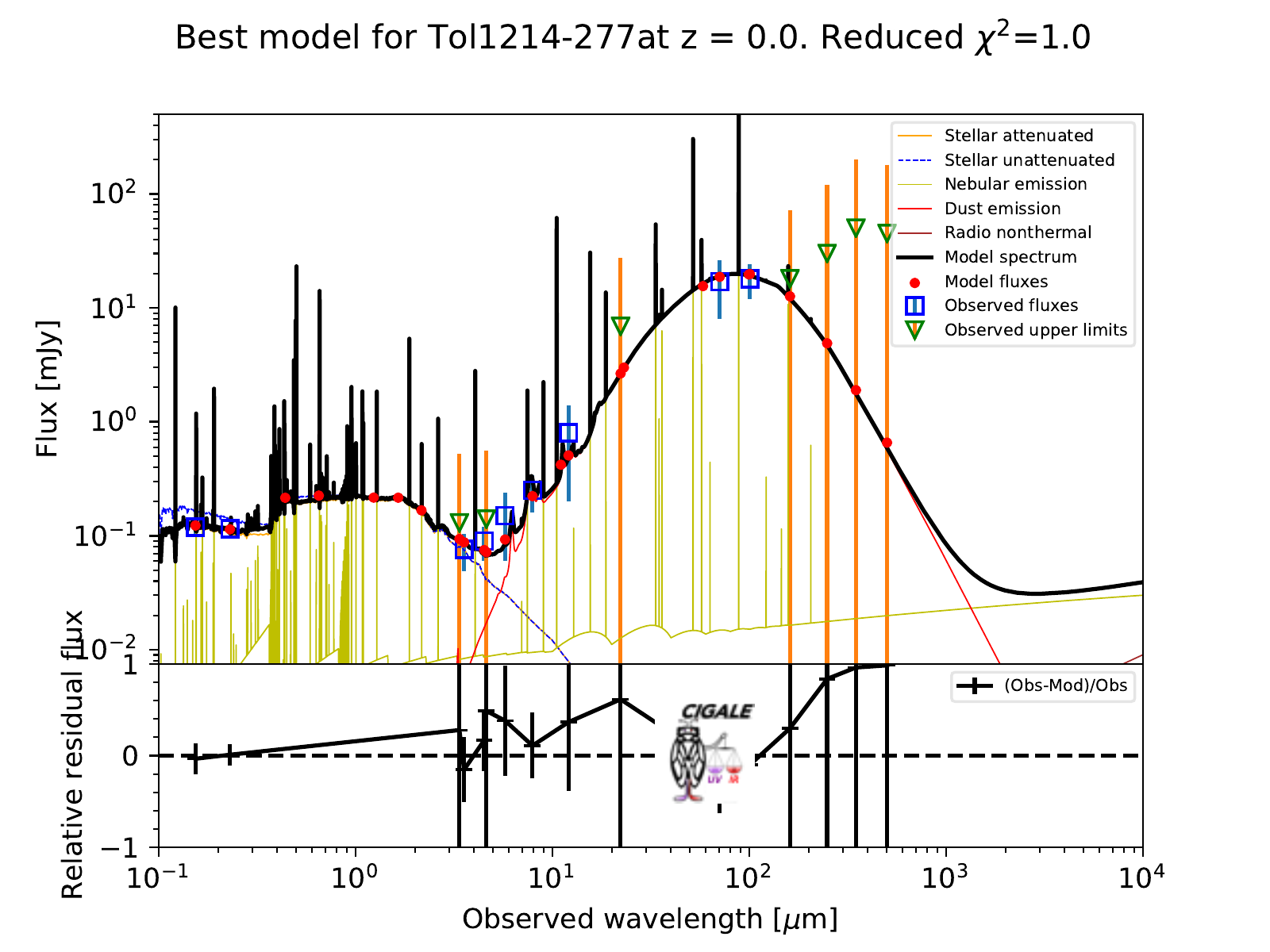}
%\caption{}
         \label{Fig:Tol1214-277}
   \end{figure}

      \begin{figure}
\centering
\includegraphics[scale=0.5]{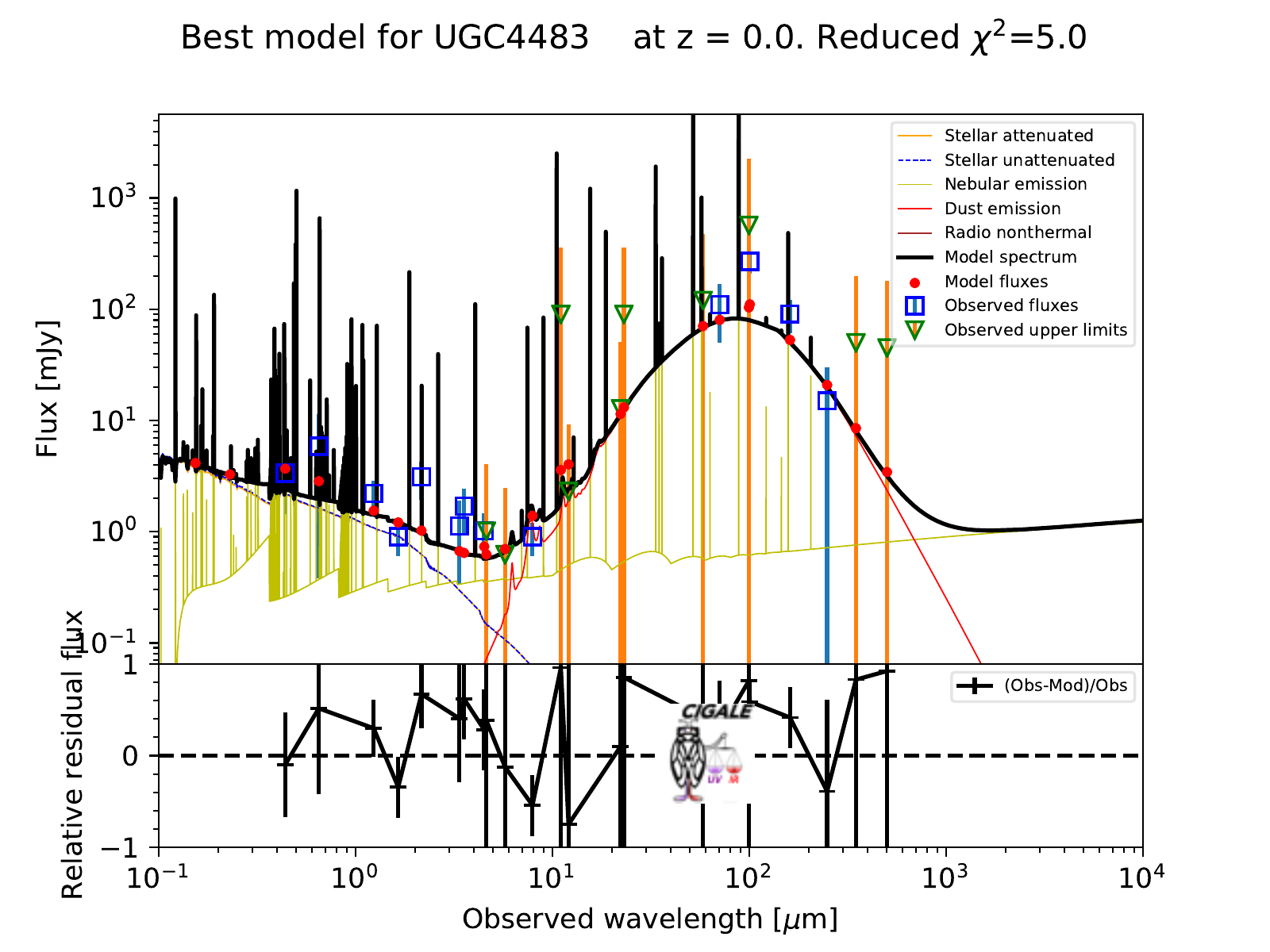}
%\caption{}
         \label{Fig:UGC4483}
   \end{figure} 
   
      \begin{figure}
\centering
\includegraphics[scale=0.5]{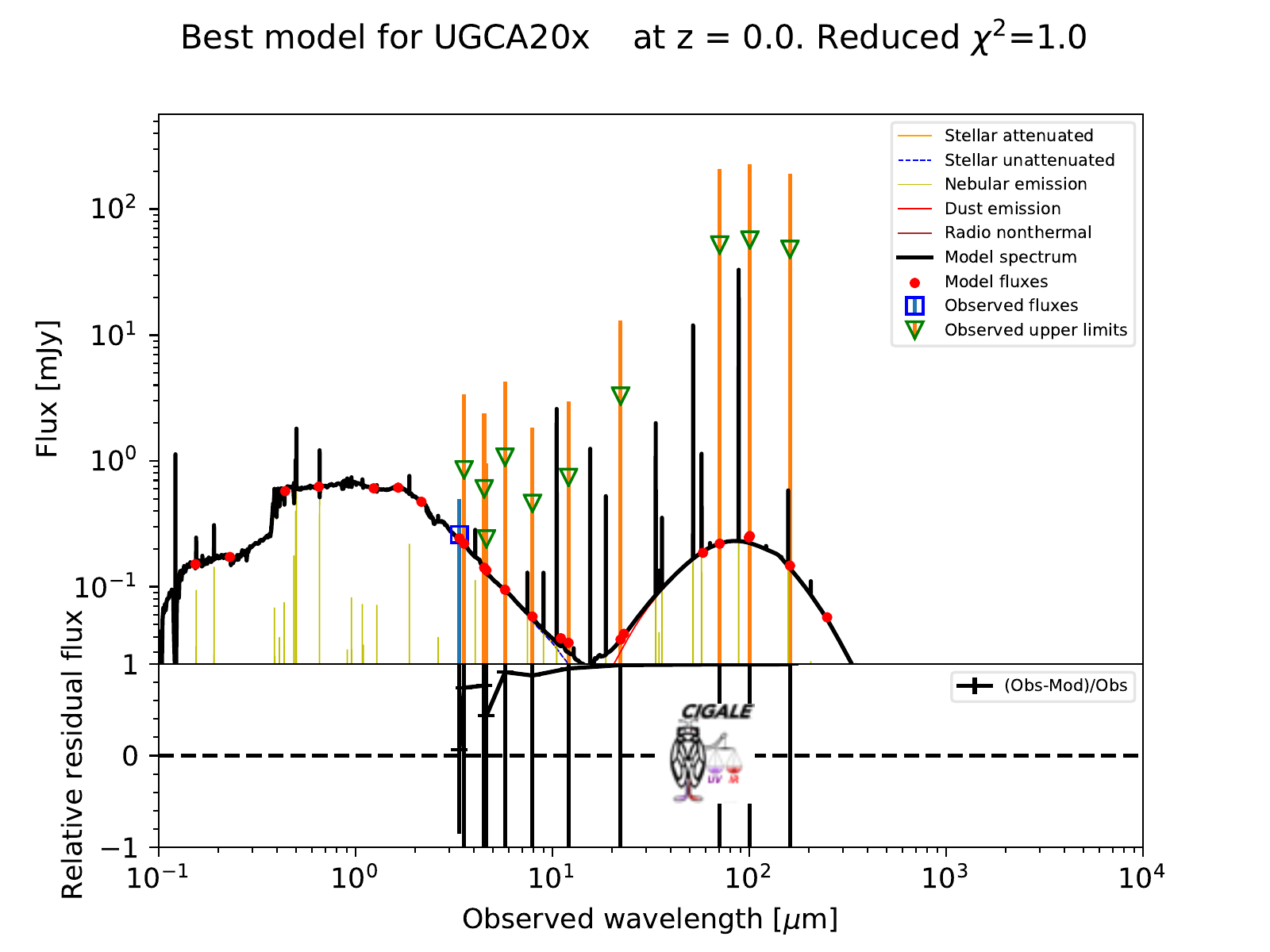}
%\caption{}
         \label{Fig:UGCA20}
   \end{figure} 
  
         \begin{figure}
\centering
\includegraphics[scale=0.5]{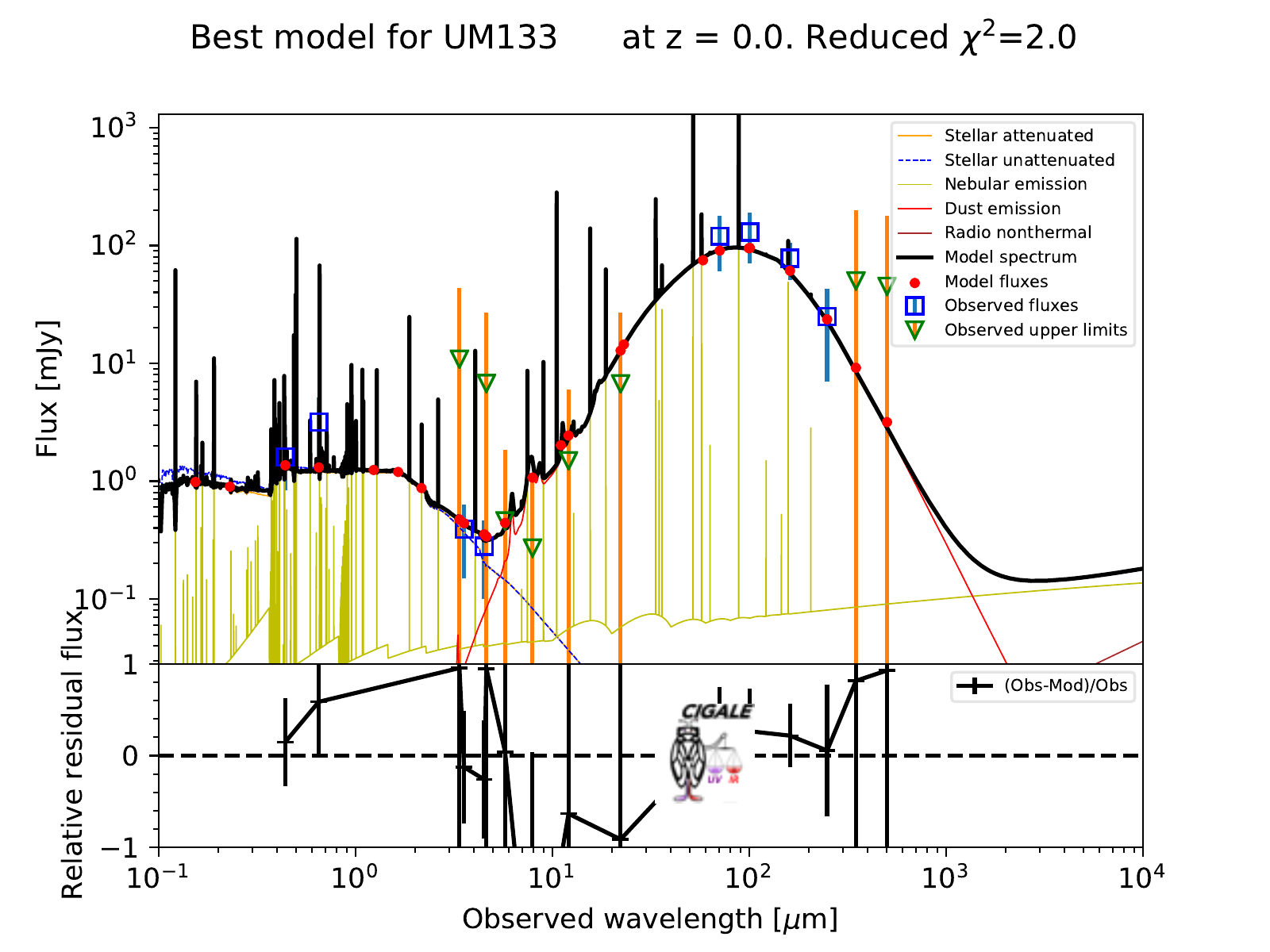}
%\caption{}
         \label{Fig:UM133}
   \end{figure} 
         \begin{figure}
\centering

\includegraphics[scale=0.5]{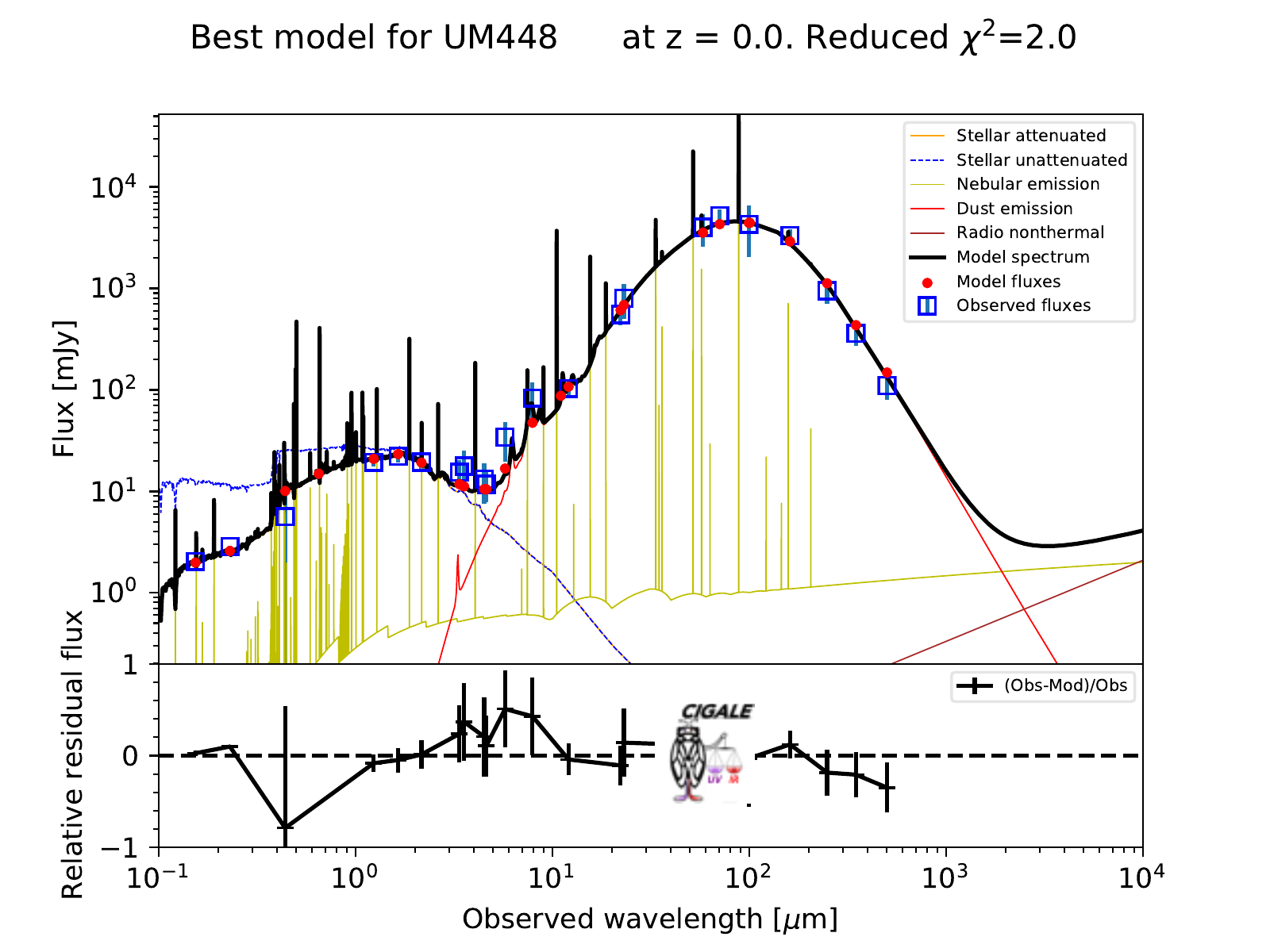}
%\caption{}
         \label{Fig:UM448}
   \end{figure} 
         \begin{figure}
\centering
\includegraphics[scale=0.5]{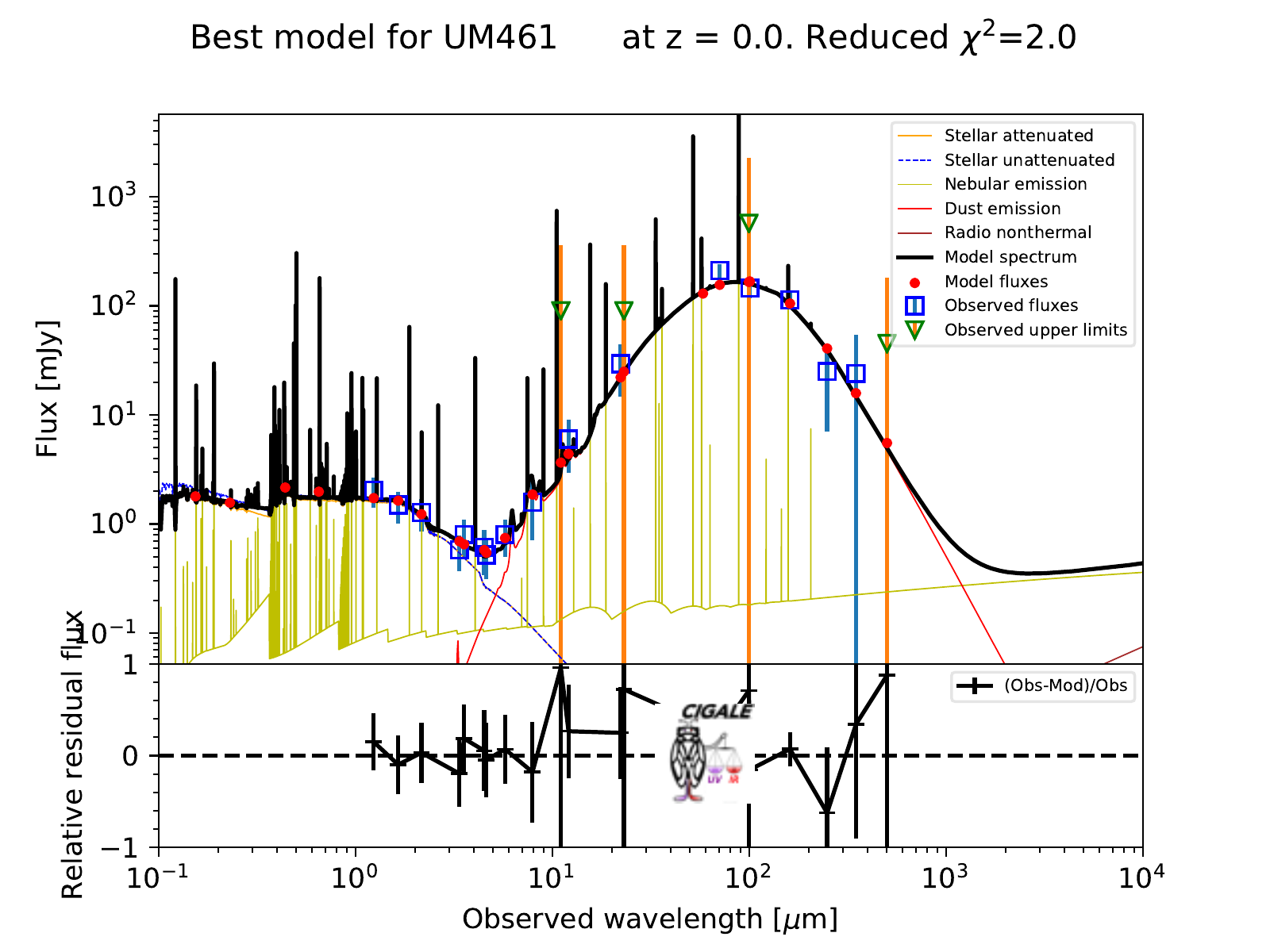}
%\caption{}
         \label{Fig:UM461}
   \end{figure} 
\label{lastpage}
\end{document}